\documentclass[12pt,a4paper]{article}
\usepackage[a4paper]{geometry}
\usepackage[english]{babel}
\usepackage[cp850]{inputenc}
\usepackage[dvips]{graphicx}
\usepackage{amsmath,amsfonts,amsthm,amssymb}
\usepackage{mathrsfs}
\usepackage[usenames, dvipsnames]{color}
\usepackage{xspace}
\usepackage[colorlinks=true,citecolor=blue,linkcolor=blue,urlcolor=blue,pdfstartview=FitH]{hyperref}
\usepackage{nameref}
\usepackage{enumitem}

\usepackage{algorithm}
\usepackage{algorithmicx}
\usepackage{algpseudocode}
\usepackage{caption}
\usepackage{geometry}
\usepackage{tikz,tkz-tab}
\usetikzlibrary{fit,calc,positioning,trees}
\tikzstyle{level 1}=[level distance=4cm, sibling distance=2.5cm]
\tikzstyle{level 2}=[level distance=8cm, sibling distance=0.6cm]
\tikzstyle{bag} = [text width=4em, text centered]
\tikzstyle{end} = [circle, minimum width=3pt,fill, inner sep=0pt]
\usetikzlibrary{decorations.pathmorphing}
\usetikzlibrary{decorations.pathreplacing}
\usetikzlibrary{decorations.shapes}
\usetikzlibrary{decorations.text}
\usetikzlibrary{decorations.markings}
\usetikzlibrary{decorations.fractals}
\usetikzlibrary{decorations.footprints}

\graphicspath{{images/}}

\DeclareGraphicsExtensions{.pdf,.png}

\providecommand{\keywords}[1]{\textbf{Keywords:} #1}

\makeatletter
\newcommand{\manuallabel}[2]{\def\@currentlabel{#2}\label{#1}}
\makeatother
\newcommand\numberthis{\addtocounter{equation}{1}\tag{\theequation}}

\algnewcommand{\Inputs}[1]{%
  \State \textbf{Inputs:}
  \Statex \hspace*{\algorithmicindent}\parbox[t]{.8\linewidth}{\raggedright #1}
}
\algnewcommand{\Initialize}[1]{%
  \State \textbf{Initialize:}
  \Statex \hspace*{\algorithmicindent}\parbox[t]{.8\linewidth}{\raggedright #1}
}

\usepackage{amsmath}
\DeclareMathOperator*{\argmax}{arg\,max}

\def\Esp{\mathbb{E}}

\def\é{\'{e}}
\def\è{\`{e}}
\def\ê{\^{e}}
\def\à{\`{a}}
\def\ô{\^{o}}
\newtheorem{theo}{Theorem}
\newtheorem{prop}{Proposition}
\newtheorem{lem}{Lemma}
\newtheorem{Assumption}{Assumption}

\newtheorem{defi}{Definition}
\newtheorem{rem}{Remark}

\title{Optimal liquidity-based trading tactics}
\author{Charles-Albert Lehalle\thanks{Capital Fund Management, Paris and Imperial College, London}{}, Othmane Mounjid\thanks{\'{E}cole Polytechnique, CMAP} ~%
  and %
  Mathieu Rosenbaum\footnotemark[2]}
\date{\today\\}
\begin{document}
\maketitle

\begin{abstract}
\noindent
We consider an agent who needs to buy (or sell) a relatively small amount of asset over some fixed short time interval. We work at the highest frequency meaning that we wish to find the optimal tactic to execute our quantity using  limit orders, market orders and cancellations. To solve the agent's control problem, we build an order book model and optimize an expected utility function based on our price impact. We derive the equations satisfied  by the optimal strategy and solve them numerically. Moreover, we show that our optimal tactic enables us to outperform significantly naive execution strategies. 
\end{abstract}



\section{Introduction}
\label{sec:Introduction}
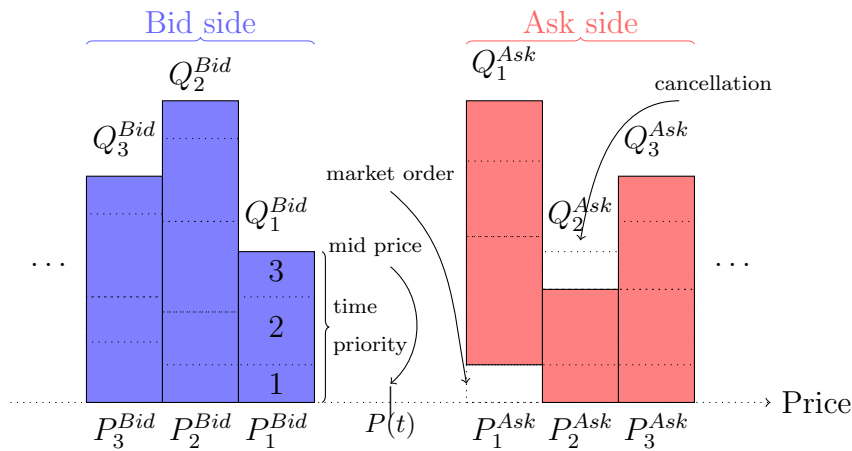
\begin{figure}[H]
\centering
\begin{tikzpicture}
\draw [fill=blue!50] (1,-2) rectangle (2,-5);
\draw  [dotted](1,-2.5) rectangle (2,-3.6);
\draw  [dotted](1,-3.6) rectangle (2,-4.2);
\draw [fill=blue!50](2,-1) rectangle (3,-5);
\draw  [dotted](2,-1.5) rectangle (3,-2.6);
\draw  [dotted](2,-2.6) rectangle (3,-3.8);
\draw  [dotted](2,-3.8) rectangle (3,-4.5);
\draw  [fill=blue!50](3,-3) rectangle (4,-5);
\draw  [dotted](3,-3.6) rectangle (4,-4.5);
\draw (3.5,-4.75) node[sloped]{1};
\draw (3.5,-4) node[sloped]{2};
\draw (3.5,-3.25) node[sloped]{3};
\draw[decorate,decorate,decoration={brace}](4.1,-3) -- ++(0,-2);
\draw (4.1,-4) node[right,align=left] {\scriptsize time \normalsize\\ 
\scriptsize priority \normalsize};
\draw [fill=red!50](6,-1) rectangle (7,-4.5);
\draw  [dotted](6,-1.8) rectangle (7,-2.8);
\draw  [dotted](6,-2.8) rectangle (7,-4.5);
\draw  [dotted](6,-4.5) rectangle (7,-5);
\draw [->](5,-2.2) to[out=-40,in=100] (6,-4.75);
\draw (5,-2.2) node[above]{\scriptsize market order \normalsize} ;
\draw [fill=red!50](7,-3.5) rectangle (8,-5);
\draw  [dotted](7,-3.5) rectangle (8,-4.5);
\draw  [dotted](7,-3) rectangle (8,-3.5);
\draw [->](8.8,-1) to[out=180,in=80] (7.5,-2.85);
\draw (9.25,-1) node[above]{\scriptsize cancellation \normalsize} ;
\draw [fill=red!50](8,-2) rectangle (9,-5);
\draw  [dotted](8,-2.6) rectangle (9,-3.5);
\draw  [dotted](8,-3.5) rectangle (9,-4.5);
\draw[color=blue!60,decorate,decorate,decoration={brace}]
(1,-0.25) -- ++(3,0) node[above,pos=0.5] {Bid side};
\draw[color=red!60,decorate,decorate,decoration={brace}]
(6,-0.25) -- ++(3,0) node[above,pos=0.5] {Ask side};
\draw [dotted][->](0,-5) -- ++(10,0);
\draw [->](5,-3.2) to[out=-40,in=40] (5,-4.75);
\draw (5,-5) node[sloped]{$|$};
\draw (4.8,-3.2) node[above]{\scriptsize mid price \normalsize} ;
\draw (3.5,-5) node[below]{$ P_1^{Bid} $} ;
\draw (2.5,-5) node[below]{$ P_2^{Bid}  $} ;
\draw (1.5,-5) node[below]{$ P_3^{Bid}  $} ;
\draw (0.5,-3) node[below]{$ \ldots  $} ;
\draw (6.5,-5) node[below]{$ P_1^{Ask}  $} ;
\draw (7.5,-5) node[below]{$ P_2^{Ask}  $} ;
\draw (8.5,-5) node[below]{$ P_3^{Ask}  $} ;
\draw (9.5,-3) node[below]{$ \ldots  $} ;
\draw (6.5,-0.5) node {$Q_1^{Ask}$} ;
\draw (7.5,-2.5) node {$Q_2^{Ask}$} ;
\draw (8.5,-1.5) node {$Q_3^{Ask}$} ;
\draw (3.5,-2.5) node {$Q_1^{Bid}$} ;
\draw (2.5,-0.65) node {$Q_2^{Bid}$} ;
\draw (1.5,-1.5) node {$Q_3^{Bid}$} ;
\footnotesize
\draw (5,-5) node[below]{$ P(t) $} ;
\normalsize
\draw (10,-5) node[right]{Price} ;
\end{tikzpicture}
\caption{Order book representation at a given time. Here $P^{Ask}_i$ (resp. $P^{Bid}_i$) with $i\geq1$ are the sellers (resp. buyers) limit prices and they are increasingly (resp. decreasingly) ordered. For a given price $P^{Ask}_i$ (resp. $P^{Bid}_i$), the limit $Q^{Ask}_i$ (resp. $Q^{Bid}_i$) is the available selling (resp. buying) quantity.}
\label{fig:RepOrderBook}
\end{figure}

Most electronic exchanges use an order book mechanism. In such markets, buyers and sellers send their orders to a continuous-time double auction system. These orders are then matched according to price and time priority. Every submitted order has a specific price and size and the order book is the collection of all submitted and unmatched limit orders. This is illustrated in Figure \ref{fig:RepOrderBook}, which shows a classical representation of an order book at a given time.\\
In this limit order book setting, we consider the following issue: an agent has to buy or sell a given quantity of asset before a fixed horizon time. During the execution process, the agent can take four elementary decisions:
\begin{itemize}
\item Insert limit orders in the order book, hoping to get execution at the best ask/bid (we will assume that the agent does not place limit orders above the best limits). 
\item Stay in the order book with an already existing limit order, to keep his tactical placement.
\item Cancel existing limit orders.
\item Send market orders to get immediate execution.
\end{itemize}
Note that this is the microstructural version of the classical Almgren-Chriss optimal scheduling problem for the liquidation of a large quantity of asset over a time interval $[0,T] $, see \cite{almgren2000optimal,RePEc:eee:finmar:v:1:y:1998:i:1:p:1-50,Grinold:1110267} and \cite{cartea15book,gueant2016financial} for various extensions. In the setting of \cite{almgren2000optimal}, $[0,T]$ is split in sub time windows (typically a few minutes per window) and one derives the number of shares to be executed in each window. In our case, we want to specify how to act optimally within each window. Indeed, the buyer or seller reacts to every order book move and handles reasonably small quantities during short periods of time.\\

In order to solve this problem, we of course need to model the order book dynamic. There are essentially two order book modelling approaches in the litterature. First, ``general equilibrium models", based on interactions between rational agents who take optimal decisions, see \cite{RePEc:eee:finmar:v:2:y:1999:i:2:p:99-134,citeulike:1204538,Rosu2009}. Second, ``statistical models" where the order book is seen as a suitable random process, see \cite{citeulike:10868255,LOBModHawkes,BayerHorstQiu2015,citeulike:8531765,citeulike:7341957,citeulike:12386824,citeulike:12514252,farmer03a}. Statistical models focus on reproducing many salient features of real markets rather than individual agents behaviours and interactions between them. In this paper, we use a statistical model. In such models, the arrival and cancellation flows often follow independent Poisson processes. The Poisson assumption allows for the derivation of simple, and often closed form, formulas, for example for the probabilities of various order book events, see \cite{citeulike:10868255,citeulike:7341957,citeulike:12386824,citeulike:13530453}.\\

However, as clearly shown in \cite{citeulike:12810809}, this assumption is not realistic and it is necessary to take into account accurately the local state dependent behaviour of the order book. So in \cite{citeulike:12810809,citeulike:13675327}, the authors introduce the Queue-Reactive order book  model where order flows follow a Markov jump process. They also provide ergodicity conditions and model parameters calibration methodology. Here we refine this model to make it compatible with a stochastic control framework enabling us to solve important practical issues. To do so, we only consider the best bid and ask limits to work with a reasonably small state space. Furthermore, in order to get a truly good fit of the order book dynamic, we focus on the so-called regeneration process which models the order book state right after the total depletion of a limit. Indeed, in our setting, when one limit is totally depleted, the order book is regenerated in a new state whose regeneration law depends on the order book state just before the depletion. In general, order book models consider several bid and ask limits and use a regeneration process independent from the order book state, see \cite{LOBModHawkes,citeulike:7341957,citeulike:12810809}. Here, we model the order book by a three-dimensional Markov jump process  $\left(Q^{1}_t, Q^{2}_t, P_t\right)$  where $Q^{1}_t$ is the available quantity at the best bid, $Q^{2}_t$ is the available quantity at the best ask and $P_t$ is the mid price. Furthermore, we focus on large tick assets (see \cite{RePEc:arx:papers:1207.6325}) and so we fix the spread as a constant\footnote{Note that this assumption can be relaxed by enlarging the state space.}.\\
In this work, we deal with an optimal execution problem. Actually, the question of what a good execution means is not trivial. This is because it is difficult to define a suitable benchmark price. Indeed,  agents need a benchmark to compare it with the execution price of their own trading strategy. In our work, we place ourselves in a setting where we can define a notion of long term value of the price $P_{\infty} = \underset{t \rightarrow \infty}{\lim} P_t$. We use it as a benchmark since it represents the asset future value. In practice, if the agent is able to buy the asset at a price lower than $P_{\infty}$, he can, in principle, make profit by selling it back in the future.\\

Let us now introduce the agent's control problem. We express it for a buy order of size $q^a$ (it can be changed to a sell order in an  obvious way). From time zero to the final time $T$, we assume that, at every decision time, the buyer can do nothing or use one of the three following actions: insert the remaining quantity to buy (if not already inserted) at the top of the bid queue (decision l), cancel the already inserted limit orders (decision c) or send a market order (decision m). We suppose that actions involve the whole inventory of the agent. It will not be possible to apply a decision to part of the inventory and another to the remaining part. If the agent does not obtain the total execution of $q^a$ at time $T$, he cancels the remaining quantity in the order book and send a market order. The agent aims  at determining the optimal sequence of decisions to outperform the benchmark $P_{\infty}$.\\

Let $t$ be the current observation time, $\mu_t$ the agent's control and $I^{\mu}_t$ the agent's inventory, that is the remaining quantity he has to buy at time $t$. We view the agent's control $\mu = \{\mu_t, t \leq T \}$ as a process valued in $\{l,c,m\}$ which  remains constant when the user does nothing\footnote{If the agent has no order in the order book and does nothing at the beginning of the period, we consider he starts with control $c$.}. Recall that we need to handle the agent's inventory since we may have a partial execution of $q^a$ but the agent does not do any splitting of $q^a$\footnote{This could also be relaxed by enlarging the dimension of the control.}. The benchmark being $P_\infty$, the agent wishes to get the quantity $\Esp \big[ P_{\infty} - P^{Exec,\mu}_{T^{\mu}_{Exec}}\big]$ high, where $P^{Exec,\mu}_t$ is the acquisition price of the quantity $q^a- I^{\mu}_t$ and $T^{\mu}_{Exec}$ is the time where $q^a$ is totally executed. An important point is that in our setting, our trading has an impact on prices. In particular, trying to get $P^{Exec,\mu}_{T^{\mu}_{Exec}}$ small means that we want to minimize our transient market impact, that is the price impact of our trading during the execution. Note that our trading also has an influence on  $P_{\infty}$ that we will be able to compute and therefore we write $P^{\mu}_{\infty}$ instead of $P_{\infty}$.
To take into account the waiting cost, the sensitivity to the price impact and to work in a slightly more general setting, we consider the following optimisation problem:
\begin{equation*}
\underset{\mu}{\sup}\, \Esp \big[ f\big(\Esp \big[ P^{\mu}_{\infty} - P^{Exec,\mu}_{T^{\mu}_{Exec}}/ \mathcal{F}_{T^{\mu}_{Exec}} \big]\big) - c q^a T^{\mu}_{Exec}\big],
\end{equation*}
where $f: \mathbb{R}\rightarrow \mathbb{R}$ is a Lipschitz function and $c$ is an homogenization non-negative constant representing the waiting cost. Note that we use a conditional expectation to account for the fact that agents collect information along their own trading.\\

We study this problem in two cases. First, when agent's decisions are taken at fixed frequency $\Delta^{-1}$. This enables us to investigate latency effects and moderately high frequency trading issues. Second, when agent's decisions are taken at any time, to handle the situation where one has access to ultra high frequency trading technology. \\

Note that this paper is obviously not the first work where a stochastic control framework involving limit orders, market orders and cancellations is used to solve a high frequency trading problem. For example, in \cite{citeulike:10160160,DiscrOptTrad} the authors consider the problem of optimal posting of a limit order while market making issues are adressed in \cite{avellaneda2008high,cartea15book,citeulike:9304794,guo2017optimal}. However, to our knowledge, this is the first approach where the interactions between market participants decisions, liquidity and behaviour of the order book are accurately taken into account, see also the complementary paper \cite{OptInventManageLOB}.\\

The paper is organized as follows. We introduce our order book model in Section \ref{sec:LOBmodel}. In Section \ref{sec:dpp}, we formulate the agent's control problem. Our main results including the computation of $P_{\infty}$ and the equations satisfied by the value functions are provided in Section \ref{sec:TheoRes}. Finally, the numerical methodology to solve these equations is given in Section \ref{sec:understanding}. The proofs and additional results are relegated to an appendix.
\section{Order book modelling}
\label{sec:LOBmodel}
In this section, we first confirm on data that agents behaviours depend on order book liquidity, see \cite{citeulike:12810809,DiscrOptTrad,SignOptTrad} for closely related results. Then, we describe the order book dynamic.
\subsection{Preliminary: Empirical evidences}
\manuallabel{subsec:EmpEvid}{2.1}
One specificity of our work is that we wish to carefully model the interactions between market participants and liquidity. We first show on real data that market participants act differently when facing different liquidity conditions. 
\paragraph{Database presentation.} Data used here are from Bund futures on Eurex exchange Frankfurt. We focus on Bund futures since they are a good example of a very liquid and large tick asset. The database records, during one week from 1 to 5 September 2014, the state of the order book (i.e available quantities and prices at best limits) event by event with microsecond accuracy. For each day, our data cover the time period from 8 a.m to 10 p.m Frankfurt time. Each event has a type, a side (i.e bid/ask) and a size. We consider three types of events: insertion of limit orders, cancellation of existing limit orders and market orders. The database accounts for 3 407 574 events.\\

Let $t$ be the time where an event happens in the order book. We define the imbalance $\text{Imb}_t$ and the mid price move $\delta$ seconds after the event time $t$, $\Delta P^{mid}_{\delta }(t)$, by
\footnotesize
\begin{equation*}
\left\{
\begin{array}{ll}
\text{Imb}_t =  \epsilon_t \cfrac{Q^{1}_{t}-Q^{2}_{t}}{Q^{1}_{t}+Q^{2}_{t}}, &\\ 
\Delta P^{mid}_{\delta }(t) = \epsilon_t \cfrac{P_{\delta +t} - P_{t}}{\psi_t}, & \\
\end{array}
\right.
\label{Eq:ImbAfterExec}
\end{equation*}
\normalsize
where $Q^{1}_{t}$ (resp. $Q^{2}_{t}$) is the available quantity at the best bid (resp. ask), $P_t$ is the mid price, $\epsilon_t$ is the event sign (i.e $\epsilon_t = 1$ when it is a buy order and -1 otherwise) and $ \psi_t$ is the spread (i.e $\psi_t = P^{Ask}_t - P^{Bid}_t$ with $P^{Ask}_t$ the best ask price and $P^{Bid}_t$ the best bid price).\\

\begin{figure}[H]
  \centering
  \begin{tabular}{c}
  \hfill (a) Average imbalance before\\
  Limit/Cancel/Market order \\
  \end{tabular} \hfill \hfill \hfill \hfill \hspace{3cm}
  \begin{tabular}{c}
  (b) Imbalance density before\\
  liquidity provision/consumption event\\
  \end{tabular}  \hfill~\\
  \includegraphics[width=.4\linewidth]{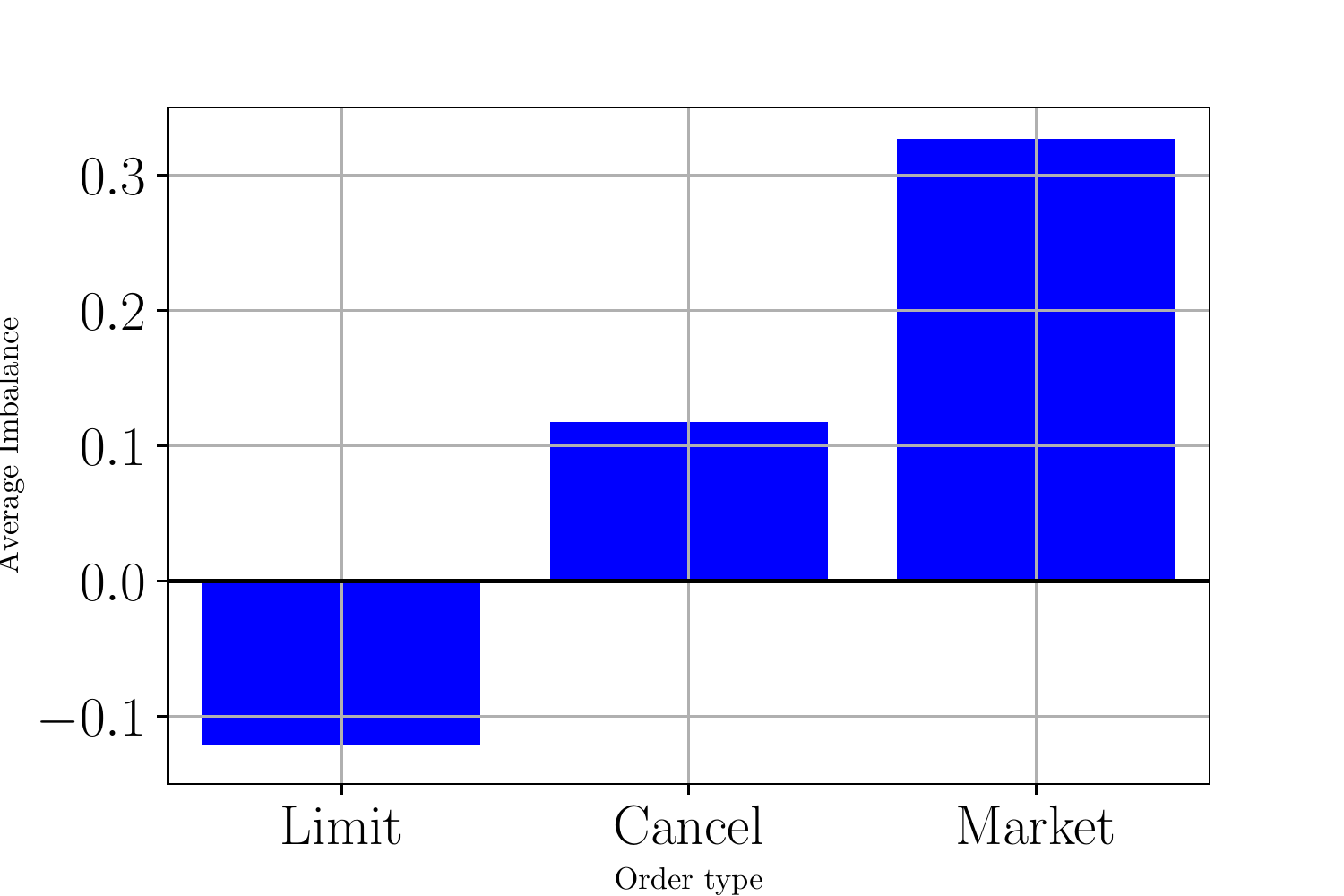}\hfill
  \includegraphics[width=.4\linewidth]{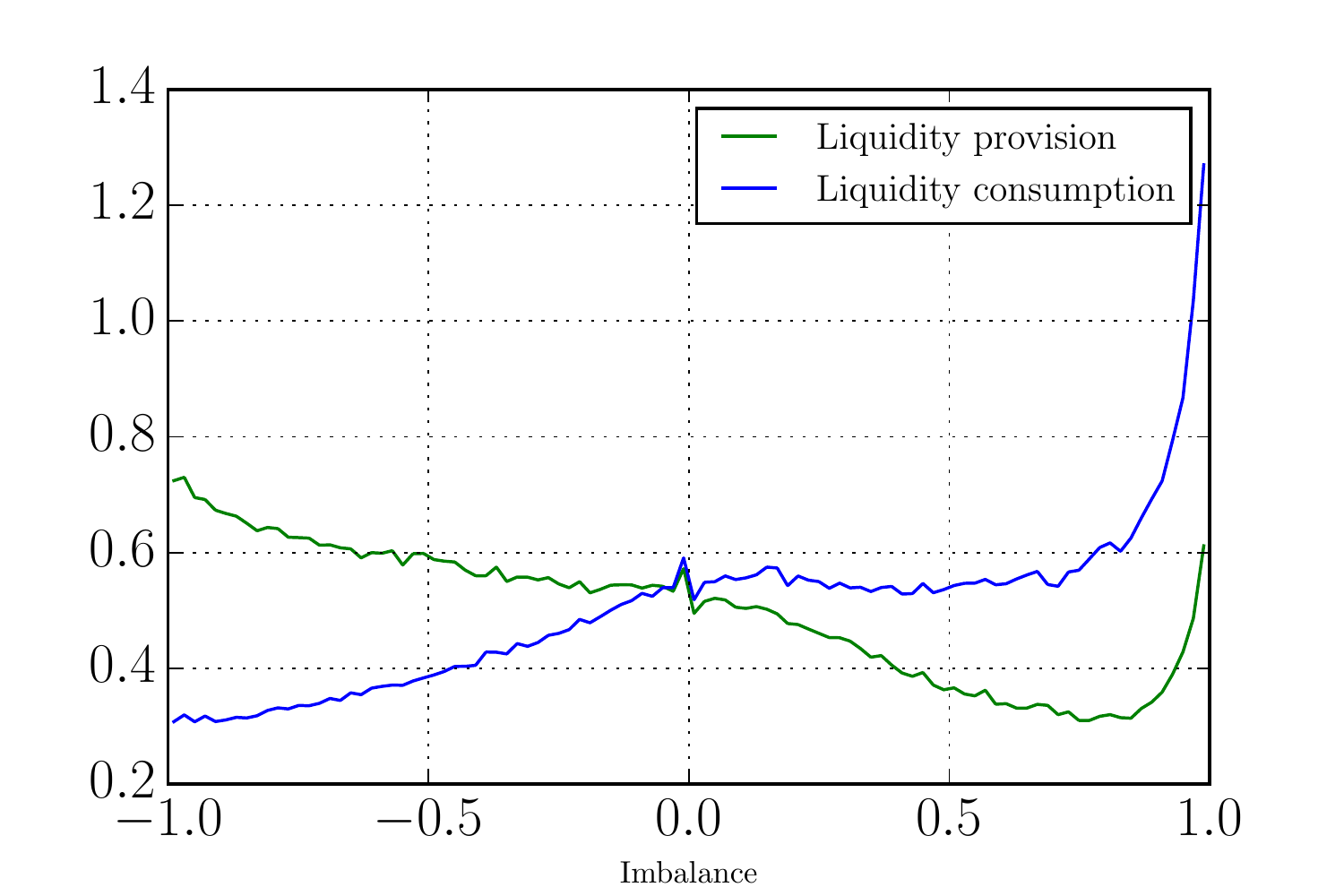}\\
  \hfill (c) Average price move after 2 minutes against imbalance \hfill~\\
  \includegraphics[width=.4\linewidth]{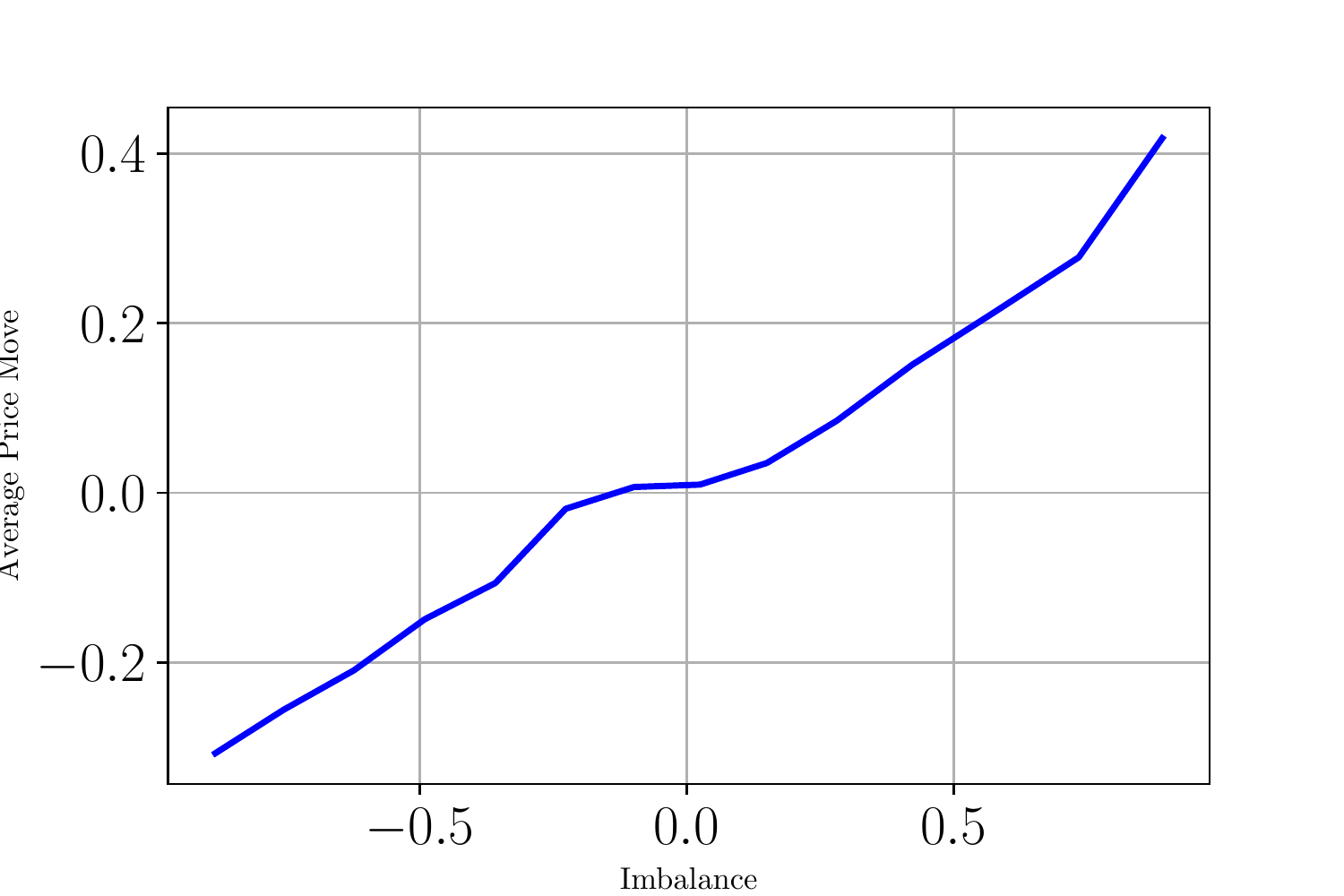}
  \caption{Some statistics about the imbalance.}
  \label{fig:predpowimb}
\end{figure}

We want to confirm that agents decisions depend on the order book liquidity. A simple way to do it is to summarize the state of the order book liquidity through the imbalance. Figure \ref{fig:predpowimb}.a shows the average imbalance value for each event type. We give the interpretation of Figure \ref{fig:predpowimb}.a in the case of a buy limit/cancellation/market order, since the event sign is taken into account in the expression of $ \text{Imb}_t$. We see that market participants insert limit orders when imbalance is negative (execution highly probable), cancel orders when imbalance is positive (less chance to be executed) and use market orders when imbalance is highly positive (rushing for liquidity when it is scarse).\\

Figure \ref{fig:predpowimb}.b shows the distribution of imbalance just before a liquidity provision event (i.e insertion of limit orders) and a liquidity consumption event (i.e cancellation of limit orders or market orders). We see that agents are highly active at extreme imbalance values\footnote{The high rate of liquidity provision for very positive imbalance can be surprising at first sight. However it may be due to orders inserted within the spread creating a new best limit.}. Indeed, in these cases, they identify a profit opportunity to catch or on the contrary an adverse selection effect to avoid (for example buying just before a price decrease). This is related to the predictive power of the imbalance. As can be seen in Figure \ref{fig:predpowimb}.c, $\Delta P^{mid}_{\delta }(t)$ after 2 minutes (i.e $ \delta = 2 \min$) is highly correlated to the imbalance. Hence, market participants use the imbalance as a signal to anticipate next price moves\footnote{Quoting Sasha Stoikov: Imbalance is the least well hidden secret of high frequency trading.}. \\

Hence, our empirical results clearly confirm that agents decisions depend on the order book liquidity.\\
\subsection{Order book framework}
\manuallabel{subsec:LOBPresentation}{2.2}
Let $\left(\Omega, \mathcal{F},(\mathcal{F}_t),\mathbb{P} \right)$ be a filtered probability space with $\mathcal{F}_0 $ the trivial $\sigma$-algebra. The order book state is modelled by the Markov process  $U_t = \left(Q^{1}_t, Q^{2}_t, P_t \right)$  where $Q^{1}_t$ (resp. $Q^{2}_t$) is the best bid (resp. ask) quantity and $P_t$ is the mid price. We focus on large tick assets and fix the spread $\psi$ as a constant.
\paragraph{Order book regeneration.} When one limit is totally depleted, the order book is regenerated in a new state whose law  depends on the order book state just before the depletion and the depleted side (i.e best bid/ask). We denote by $d^{1}_{u}$  (resp. $d^{2}_{u}$) the probability distribution of the regenerated state when the best bid (resp. ask) is totally depleted, with $u\in (\mathbb{N}^{*})^2\times \mathbb{R}$ the order book state before the depletion. For $i \in \{1,2\}$ and $u\in (\mathbb{N}^{*})^2\times \mathbb{R}$, the regeneration law $d^{i}_{u}$ has support in $(\mathbb{N}^{*})^2\times \mathbb{R}$. A simple choice is to consider for example the case where the mid price decreases (resp. increases) by one tick when the best bid (resp. ask) is totally depleted and to draw new best bid and ask quantities from a fixed stationary distribution, see \cite{citeulike:8531765}.\\

For $q = (q^1,q^2) \in (\mathbb{N}^{*})^2$, $q' = (q'^1,q'^2) \in (\mathbb{N}^{*})^2$, $(p,p')\in \mathbb{R}^2$, $u = (q^1,q^2,p)$, $n \in \mathbb{N}^* $,  $e_1=(1,0)$, $e_2=(0,1)$ and $i \in \{1,2\}$, the infinitesimal generator  $\mathcal{Q}$ of the process $U_t$ has the following form:
\begin{equation*}
\begin{array}{lcll}
\mathcal{Q}_{(q,p),(q+ne_i,p)} & =  & \lambda^{i,+}(u,n)+ \sum_{i=1}^2  \sum_{k \geq q^i}\lambda^{i,-}(u,k)d^{i}_{(q,p);(q+ne_i,p)}  &\\
\mathcal{Q}_{(q,p),(q-ne_i,p)} & =  & \lambda^{i,-}(u,n)+ \sum_{i=1}^2 \sum_{k \geq q^i}\lambda^{i,-}(u,k)d^{i}_{(q,p);(q-ne_i,p)} & \text{, if } q^i > n  \\
\mathcal{Q}_{(q,p),(q',p')} & =  & \sum_{i=1}^2 \sum_{k \geq q^i}\lambda^{i,-}(u,k)d^{i}_{(q,p);(q',p')} & \text{,  if } p' \ne p, 
\end{array}
\end{equation*} 
where
\begin{itemize}
\item $\lambda^{1,+}$ (resp. $\lambda^{2,+}$) is a non-negative function from $(\mathbb{N}^{*})^2 \times \mathbb{R} \times \mathbb{N}$ to $\mathbb{R}_{+}$ representing the arrival rate of limit orders at the best bid (resp. best ask) (i.e liquidity provision).
\item $\lambda^{1,-}$ (resp. $\lambda^{2,-}$) is a non-negative function from $(\mathbb{N}^{*})^2 \times \mathbb{R}\times \mathbb{N}$ to $\mathbb{R}_{+}$ representing the consumption rate of limit orders at the best bid (resp. best ask) (i.e liquidity consumption).
\item $\sum_{i=1}^2 \sum_{k \geq q^i} \lambda^{i,-}(u,k)d^{i}_{(q,p);(q',p')}$ represents the order book regeneration when the best bid (or ask) is totally depleted.
\end{itemize}
Moreover, without loss of generality, we can initialize the mid price at zero. Then, for every $u = (q^1,q^2,p)$,  $u '= (q'^1,q'^2,p')$ and $n \in \mathbb{N}$, we assume the following bid-ask symmetry relations:
\begin{equation}
\left\{
\begin{array}{ccl}
\lambda^{1,+}(u,n) & = & \lambda^{2,+}(u^{sym},n) \\
\lambda^{1,-}(u,n) & = & \lambda^{2,-}(u^{sym},n) \\
d^{1}_{u}(u')      & = & d^{2}_{u^{sym}}(u'^{sym}) ,
\end{array}
\right.
\label{Eq:LOBAssumption}
\end{equation}
where $u^{sym} = (q^2,q^1,-p)$. 
\subsection{Ergodicity}
We now provide a theoretical result on the ergodicity of the process $Q_t =(Q^{1}_t,Q^{2}_t)$ under four general assumptions given below. A definition of ergodicity is given in Appendix \ref{sec:ErgValFct}. 

\begin{Assumption}[Negative individual drift] There exist three positive constants $C_{bound}$, $z_0 > 1$ and $\delta>0$ such that for any $u=(q^{1},q^{2},p) \in (\mathbb{N}^{*})^2\times \mathbb{R}$ with $q^{1} \geq C_{bound}$,
\begin{equation*}
\sum_{n \geq 0} (z_0^n -1)(\lambda^{1,+}(u,n) - \lambda^{1,-}(u,n)\frac{1}{z^n_0}) \leq -\delta.
\end{equation*}
\label{Assump:Negativeindividualdrift}
\end{Assumption}
Assumption \ref{Assump:Negativeindividualdrift} ensures that the queue size of a limit tends to decrease when it becomes too large. Using Equation (\ref{Eq:LOBAssumption}), we also have 
\begin{equation*}
\sum_{n \geq 0} (z_0^n -1)(\lambda^{2,+}(u,n) - \lambda^{2,-}(u,n)\frac{1}{z^n_0}) \leq - \delta, \quad \forall q^{2} \geq C_{bound}.
\end{equation*}
\begin{Assumption}[Bound on the incoming flow] There exists a positive constant H such that for any $u = (q^{1},q^{2},p)\in (\mathbb{N}^{*})^2 \times \mathbb{R}$,
\begin{equation*}
 \sum_{n\geq 0}\lambda^{1,+} (u,n) +  \lambda^{1,-} (u,n) \leq H.
\end{equation*}
\label{Assump:Boundontheincomingflow}
\end{Assumption}
 Assumption \ref{Assump:Boundontheincomingflow} ensures no explosion in the system: the order arrival speed stays bounded for any given state of the order book. Using the symmetry relation, we also have  \newline $\displaystyle \sum_{n\geq 0}\lambda^{2,+} (u,n) +  \lambda^{2,-} (u,n) \leq H$. \\

For a state  $u = (q^{1},q^{2},p)$ and $i \in \{1,2\}$, we write $U^{Disc,i,u} = (Q^{1,i,u},Q^{2,i,u},P^{Disc,i,u}) $ for a random variable with law $d^i_{u}$.
\begin{Assumption}[Regeneration bound] There exist  three positive constants $C_{Disc}$, $L$ and $z_1>1 $ such that for any $ u = (q^1,q^2,p) \in (\mathbb{N}^*)^2 \times \mathbb{R}$ and $i \in \{1,2 \} $,
\begin{equation*}
\begin{array}{l}
\Esp \big[\sum_{j=1}^2 z_1^{|U^{Disc,i,u}_j-C^{Disc}|_{+}} \big] < L, \\
\end{array}
\end{equation*}
where $U^{Disc,i,u}_j$ is the $j$-th coordinate of $U^{Disc,i,u}$.
\label{Assump:RegenerationBound}
\end{Assumption}
Assumption  \ref{Assump:RegenerationBound} ensures no explosion as well by assuming that the probability to discover large quantities tends quickly to zero.  For example, when there is a quantity $\tilde{Q}^{max}$ such that, for any $q \in (\mathbb{N}^*)^2$ and $i \in  \{1,2\}$, $Q^{1,i,u}\leq \tilde{Q}^{max}$ and $Q^{2,i,u}\leq \tilde{Q}^{max}$ a.s, Assumption  \ref{Assump:RegenerationBound} is satisfied.
\begin{Assumption}[Jumps bound] There exist  two positive constants $L^J$ and $z_2>1 $ such that for any state $u = (q^{1},q^{2},p)\in(\mathbb{N}^*)^2  \times \mathbb{R}$ and $i \in \big\{ 1,2\big\}$,
\begin{equation*}
\sum_n z_2^{n}\lambda^{i,+}(u,n)< L^J.
\end{equation*}
\label{Assump:JumpsBound}
\end{Assumption}
Finally, Assumption  \ref{Assump:JumpsBound} means that the arrival rate of very large jumps  $\lambda^{i,+}(u,n)$ tends quickly to zero when $n$ is high. Assumptions \ref{Assump:Negativeindividualdrift}, \ref{Assump:Boundontheincomingflow}, \ref{Assump:RegenerationBound} and \ref{Assump:JumpsBound} are close to those used in \cite{citeulike:13675327} within a close setting. We have the following result.
\begin{theo}
[Ergodicity] Under Assumptions \ref{Assump:Negativeindividualdrift}, \ref{Assump:Boundontheincomingflow}, \ref{Assump:RegenerationBound} and \ref{Assump:JumpsBound}, and when the arrival rate, the consumption rate and the regeneration distribution do not depend on the mid price, the process $Q_t = (Q^1_t,Q^2_t)$ is ergodic (i.e converges towards a unique invariant distribution). Additionally, we have the following speed of convergence:
$$
||P^t_q(.) - \pi||_{TV} \leq B(q) \rho^t,
$$
with $||.||_{TV} $ the total variation norm, $P^t_q(.)$ the Markov kernel of the process $Q_t$ starting from the initial point $q=(q^1,q^2)\in (\mathbb{N}^{*})^2$, $\pi$ the invariant distribution, $\rho < 1$ and $B(q)$ a constant depending on the initial state $q$, see Appendix \ref{sec:ErgValFct}.
\label{lem:ErgodicityProc}
\end{theo}
This theorem is the basis for the asymptotic study of the order book dynamic in Section \ref{subsec:EmpEvid}, since it ensures the convergence of the order book state towards an invariant probability distribution. Thus the stylized facts observed on market data can be explained by a law of large numbers type phenomenon for this invariant distribution. The proof of this result is given in Appendix \ref{sec:ErgValFct} for sake of completeness, although it is quite inspired from \cite{citeulike:12810809,citeulike:13675327}.

\section{Optimal tactic control problem}
\label{sec:dpp}
\subsection{Presentation of the stochastic control framework}
\manuallabel{subsec:ForModel}{3.1}
We express the control problem for a buy order of a size $q^a$. It can be changed to a sell order in an obvious way. 

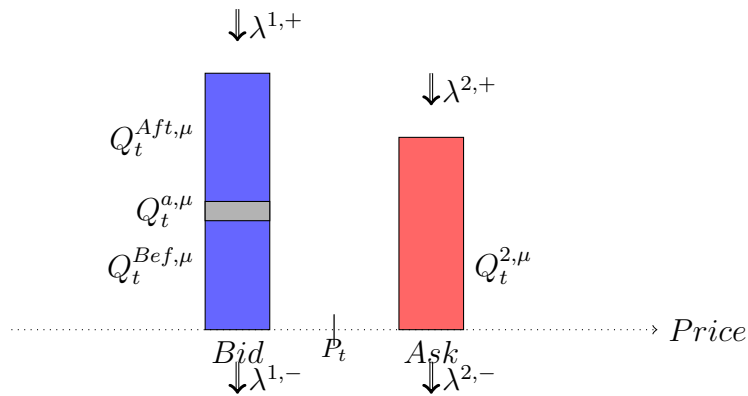
\begin{figure}[ht!]
\begin{center}
\begin{tikzpicture}[scale = 0.85]
\draw [double][->](3.5,0) -- ++(0,-0.5);
\draw [double][->](3.5,-5.5) -- ++(0,-0.5);
\draw [double][->](6.5,-1) -- ++(0,-0.5);
\draw [double][->](6.5,-5.5) -- ++(0,-0.5);
\draw [black,fill=blue!60](3,-1) rectangle (4,-5);
\draw [black,fill=gray!60] (3,-3) rectangle (4,-3.3);
\draw [black,fill=red!60](6,-2) rectangle (7,-5);
\draw [dotted][->](0,-5) -- ++(10,0);
\draw (5,-5) node[sloped]{$|$};
\draw (3.5,-5) node[below]{$ Bid $} ;
\draw (6.5,-5) node[below]{$ Ask $} ;
\draw (3,-4) node[left]{$Q^{Bef,\mu}_t$} ;
\draw (3,-3.15) node[left]{$Q^{a,\mu}_t$} ;
\draw (3,-2) node[left]{$Q^{Aft,\mu}_t$} ;
\draw (7,-4) node[right]{$Q^{2,\mu}_t$} ;
\footnotesize
\draw (5,-5) node[below]{$ P_t $} ;
\normalsize
\draw (10,-5) node[right]{$Price$} ;
\draw (3.5,-0.25) node[right]{$\lambda^{1,+ }$} ;
\draw (3.5,-5.75) node[right]{$\lambda^{1,- }$} ;
\draw (6.5,-1.25) node[right]{$\lambda^{2,+}$} ;
\draw (6.5,-5.75) node[right]{$\lambda^{2,-}$} ;
\end{tikzpicture}
\end{center}  
  \caption{Diagram of flows affecting our order book model.}
  \label{fig:lob:flows}
\end{figure}

\paragraph{Order book dynamic.} The order book state is modelled by the process \newline $U^{\mu}_t = \left(Q^{Bef,\mu}_t,Q^{a,\mu}_t,Q^{Aft,\mu}_t, Q^{2,\mu}_t,I^{\mu}_t, P^{\mu}_t ,P^{Exec,\mu}_t\right)$  where $Q^{a,\mu}_t$ is the size of agent's limit order inserted at the best bid, $Q^{Bef,\mu}_t$ is the quantity inserted before  $Q^{a,\mu}_t$, $Q^{Aft,\mu}_t$ represents orders inserted after $Q^{a,\mu}_t$ (see Figure \ref{fig:lob:flows}), $P^{Exec,\mu}_t$ is the acquisition price of $q^a - I^{\mu}_t$, $I^{\mu}_t$ is the agent's inventory and $\mu$ the control of the agent. We recall that $Q^{2,\mu}_t$ is the best ask limit and $P^{\mu}_t$ is the mid price. Then, $Q^{1,\mu}_t=Q^{Bef,\mu}_t +Q^{a,\mu}_t+ Q^{Aft,\mu}_t$ is the total volume at the best bid. It is split in three quantities to take into account the order placement. Limit orders posted by the agent have a size equal to the whole inventory and we do not handle splitting issues where a partial quantity of the inventory is inserted in the order book. We add minor changes to the order book dynamic:
\begin{itemize}
\item For the best bid, we differentiate market orders consumption rate $\lambda^{1,-}_{m}$ from limit orders cancellation rate $\lambda^{1,-}_{c}$. Cancellation orders consume $Q^{Aft,\mu}_t$ first, and market orders $Q^{Bef,\mu}_t$ first\footnote{This modelling is conservative since we delay the order execution as long as possible. It corresponds to the worst case scenario for the user.}. 
\item The regeneration process of $U^{\mu}_t$ can be deduced from that of $(Q^{1,\mu}_t, Q^{2,\mu}_t, P^{\mu}_t)$ which is unchanged. After a regeneration $Q^{a,\mu}_t = 0$ when the best bid is totally depleted and remains unchanged otherwise. Furthermore, the quantity $Q^{Aft,\mu}+ Q^{Bef,\mu}$ is given by the regenerated best bid and the position of $Q^{a,\mu}_t$  is drawn from a distribution $\iota^{i}_{u}$ depending on the order book state just before the regeneration and the depleted side (i.e best ask in our case). A natural choice is to set $Q^{Aft,\mu} = 0$ and $Q^{Bef,\mu}$ equal to the new best bid when the price moves, and keep the quantities $(Q^{Bef,\mu}_t,Q^{a,\mu}_t,Q^{Aft,\mu}_t)$ unchanged when the best ask is depleted with no price move.
\end{itemize}
The symmetry relation (\ref{Eq:LOBAssumption}) satisfied by $(Q^{1,\mu}_t, Q^{2,\mu}_t, P^{\mu}_t)$ is unchanged. A detailed description of the infinitesimal generator $\mathcal{Q}^{\mu}$ of the process $U^{\mu}_t$ is provided in Appendix \ref{sec:InftGen}. 

\paragraph{Trader's controls.} At every decision time, the trader can do nothing or  take three decisions:
\begin{itemize}
\item \emph{l} : He can insert the quantity $I^{\mu}$ at the top of the bid queue if not already inserted.
\item \emph{c} : He can cancel his already existing limit order $Q^{a,\mu}$. By acting this way, the trader can wait for a better order book state. This control will essentially be used to avoid {adverse selection}, i.e. obtaining a transaction just before a price decrease.
\item \emph{m} : He can send a market order to get immediate execution.
\end{itemize}
Thus, the trader's control $ \mu = \left\{\mu_t, t \leq T \right\}$ is a piecewise-constant c\`{a}dl\`{a}g process valued in $\left\{l,c,m\right\}$. If the agent has no order inserted in the order book and does nothing at the beginning, the initial control is $c$.
\paragraph{Trader's inventory and liquidation price.} Since  we consider a buyer with an initial inventory $q^a$, we fix $I^{\mu}_0 = q^a$. Let $q^m$ be the size of a market order sent at the best bid at time $t$ by another market participant. When $q^m> Q^{Bef,\mu}_{t^-}$, the quantity $\min(q^m-Q^{Bef,\mu}_{t^-},Q^{a,\mu}_{t^{-}})$ of our order is bought at the best bid price $P_{t^{-}}-\frac{\psi}{2}$. Thus, the dynamic of $Q^{a,\mu}_t$ and $P^{Exec,\mu}_t$  can be written  when $\mu =l$ as
\begin{equation*}
\left\{\begin{array}{lcl}
 Q^{a,\mu}_{t}  & = & Q^{a,\mu}_{t^-}- \mathbf{1}_{q^m > Q^{Bef,\mu}_{t^-}} \min(q^m-Q^{Bef,\mu}_{t^-},Q^{a,\mu}_{t^{-}})\\
 I^{\mu}_t      & = & Q^{a,\mu}_{t} \\
P^{Exec,\mu}_t & = & P^{Exec,\mu}_{t^{-}} - \Delta Q^{a,\mu}_{t}(P^{\mu}_{t^{-}}-\frac{\psi}{2}),
\end{array} 
\right.
\end{equation*}
with $\Delta X_t = X_t - X_{t^-}$ for any c\`{a}dl\`{a}g process $X$. When a market order is sent (i.e $\mu =m$), the quantity $I^{\mu}_{t^{-}}$ is bought at the best ask. A linear temporary price impact is added when the best ask is not large enough to absorb $I^{\mu}_{t^{-}}$. In this case, the dynamic of $Q^{a,\mu}_t$ and $P^{Exec,\mu}_t$  writes 
\begin{equation*}
\left\{\begin{array}{lcl}
Q^{a,\mu}_{t}  & = & 0\\
I^{\mu}_{t}  & = & 0\\
P^{Exec,\mu}_t & = & P^{Exec,\mu}_{t^{-}} + I^{\mu}_{t^-}\big[P^{\mu}_{t^{-}}+\frac{\psi}{2} + \alpha (I^{\mu}_{t^-} - Q^{2,\mu}_{t^{-}})_{+}\big],
\end{array} 
\right.
\end{equation*}
where the parameter $\alpha$ represents a linear temporary price impact and $(x)_+ = \max(x,0)$. Finally, under the control $\mu = c$, we set $Q^{a,\mu}_{t} = 0$ and keep $I^{\mu}$ and $P^{Exec,\mu}$ unchanged since the agent's order is not present in the order book. We add the final time constraint 
\begin{equation*}
\left\{\begin{array}{lcl}
Q^{a,\mu}_{T}  & = & 0\\
I^{\mu}_{T}  & = & 0\\
P^{Exec,\mu}_T & = & P^{Exec,\mu}_{T^{-}} + I^{\mu}_{T^-}\big[P^{\mu}_{T^{-}}+\frac{\psi}{2} + \alpha (I^{\mu}_{T^-} - Q^{2,\mu}_{T^{-}})_{+}\big].
\end{array}\right.
\end{equation*}
\paragraph{Optimal control problem.} We fix a finite horizon time $T<\infty$ and we want to compute  
\begin{equation*}
V_{T}(0,u) = \underset{\mu}{\sup}\, \Esp \big[ f\big( \Esp [\Delta P^{\mu}_{\infty}/ \mathcal{F}_{T^{\mu}_{Exec}}] \big) - c q^a T^{\mu}_{Exec}\big],
\end{equation*}
where 
\begin{itemize}
\item $u = (q^{bef},q^a,q^{aft},q^{2},i,p,p^{exec})$ is the initial state of the order book. 
\item $T^{\mu}_{Exec} = \inf\big\{t \geq 0, \, s.t \quad I^{\mu}_{t} = 0 \big\}\wedge T$ represents the final execution time. 
\item $\Delta P^{\mu}_{\infty} = \underset{t \rightarrow \infty}{\lim} \big(P^{\mu}_{t} - P^{Exec,\mu}_{T^{\mu}_{Exec}} \big) $ represents the price impact. We will see that $\Esp \big[ \Delta P^{\mu}_{\infty}/ \mathcal{F}_{T^{\mu}_{Exec}} \big]$ is well-defined and an explicit computation of this quantity is presented in the next section.
\item $c$ is a non-negative homogenization constant representing the waiting cost, $q^a$ is the order size, and $f: \mathbb{R}\rightarrow \mathbb{R}$ is a Lipschitz function. 
\end{itemize}
We solve the agent's control problem in two situations: when decisions are taken at fixed frequency $\Delta^{-1}$ and when they are taken at any time.

\section{Theoretical results}
In this section, we compute $\Delta P^{\mu}_{\infty}$, discuss the existence, uniqueness and regularity of the solution of our control problem and give equations satisfied by the value function.
\label{sec:TheoRes}
\subsection{Computation of $\Delta P^{\mu}_{\infty}$}
\manuallabel{subsec:CompAverPMove}{4.1}
To compute $\Esp[\Delta P^{\mu}_{\infty}/\mathcal{F}_{T^{\mu}_{Exec}}]$, we replace Assumptions \ref{Assump:RegenerationBound} and \ref{Assump:JumpsBound} by slightly less general ones.  
\begin{Assumption}[Insertion Bound] There exists a positive quantity $Q^{max}$ such that for any $u = (q^{1},q^2,p) \in (\mathbb{N}^{*})^2\times \mathbb{R}$ and $ n \geq 0$,
\begin{equation*}
\left\{
\begin{array}{ll}
\lambda^{1,+}(u,n) = 0, & \text{ \normalfont when } \quad q^1+n > Q^{max} \\ 
\lambda^{1,-}(u,n) = 0, & \text{ \normalfont when } \quad q^1 > Q^{max}. \\ 
\end{array}
\right.
\end{equation*}
\label{Assump:InsertionBound}
\end{Assumption}
This assumption is not restrictive since available quantities in the best limits remain essentially bounded. Using the symmetry relation, we have as well, for any $u = (q^{1},q^2,p) \in (\mathbb{N}^{*})^2 \times	\mathbb{R}$ and $n\geq 0$ 
\begin{equation*}
\left\{
\begin{array}{ll}
\lambda^{2,+}(u,n) = 0, & \text{ \normalfont when } \quad q^2+n > Q^{max} \\ 
\lambda^{2,-}(u,n) = 0, & \text{ \normalfont when } \quad q^2 > Q^{max}. \\ 
\end{array}
\right.
\end{equation*}
For a state  $u = (q^{1},q^{2},p)$ and $i \in \{1,2\}$, we write $U^{Disc,i,u} = (Q^{1,i,u},Q^{2,i,u},P^{Disc,i,u})$ for a random variable with a distribution $d^i_{u}$. We now give a final boundedness assumption.
\begin{Assumption}[Regeneration bound] The mid price $P^{\mu}_t$  lives in the space $\tau_0 \mathbb{Z}$, with $\tau_0 \in \mathbb{R}^+$ the tick value. Additionally, there exist two positive constants $\tilde{P}^{max}$ and $\tilde{Q}^{max}$  such that 
\begin{equation*}
\begin{array}{l}
|P^{Disc,i,u}|\leq  \tilde{P}^{max}, \quad \; Q^{1,i,u}\leq  \tilde{Q}^{max} \;  \text{ and } \; Q^{2,i,u}\leq  \tilde{Q}^{max} \qquad a.s.
\end{array}
\end{equation*}
\label{Assump:RegenerationLimit}
\end{Assumption}
\vspace{-1.25cm}
\paragraph{Computation of the long term price impact:} For every state $U_i$ at the end of the execution, we split $\Delta P^{\mu}_{\infty}$ in two quantities 
\begin{equation*}
\Delta P^{\mu}_{\infty} = \Delta P^{',\mu}_{\infty} + \big(P^{\mu}_{T^{\mu}_{Exec}} - P^{Exec,\mu}_{T^{\mu}_{Exec}} \big),
\end{equation*} 
where
\begin{itemize}
\item $P^{\mu}_{T^{\mu}_{Exec}}$ is the mid price at the execution (i.e $P^{\mu}_{T^{\mu}_{Exec}}$ and $P^{Exec,\mu}_{T^{\mu}_{Exec}}$ are known at the execution).
\item $ \Delta P^{',\mu}_{\infty} = \underset{t \rightarrow \infty}{\lim} \big(P^{\mu}_{t} - P^{\mu}_{T^{\mu}_{Exec}} \big)$ is the long term mid price move after the execution.
\end{itemize}

Thus, we only need to compute $ \Delta P^{',\mu}_{\infty}$. Since we place ourselves after the execution of $I^{\mu}$, we have $Q^{a,\mu}_t = 0 $ and $Q^{Aft,\mu}_t = 0$. We can then write $U_i = \big(Q^{1}_i,Q^{2}_i,P_i\big)$ by using a slight abuse of notation and forget the dependence on the control. Let $t_1>T^{\mu}_{Exec}$ be the first time where the best bid is totally consumed after the final execution time and $t_2>T^{\mu}_{Exec}$ the first time where the best ask is depleted. When the best bid (resp. ask) is totally consumed, the price moves on average by $\alpha_i^{-} = \Esp_{U_i} \big[\Delta P_{t_1}\big] $ (resp. $\alpha_i^{+} = \Esp_{U_i} \big[\Delta P_{t_2}\big] $) and the order book is regenerated according to a measure $d^1_{i}$ (resp. $d^2_{i}$). The index $i$ is associated to the state $U_i$. We define
\begin{itemize}
\item $q^{-}_ {ii'} = \mathbb{P}_{U_i}[\{t_1 <t_2\} \cap \{U_{t^-_1} = U_{i'} \} ]$ and $q^{+}_ {ii'} = \mathbb{P}_{U_i}[\{t_2 \leq t_1\} \cap \{U_{t^-_2} = U_{i'} \}]$. They represent respectively the probability that the best bid is consumed before the best ask or conversely and the exit state is $U_{i'}$.
\item $d^{1}_{i,k}$ (resp. $d^{2}_{i,k}$) are transition probabilities from the state $U_i$ to $U_k$ when the best bid (resp. ask) is consumed. 
\item $q_i = \sum_{i'} \big(q^{+}_ {ii'}\alpha_{i'}^{+}+q^{-}_ {ii'} \alpha_{i'}^{-}\big)$ and $p_{i,k} = \sum_{i'}\big(q^{+}_ {ii'}d^{2}_{i',k}+q^{-}_ {ii'}d^{1}_{i',k}\big)$ represent respectively the average mid price move after the first regeneration and the probability to reach the state $U_k$ starting from the initial point $U_i$ right after the first regeneration.
\item $U^{sym} = (q^2,q^1,-p)$ is the symmetric state of $U = (q^1,q^2,p)$. 
\item $D$ is a vector satisfying $D_{i} =  \Esp_{U_i}[\Delta P^{'}]$ for every state $U_i = (q^1,q^2,p)$ such that $p>0$ or $p=0$ and $ q^1 \geq q^2$. Note that $\Esp_{U_i}[\Delta P^{'}] = - \Esp_{U_i^{sym}}[\Delta P^{'}] $.
\end{itemize}

For a state $U_i$, $i^{sym}$ is the index of the symmetric state $U^{sym}_i$. We have the following result.
\begin{prop}[Average mid price move] For an irreducible process $U_t$, see Section \ref{subsec:LOBPresentation}, satisfying Assumptions \ref{Assump:InsertionBound} and \ref{Assump:RegenerationLimit}, the vector $D$ satisfies
\begin{equation*}
D = (I-A)^{-1}F.
\end{equation*}
The matrix $A$ is defined by $A_{i,k} =\frac{p_{i,k} -  p_{i,k^{sym}}}{1-(p_{i,i} - p_{i,i^{sym}})}$ when $i\ne k$ and $A_{i,i}=0$ and the vector $F$ satisfies $F_i = \frac{q_{i}}{1-(p_{i,i} - p_{i,i^{sym}})}$. The matrix $I-A $ is invertible.
\label{Theo:AvgPriceMoveComput}
\end{prop}
The proof of this result is given in Appendix \ref{sec:AveragePriceMoveAfterExec}\ref{subsec:CompPinft1}.\vspace{2mm}\\
To compute $D$, we need to estimate the regeneration distributions $d^{1}_.$, $d^{2}_.$, $\alpha^{\pm}_{.}$  and 
$q^{\pm}_{.}$. The quantities $d^{1}_.$, $d^{2}_.$ and $\alpha^{\pm}_{.}$ can be estimated from the empirical distribution of order book states after a depletion. Then, we only need to estimate $q^{\pm}_{ii'}$. We now give a result on the computation of $q^{\pm}_{ii'}$.
\begin{lem}[Computation of $q^{\pm}_{ii'}$]Let $R= [R^-,R^+]$ be the matrix such that $R^-_{ii'} = q^{-}_{ii'}$ and $R^+_{ii'} = q^{+}_{ii'}$. Then, $R$ is a solution of the equation 
\begin{equation*}
\tilde{Q}^{*} \tilde{R} = - z^1 \quad \text{ and } \quad R = M \tilde{R},
\end{equation*}
where $\tilde{Q}^{*}$, $z^1$ and $M$ are defined in Appendix \ref{sec:AveragePriceMoveAfterExec}\ref{subsec:CompPinft2}, see Equations (\ref{eq:MatrixA}) and (\ref{Eq:HitTime}). The solution of this equation is unique since $\tilde{Q}^{*}$ is invertible, see Appendix \ref{sec:AveragePriceMoveAfterExec}\ref{subsec:CompPinft3}. 
\label{lem:AveragePriceMoveAfterExec}
\end{lem}

When the dynamic of the best bid is independent from the one of the best ask, $\tilde{Q}^{*}$ is even diagonalisable. In the simple case of constant intensities as in \cite{citeulike:8531765}, $\tilde{Q}^{*}$ diagonalisation can be computed easily with closed form formulas, see Appendix \ref{sec:AveragePriceMoveAfterExec}\ref{subsec:CompPinft3}. The proof of this lemma is given in Appendix \ref{sec:AveragePriceMoveAfterExec}\ref{subsec:CompPinft2}.  A numerical computation of the vector $P$ is given in Appendix \ref{sec:ModelparamEstim}, Figure \ref{fig:AveragePriceMoveValue}. 

\subsection{Existence and uniqueness of the optimal strategy, regularity properties}
\manuallabel{subsec:ExisUniqReg}{4.2}
In the rest of the article, Assumptions \ref{Assump:Boundontheincomingflow}, \ref{Assump:InsertionBound} and \ref{Assump:RegenerationLimit} are in force. In this section, we discuss existence and uniqueness of the optimal strategy and show regularity results for the state process $U^\mu$ and the value function. First, for a finite horizon time $T$, we define the value function 
\begin{equation*}
V_T (t,u) = \underset{\mu}{\sup}\, \Esp \big[ f\big(\Esp\big[\Delta P^{\mu}_{\infty} / \mathcal{F}_{T^{t,\mu}_{Exec}}\big]\big) - cq^a (T^{t,\mu}_{Exec}-t) | U^{\mu}_t = u\big],
\label{Eq:ValFctDef}
\end{equation*}
with $0 \leq t \leq T$, $u \in \mathbb{N}^5 \times \mathbb{R}^2$ and $T^{t,\mu}_{Exec} = \inf\big\{s \geq t, \, s.t \quad I^{\mu}_{s} = 0 \big\}\wedge T$.
\paragraph{Existence - uniqueness of the optimal control.} The optimal strategy exists in the two frameworks (i.e decisions taken at fixed frequency and at any time) but for different reasons. When decisions are taken at fixed frequency  $\Delta^{-1}$ the optimal strategy exists since we have a finite number of available strategies. When decisions are taken at any time, the sequence of optimal controls $(\tau_i,\upsilon_i)$, where $\tau_i$ is the optimal decision time and $\upsilon_i$  the optimal decision, satisfies $\tau_0=0$ and $\upsilon_{0} =  \underset{r\in\{l,c,m\}}{\text{argmax}} \left\{ \Esp\big[V_T(0, U^{r}_{0})\big]\right\}$ with $u^r$ the new state when the agents takes the decision $r$ and
\begin{equation}
\left\{
\begin{array}{l}
\tau_{i+1} = \inf \big\{ t> \tau_i; V_T(t,U^{\hat{\mu}}_{t^-}) = \mathcal{H}^{-\upsilon_i}V_T(t,U^{\hat{\mu}}_{t^-})\big\} \\
\upsilon_{i+1} = \underset{r\in\{l,c,m\},\upsilon_i\ne r}{\text{argmax}} \left\{ \Esp\big[V_T(\tau_{i+1}, (U^{\hat{\mu}}_{\tau_{i+1}^-})^{r})\big]\right\},
\end{array}
\right.
\label{Eq:OptimalStratEqu}
\end{equation}
where, for a given state $u$ and control $r$, $\mathcal{H}^{-r}V_T(t,u) = \underset{o\in\{l,c,m\},o\ne r}{\max} \Esp\big[V_T(t, u^{o})\big] $ and  $\hat{\mu} = (\tau_j,\upsilon_j)_{j\leq i}$. Since $V_T$  is continous, the optimal control is well-defined, see Equations (\ref{Eq:LipschitzValFct}) and
(\ref{Eq:HolderRegValFct}). The proof of (\ref{Eq:OptimalStratEqu}) is given in Appendix \ref{sec:OptiProDynEq}. However, there is a priori no uniqueness of the optimal strategy in the two frameworks.
\paragraph{Regularization of the problem.} To force the uniqueness of the optimal strategy, we present a practical criterion. We define an order relation between trader's decisions $c < l < m$. 
The intuition behind this relation is that $m$ is the least risky decision because we get direct execution, $s$ is riskier than $m$ but less risky than $c$ because there is no delay of the execution. Hence,  we can choose the least risky decision among the optimal ones in the above sense. 
\paragraph{Regularity of the state process and the value function.} The value function $V_T$ is Lipschitz in time, see Appendix \ref{sec:RegValFctProofExecTimeVSInitState}. Results of Appendix \ref{sec:RegValFctProofExecTimeVSInitState} are provided in the more general framework where we allow the state process to be valued in $\mathbb{R}_{+}^5\times \mathbb{R}^2$. In this setting, we study the regularity of the process $U^{\mu}$, see Theorem \ref{Th:Regstateprocess}, which enables us to recover the Lipschitz property of $V_T$. 

\subsection{Decisions taken at fixed frequency $\Delta^{-1}$: dynamic programming equation }
\manuallabel{subsec:EqDynFixedFreq}{4.3}

In this section, we provide and solve the system of equations satisfied by the value function $V$ of the optimal control problem. We have the following result.
\begin{theo}
Let $u = (q^{bef},q^a,q^{aft},q^{2},i,p,p^{exec})$ be an initial state and $t\in [0,T]$. Then $V(t,u)$ satisfies:
\begin{itemize}
\item When $i > 0$:
\begin{itemize}
\item At the decision time $t = k \Delta < T$:
\begin{equation}
V(k\Delta,.) = \max \left(
\begin{array}{l}
V^l((k\Delta)_+,.) \\
V^c((k\Delta)_+,.) \\
 g(.)\\
\end{array}
\right),
\label{Eq:EqDynProgBetStep0} 
\end{equation}
where $V(t,.)$, $g(.)$, $V^{c}(t,.)$ and $V^{l}(t,.)$ are vectors such that $V(t,.)_i = V(t,u_i)$, $g(.)_i = f\big(\Esp[\Esp_{u_i^m}[\Delta P_{\infty}]]\big)$, $V^{c} (t,.)_i= \Esp[V(t,u^{c}_i)]$ and $V^{l} (t,.)_i= V(t,u^{l}_i)$, where $u^{r}_i$ is the new order book state when the decision $r \in \{l,c,m\}$ is taken. We keep in mind that the controls $c$ and $m$ may lead to several order book states because of the regeneration. Equation \ref{Eq:EqDynProgBetStep0} should be understood coordinate by coordinate. 
\item At $t \ne k \Delta < T$:
\begin{equation}
0 = - cq^a \mathbf{1} + \mathcal{A}V(t,.),
\label{Eq:EqDynProgBetStep} 
\end{equation}
where $\mathcal{A}	= \partial_t + Q$ is the infinitesimal generator of the process $U^{\mu}_t$. The expression of $Q$ is given in Appendix \ref{sec:InftGen}.
\end{itemize}
\item When $i = 0$ (execution time condition):
\begin{equation*}
V(t,u) = g(u), \, \forall t < T,
\label{Eq:ExecTime}
\end{equation*}
with $g(u) = f\big(\Esp[\Esp_{u^m}[\Delta P_{\infty}]]\big)$.
\item The terminal condition is:
\begin{equation}
V(T,u) = g(u).
\label{Eq:EqDynProgPrinFinalT}
\end{equation}
\end{itemize}
\label{Eq:EqDynProgPrin}
\end{theo}
The proof of this result is given in Appendix \ref{sec:OptiProDynEq}.
\begin{rem} At every decision time, as long as the order is not executed, the agent compares the value function given by each control and takes the highest one, see Equation (\ref{Eq:EqDynProgBetStep0}). When, the order is executed, the agent gain is $g(U)$. If the order is not executed before time $T$, the agent send a market order to obtain immediate execution and earn $g(U)$.
\end{rem}
\begin{rem}Without the control $c$ and $l$, Equations (\ref{Eq:EqDynProgBetStep0}), (\ref{Eq:EqDynProgBetStep}) and (\ref{Eq:EqDynProgPrinFinalT}) are equivalent, in dimension 1, to the classical problem of finite horizon Bermudean options. The above system can be solved explicitely, see Appendix \ref{sec:OptiProDynEq}.
\end{rem}
\subsection{Second approach: Decisions taken at any time}
\manuallabel{subsec:EqDynLOBEvt}{4.4}
Let us now consider the case where the agent takes a decision at any time. In this section, we provide the system of equations satisfied by the value function and we also introduce a simplified control problem whose value function can be easily computed numerically, and converges towards the one of the initial optimal control problem.
\subsubsection{Dynamic programming equation}
We keep the same notations as in Theorem \ref{Eq:EqDynProgPrin}. We have the following result for the value function in this setting.
\begin{theo}
Let $u = (q^{bef},q^a,q^{aft},q^2,i,p,p^{exec})$ be an initial state and $t \in [0,T] $. Then $V(t,u)$ satisfies in the viscosity sense and almost everywhere 
\begin{itemize}
\item When $i > 0$:
\begin{equation}
\max \left(
\begin{array}{l}
\mathcal{A}V(t,.) - cq^a\mathbf{1} \\
 V^{l}(t,.) - V(t,.)\\
 V^{c}(t,.) - V(t,.)\\
  g(.) - V(t,.)\\
\end{array}
\right) = 0.
\label{Eq:ContEqDynProgBetStep0} 
\end{equation}
\item When $i = 0$ (execution time condition):
\begin{equation*}
V(t,u) = g(u) , \, \forall t \leq T.
\end{equation*}
\item The terminal condition is:
\begin{equation}
V(T,u) = g(u).
\label{lem:EqContDynProgFinT}
\end{equation}
\end{itemize}
\label{lem:EqContDynProg}
\end{theo}
The proof of the result is given in Appendix \ref{sec:OptiProDynEq}. Since $\partial_t V$ is a priori not continuous, we use the notion of viscosity solution. However we show that $\partial_t V$ is continuous except on the boundary of $\{V=g\}$ and the above equations are satisfied pointwise except on this boundary, see Appendix \ref{sec:OptiProDynEq}.
\begin{rem} When there is no control $c$ and $l$, Equations (\ref{Eq:ContEqDynProgBetStep0}) and (\ref{lem:EqContDynProgFinT}) are equivalent, in dimension 1, to the classical problem of finite horizon American option.
\end{rem}
\subsubsection{Numerical resolution of the optimal execution problem}
\manuallabel{subsec:ResPrincOptiExec}{4.4.3}
To solve numerically the preceding optimal control problem, we consider a discrete framework. We show here how this discrete framework can be used to approximate the solution of the continuous control problem. Furthermore, an error estimate is provided.
\paragraph{Discrete-time Markov chain approximation.} Let $q = (q^1,q^2) \in (\mathbb{N}^{*})^2$, $p\in \mathbb{R}$, $n \in \mathbb{N} $,  $e_1=(1,0)$, $e_2=(0,1)$ and $i \in \{1,2\}$. Let $U^\Delta_n$ be a Markov chain with transition matrix $P$ defined by  
\begin{equation*}
\begin{array}{lcll}
P_{(q,p),(q',p')} & =  & \mathbb{P}\big[U_{\Delta}=(q',p')| U_{0}=(q,p)\big] ,
\end{array}
\label{Eq:DiscProbTransRel}
\end{equation*} 
with $U_t$ the process defined in Section \ref{subsec:LOBPresentation}. Given the infinitesimal generator $\mathcal{Q}$ of $U$, the transition matrix $P$ can be easily computed since $P = (e^{\Delta Q})$, see Appendix \ref{sec:InftGen}. In this approximation, $U^\Delta_n$ is viewed as the market evolution without the intervention of the agent. Associated to this new market, we introduce the controlled discrete-time Markov chain $U^{\Delta,\mu}_n = \left( Q^{Bef,\mu}_n, Q^{a,\mu}_n,Q^{Aft,\mu}_n, Q^{2,\mu}_n,I^{\mu}_n,P^{\mu}_n ,P^{Exec,\mu}_n\right)$ by using the same construction as in Section \ref{subsec:ForModel}. Additionally, we can also compute $\Delta P^{\mu}_{\infty}$ in this discrete-time approximation by following the same approach as in Section \ref{subsec:CompAverPMove}.
\paragraph{Solving numerically the optimal control problem in the discrete framework.} We denote by $ V^\Delta(n,u)$ the value function associated to the discrete control problem, with $n$ the period and $u$ the order book state.
The dynamic programming principle reads 
\begin{equation*}
 V^\Delta(i,u) = \underset{\mu \in \mathcal{U}}{\sup} \, \Esp\left[ V^\Delta\left((i+1),U^{\mu}_{i+1}\right) -cq^a\Delta | U^{\mu}_{i} = u   \right].
\end{equation*}
Consequently, we have
\begin{equation}
 V^\Delta(i,u) = \max 
\left\{
\begin{array}{ll}
\sum_{u'} P_{u^l,u'}V^\Delta(i+1,u')-cq^a\Delta & \text{ control } l\\
\sum_{u'} P_{u^c,u'}V^\Delta(i+1,u')-cq^a\Delta & \text{ control } c\\
g(u) & \text{ control } m,
\end{array}
\right.
\label{Eq:SimDiscretOptiCtrlPbm}
\end{equation}
with the terminal constraint $ V^\Delta(n_f,u) = g(u)$ where $u^r$ is the new order book state after the control $r\in\{l,c,m\}$ and $n_f$ is the final period. Equation  (\ref{Eq:SimDiscretOptiCtrlPbm}) provides a numerical scheme to compute $ V^\Delta(0,u)$. At the final time $T$, we can compute $V^\Delta(n_f,u)$ for each reachable state. Using the backward Equation (\ref{Eq:SimDiscretOptiCtrlPbm}), we can compute $V^\Delta(i,u)$ knowing $V^\Delta(i+1,u)$ to get the initial value $V^\Delta(0,u)$. The numerical results of simulations are presented in Section \ref{sec:understanding}. To compute efficiently the value function, the dynamic programming scheme can be parallelized.
\begin{rem}
Note that applying a finite difference scheme to the equations of Theorem \ref{lem:EqContDynProg} provides the same result as in the discrete-time approximation given by the Markov chain $\tilde{U}^{\Delta}_n$ with transition matrix $\tilde{P} = I + Q\Delta$. When $\Delta$ is small, our discrete-time approximation is almost equivalent to the finite difference scheme since $P = e^{\Delta Q} = I + Q\Delta + o(\Delta) $.
\label{rem:FinDiffScheme}
\end{rem}
Finally, for every $k \geq 0$, we define the piecewise constant process $\tilde{U}^{\Delta}$ associated to $U^{\Delta}_n$ such that:
\begin{equation*}
\tilde{U}^{\Delta}_t = U_k , \qquad \forall t \in [k\Delta,(k+1)\Delta).
\end{equation*}
We denote by $\tilde{V}^{\Delta}(t,U)$ the value function of the control problem where the state process is $\tilde{U}^{\Delta}$.
Then we have the following error estimate result.
\begin{theo} For every state $u = \left(q^{bef},q^a,q^{aft},q^{2},i,p,p^{exec}\right)$, we have
\begin{equation}
|\tilde{V}^{\Delta}(t,u)-V(t,u)| \leq R (T-t) \Delta,
\label{Eq:ErrEstimPartCont} 
\end{equation}
with $R = 4cq^aH $ and $H$ is defined in Assumption \ref{Assump:Boundontheincomingflow}. This ensures the convergence of our discrete approximation:
\begin{equation*}
\tilde{V}^{\Delta}(t,u)   \underset{\Delta \rightarrow 0}{\rightarrow} V(t,u). 
\end{equation*}
Moreover, the sequence $\mu^{Opti,\Delta}$ associated to the process $\tilde{U}^{\mu}_t$ satisfies
\begin{equation}
\mu^{Opti,\Delta} \underset{\Delta \rightarrow 0}{\rightarrow} \mu^{Opti}, \qquad a.s,
\label{Eq:ConvergenceControlDiscrete}
\end{equation}
where $ \mu^{Opti} $ is the optimal control associated to the continuous time control problem.
\label{lem:ConvergenceDiscreteFrameRes}
\end{theo}

The proof of this result is given in Appendix \ref{sec:ProofConvergenceDiscreteFrameRes}. When $\Delta$ is small, the above error estimate remains valid for $P = I + Q\Delta $ (i.e finite difference scheme), see Appendix \ref{sec:ProofConvergenceDiscreteFrameRes}.
\section{Numerical experiments }
\label{sec:understanding}
In this section, we show the relevance of the optimal strategy in both frameworks : when decisions are taken at fixed frequency $\Delta^{-1}$ and when they are taken at any time. To do so, we compare the optimal gain given by our strategy and the one given by the standard strategy join the bid: stay in the order book at the best bid until the final time. Here, we write $Q^1$ (resp. $Q^2$) for the best bid (resp. ask) limit.
\subsection{Computation of the optimal gain : decisions taken at a fixed frequency $\Delta^{-1}$}
\begin{figure}[!ht]
	\centering
	\includegraphics[width=.7\linewidth]{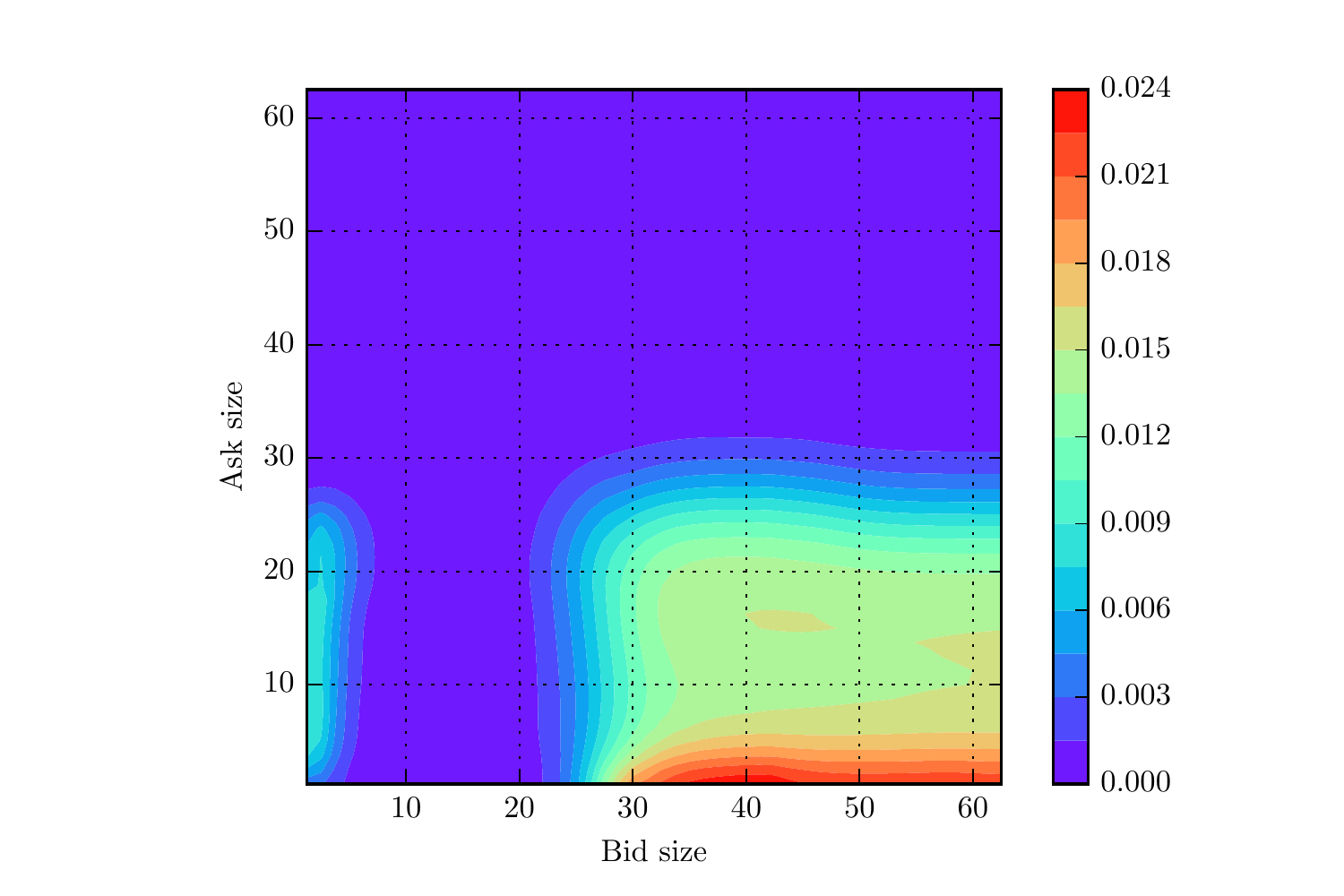}
  \caption{Difference between the optimal gain of the optimal strategy and the one of the strategy stay at the best bid.  The initial parameters are fixed as follows: the time frequency is equal to $\Delta = 10 $ seconds, the final time $ T= 100$ seconds, arrival and consumption rates are estimated on data (see Appendix \ref{sec:ModelParamEstim}), the new bid (resp. ask) is set to $5$ and the new ask (resp. bid) to $3$ after the total depletion of the bid (resp. ask) limit, the quantity $q^a = 1$, the waiting cost $c=0$, the price increases (resp. decreases) by $\delta = 0.01$ when the ask limit (resp. bid limit) is totally consumed and the function $f$ is equal to the identity.}
  \label{fig:MeanPriceMoveImbalanceOpti1}
\end{figure}

Figure \ref{fig:MeanPriceMoveImbalanceOpti1} shows for an order of size $1$ the difference between the average gain (i.e the initial value function) of the optimal strategy and the one of the strategy join the bid for different values of the initial $Q^{1}$ and $Q^{2}$. The gain of the optimal strategy is obviously always higher than that of the strategy stay in the order book. However, because of the priority value, that is the advantage of a limit order compared with another limit order standing at the rear of the same queue, is important, it is more useful to be active (i.e cancel the order or send a market order) when imbalance is highly positive than when it is negative. Finally, note that the optimal strategy reaches the maximum value of 2.4 ticks (since the tick $\delta = 0.01$).
\subsection{Computation of the optimal gain : decision taken at any time}
\begin{figure}[!ht]
	\centering
	\includegraphics[width=.7\linewidth]{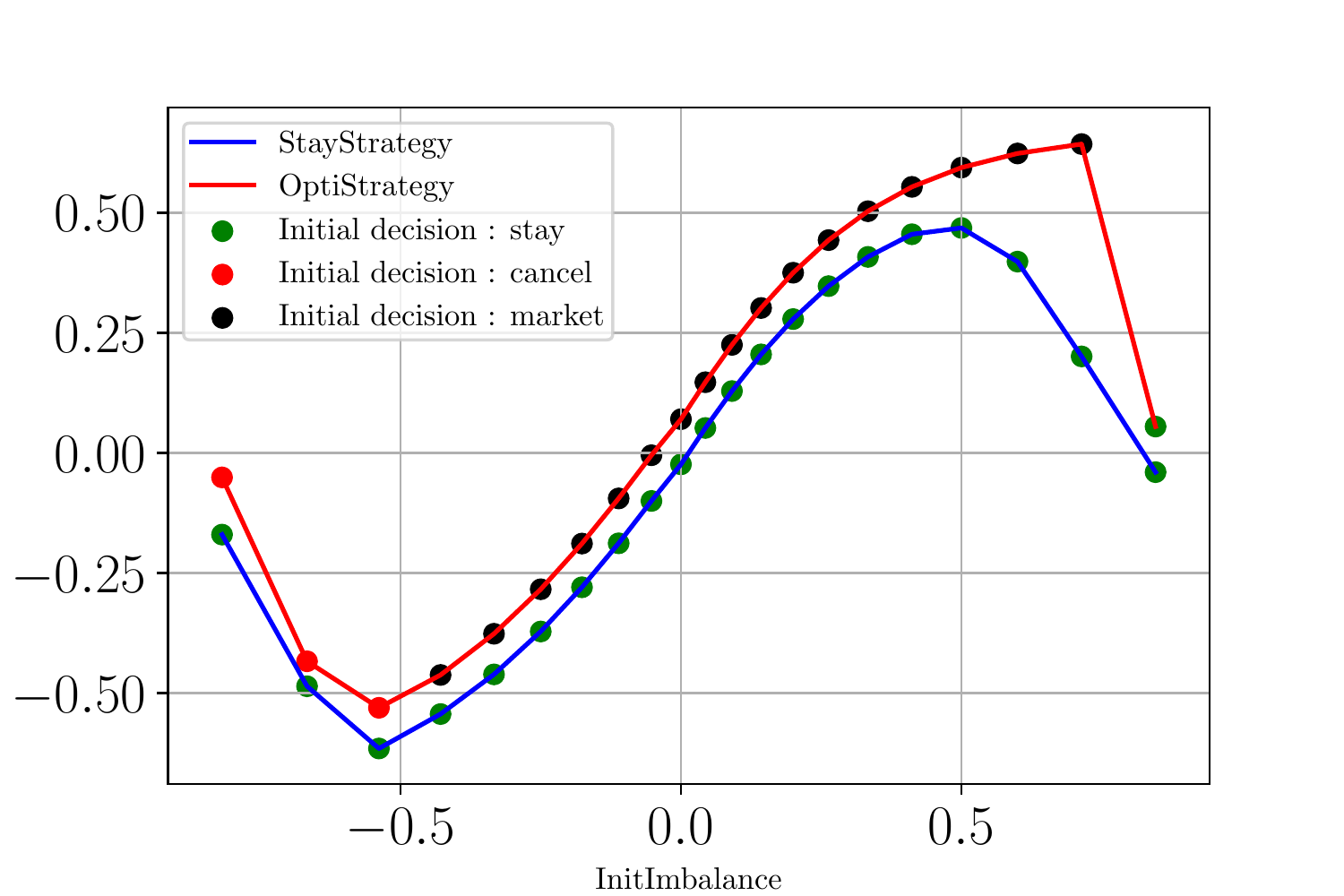}
  \caption{The gain per tick of the optimal strategy in red and the one of the strategy join the bid in blue for different values of the initial imbalance. Initial imbalances are obtained with $Q^{Bef} = 0$, $Q^1 = 11$ and $Q^{2}$ from $1$ to $11$, and $Q^{2} = 11$ and $Q^{1}$ from $10$ to $1$. Initial parameters are as follows: the time step is equal to $\Delta = 1 $ second, there are 10 periods, arrival and consumption rates are constant $\lambda^{1,+} = \lambda^{2,+} = 0.06 $  and $\lambda^{1,-} = \lambda^{2,-} = 0.12 $, the new bid (resp. ask) is set to $5$ and the new ask (resp. bid) to $3$ after the total depletion of the bid (resp. ask) limit, the quantity $q^a = 1$, the waiting cost $c=0.0085$, the price increases (resp. decreases) by $\delta = 0.01$ when the ask limit (resp. bid limit) is totally consumed and the function $f$ is the identity.}
  \label{fig:MeanPriceMoveImbalanceOpti2}
\end{figure}

Figure \ref{fig:MeanPriceMoveImbalanceOpti2} shows  the value function at time zero (i.e trader's gain) of the optimal strategy in red and the one of the strategy stay at the best bid in blue in percentage of the tick $\delta = 0.01$ using the discrete approximation. The points colors refer to the initial decision given by the strategy: green points means stay in the order book at the beginning is the best decision, red points means cancel is the best initial decision and black points means send a market order is the best initial decision. When imbalance is highly negative, it is optimal to cancel the order to avoid adverse selection, when imbalance is highly positive it is optimal to send a market order or stay in the order book. In our case, stay in the order book is interesting when imbalance is highly positive since the priority value is important (i.e $Q^{Bef}$ is fixed equal to 1). In mid cases (i.e imbalance close to 0), it is optimal to send a market order to reduce the waiting cost. We note that the gain of the optimal strategy is significanlty better than the one of the strategy join the bid. 
\paragraph{Acknowledgements.}
The authors gratefully acknowledge the financial support of the ERC grant 679836 Staqamof and the Chair Analytics and Models for Regulation.
\newpage 

\bibliographystyle{plain}
\bibliography{othmane}

\newpage

\appendix

\section{Model parameters estimation }
\label{sec:ModelparamEstim}
The estimation methodology of the arrival and cancellation rates of limit orders is similar to that in \cite{citeulike:12810809}. The regeneration distribution of the order book is estimated from the empirical distribution of order book states after a depletion.\\

\begin{figure}[!ht]
	\centering
	\hspace{1cm}(a) $\lambda^{1,+}$ \hfill \hfill (b) $\lambda^{1,-}$\hfill~\\
	\includegraphics[width=.4\linewidth]{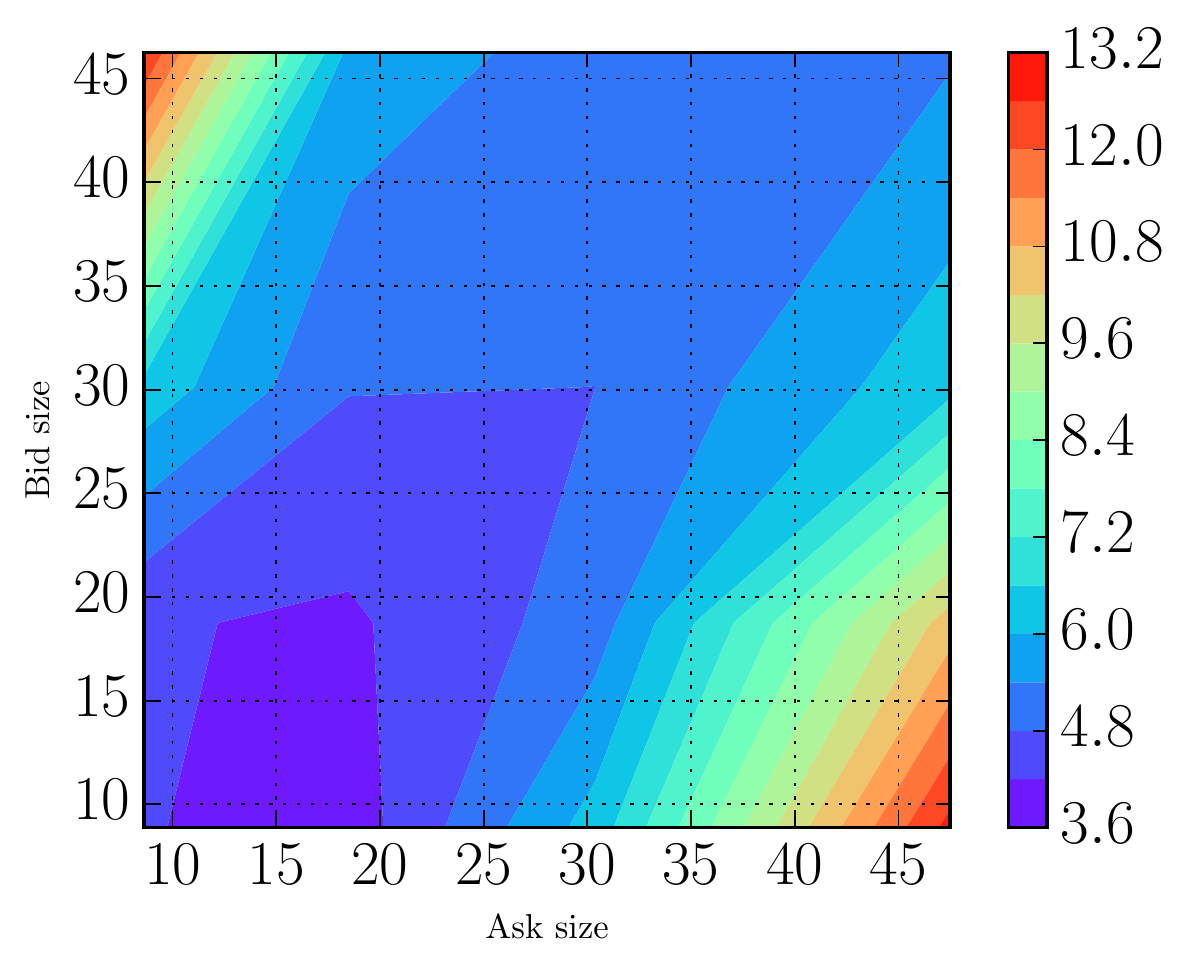}\hfill
	\includegraphics[width=.4\linewidth]{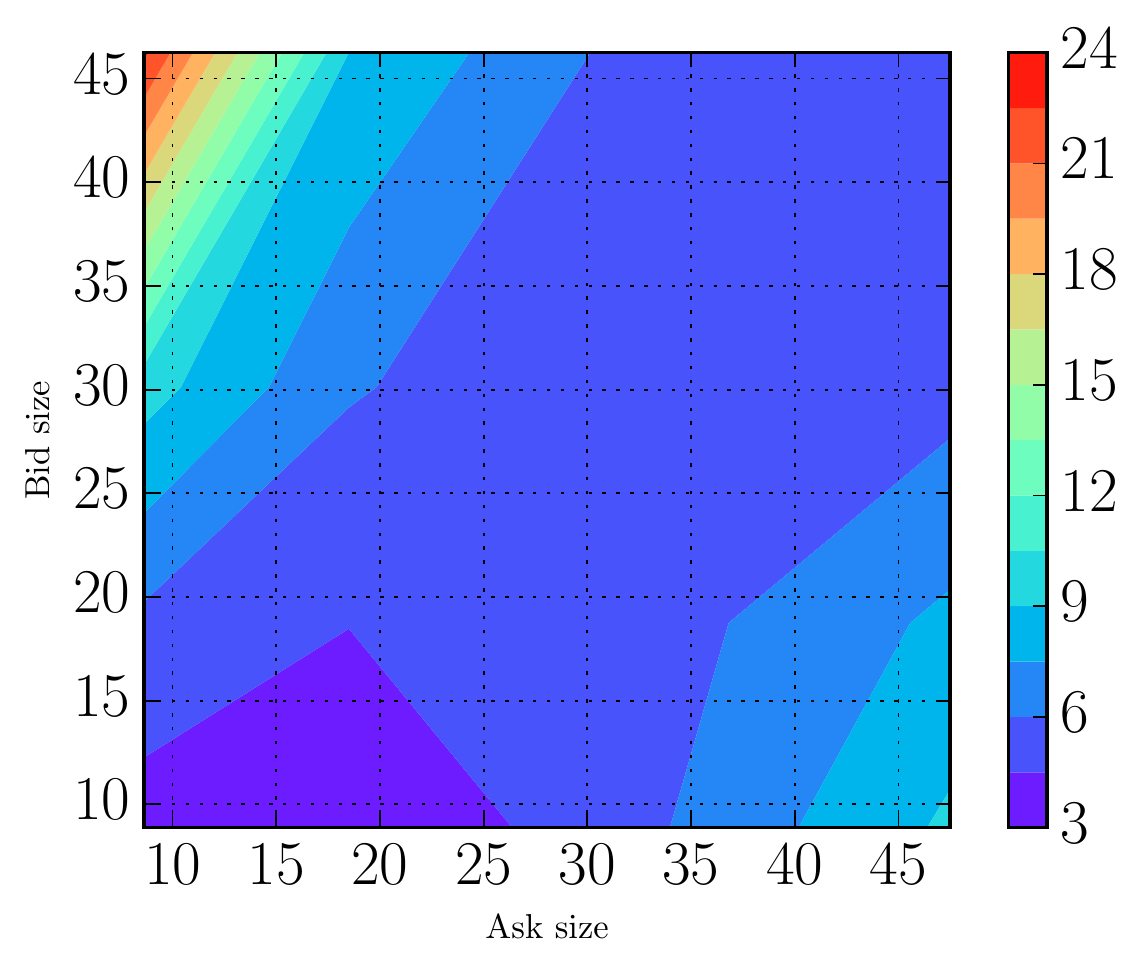}\\
  	\hspace{1cm}(c) $\tau^{1,+}$ \hfill \hfill (d) $\tau^{2,+}$\hfill~\\
	\includegraphics[width=.4\linewidth]{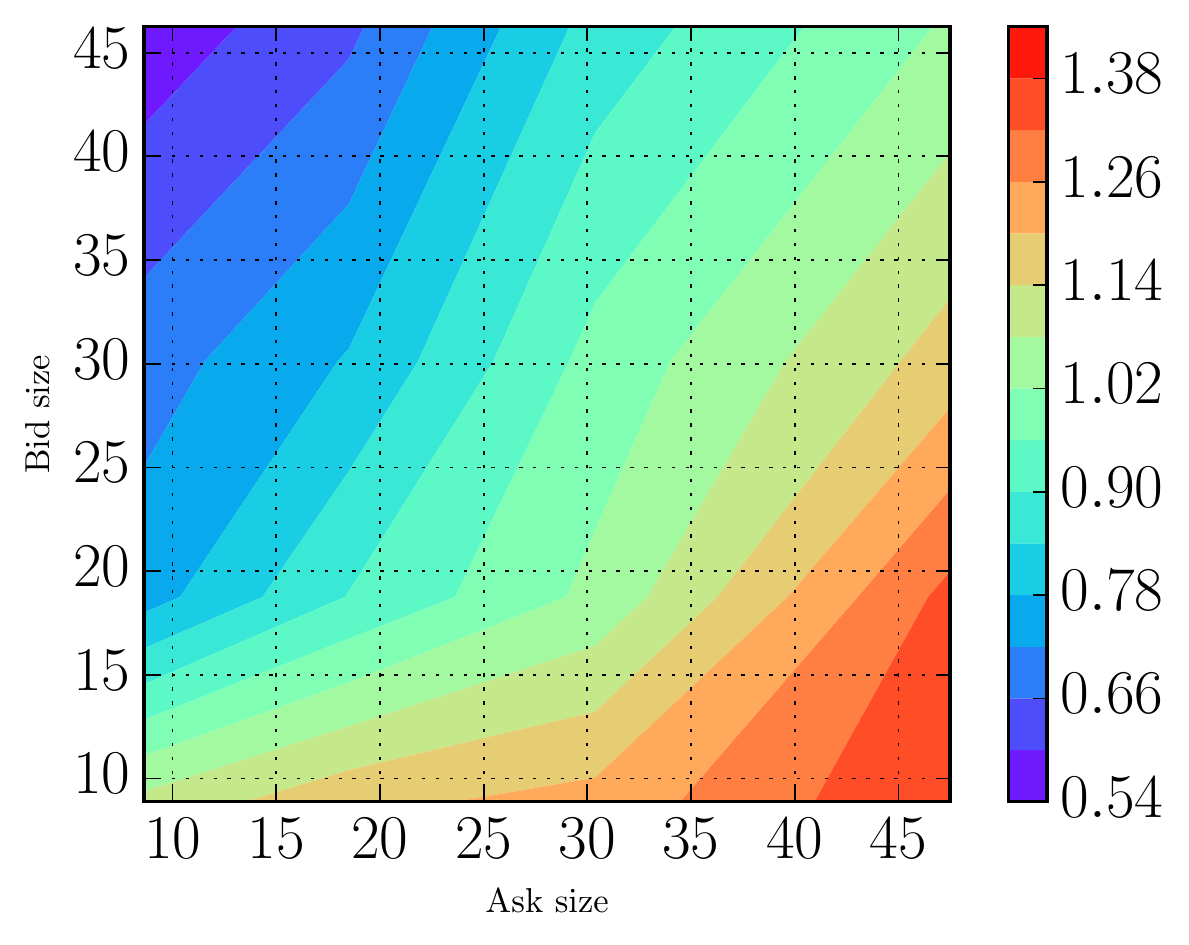}\hfill
	\includegraphics[width=.4\linewidth]{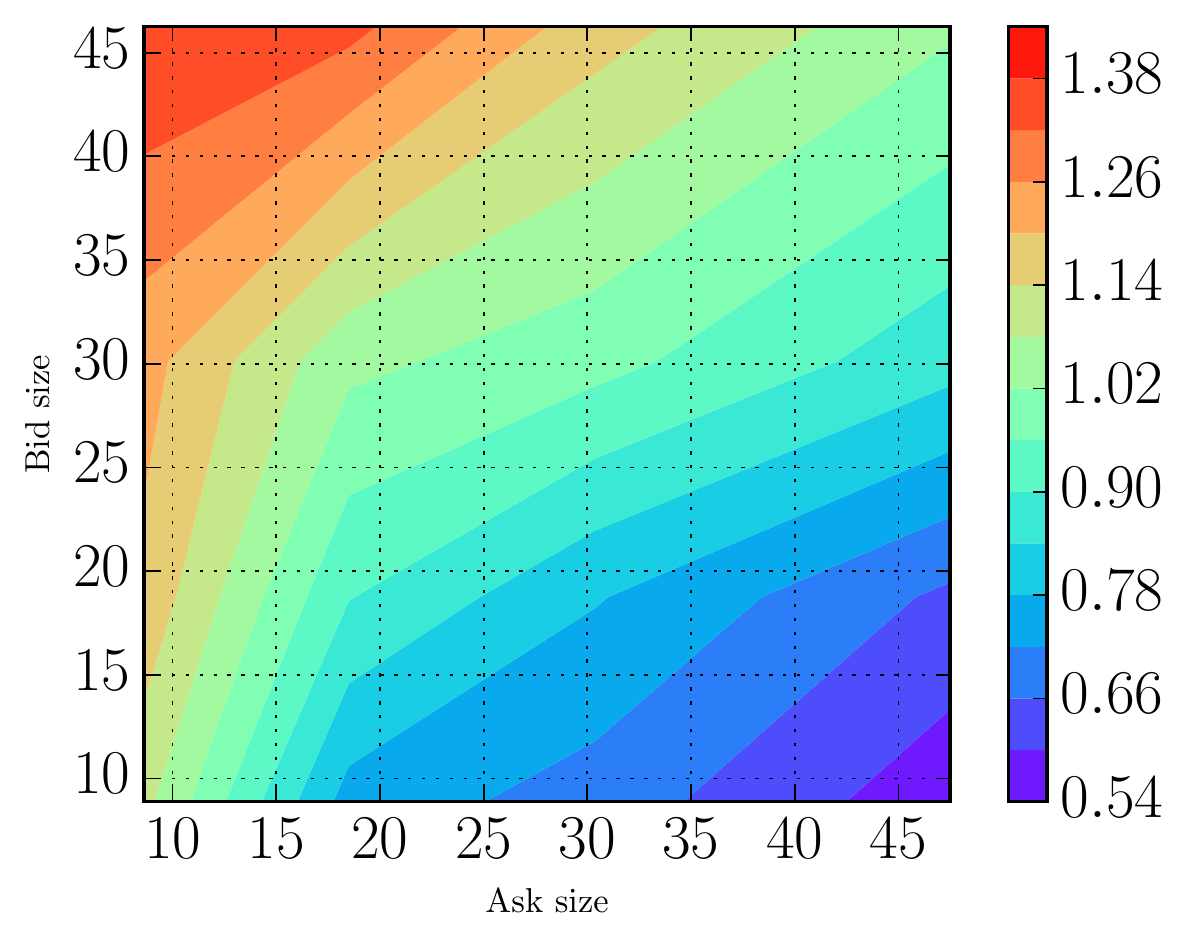}
 	\caption{(a) $\lambda^{1,+}$, (b) $\lambda^{1,-}$, (c) $\tau^{1,+}$  and (d) $\tau^{2,+}$ for different values of $(Q^{1},Q^{2})$. $Q^{1}$ and $Q^{2}$ are divided by the average event size.}  	
  	\label{fig:InsertCancelQRatioAgainstQSameQOpp}
\end{figure}

In what follows, we provide the calibration results of our order book model using the database described in Section \ref{subsec:EmpEvid}. Here, we write $Q_t =(Q^1_t,Q^2_t)$ with $Q^{1}_t$ (resp. $Q^{2}_t$) the best bid (resp. ask) quantity and consider that intensities and regeneration distributions depend only on $Q_t$. 
\paragraph{Intensities estimation.} For every $Q =(Q^1,Q^2)$, we write $\tau^{1,+}(Q) = \lambda^{1,+}/\lambda^{1,-}$ and $\tau^{2,+}(Q) = \lambda^{1,+}/\lambda^{1,-}$ respectively for the bid and ask side growth ratios. Given the bid-ask symmetry relation, we can aggregate data and focus on the bid side only. Figures \ref{fig:InsertCancelQRatioAgainstQSameQOpp}.a, \ref{fig:InsertCancelQRatioAgainstQSameQOpp}.b, \ref{fig:InsertCancelQRatioAgainstQSameQOpp}.c and \ref{fig:InsertCancelQRatioAgainstQSameQOpp}.d show respectively $\lambda^{1,+}$, $\lambda^{1,-}$, $\tau^{1,+}$ and $\tau^{2,+}$  for different values of $Q$. As expected, we can see that participants insert more limit orders when the imbalance is negative (see Figure \ref{fig:InsertCancelQRatioAgainstQSameQOpp}.a when $Q^{2} \gg Q^{1}$) while they cancel more when the imbalance is positive (see Figure \ref{fig:InsertCancelQRatioAgainstQSameQOpp}.b when $Q^{1} \gg Q^{2}$). Finally, Figure \ref{fig:InsertCancelQRatioAgainstQSameQOpp}.c (resp. Figure \ref{fig:InsertCancelQRatioAgainstQSameQOpp}.d) shows that $\tau^{1,+}$ (resp. $\tau^{2,+}$) is high when imbalance is negative (resp. positive) and becomes low when imbalance is positive (resp. negative) which means that the bid limit (resp. ask limit) tends to increase (resp. decrease) when  $Q^{1} \ll Q^{2}$ and tends to decrease (resp. increase) when $Q^{1} \gg Q^{2}$.

\begin{figure}[!ht]
	\centering
	\hspace{1cm}(a) $Q^{New,1}$ \hfill \hfill \hspace{0.6cm} (b) $Q^{New,2}$ \hfill~\\
	\includegraphics[width=.4\linewidth]{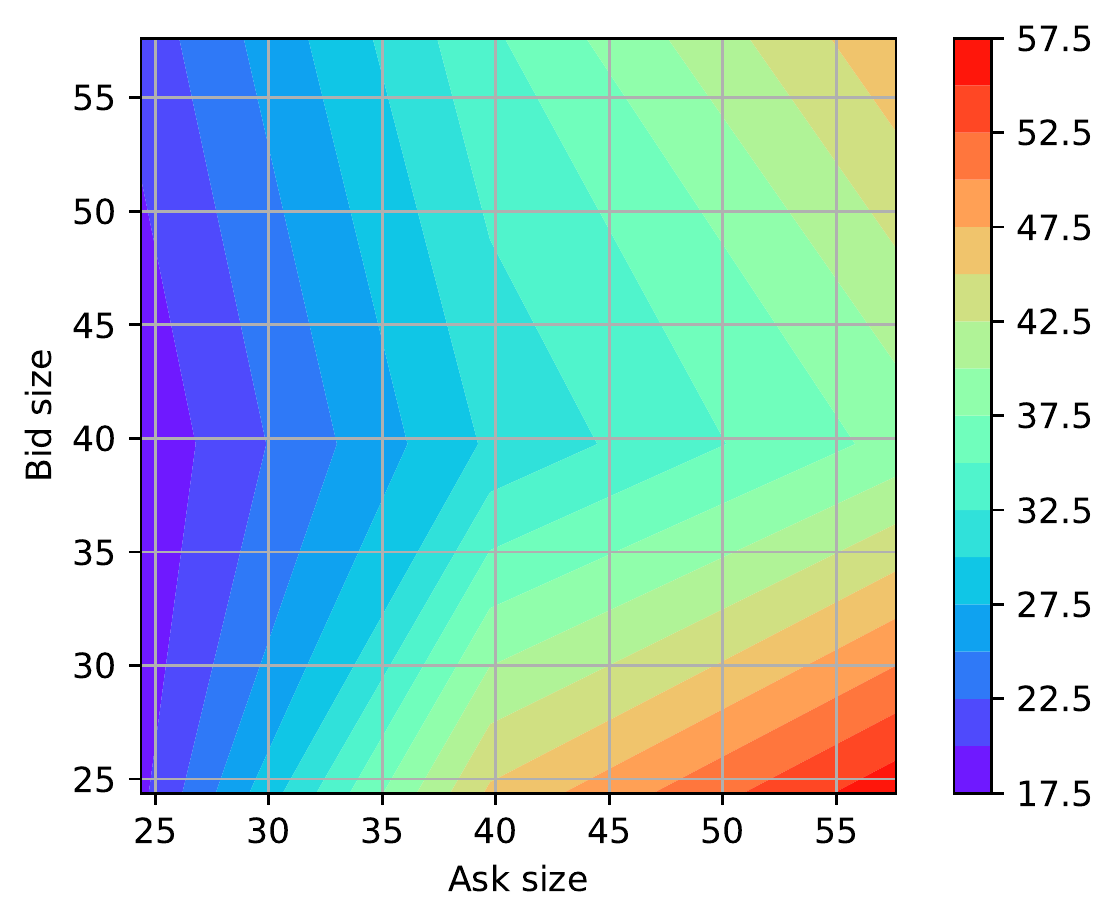}\hfill
	\includegraphics[width=.4\linewidth]{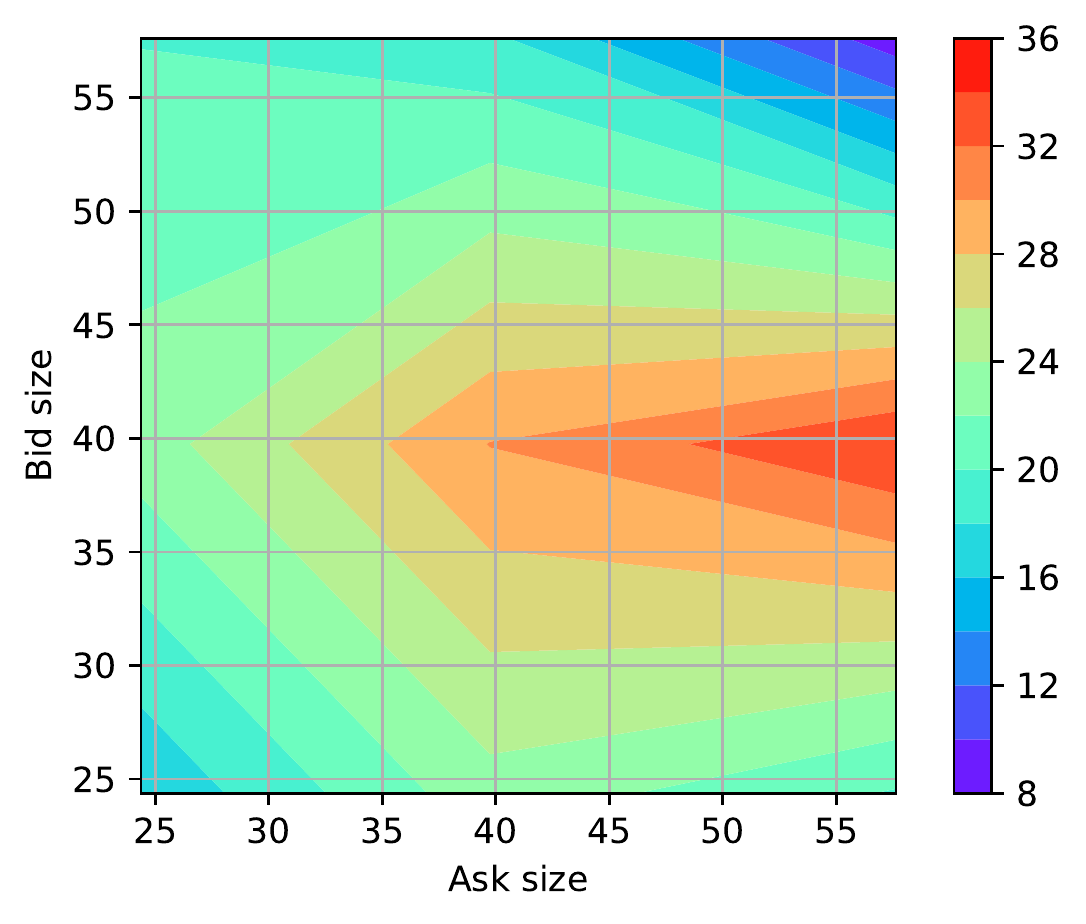}\\
	\hfill (c) $r^{+}$ \hfill ~\\
	\includegraphics[width=.4\linewidth]{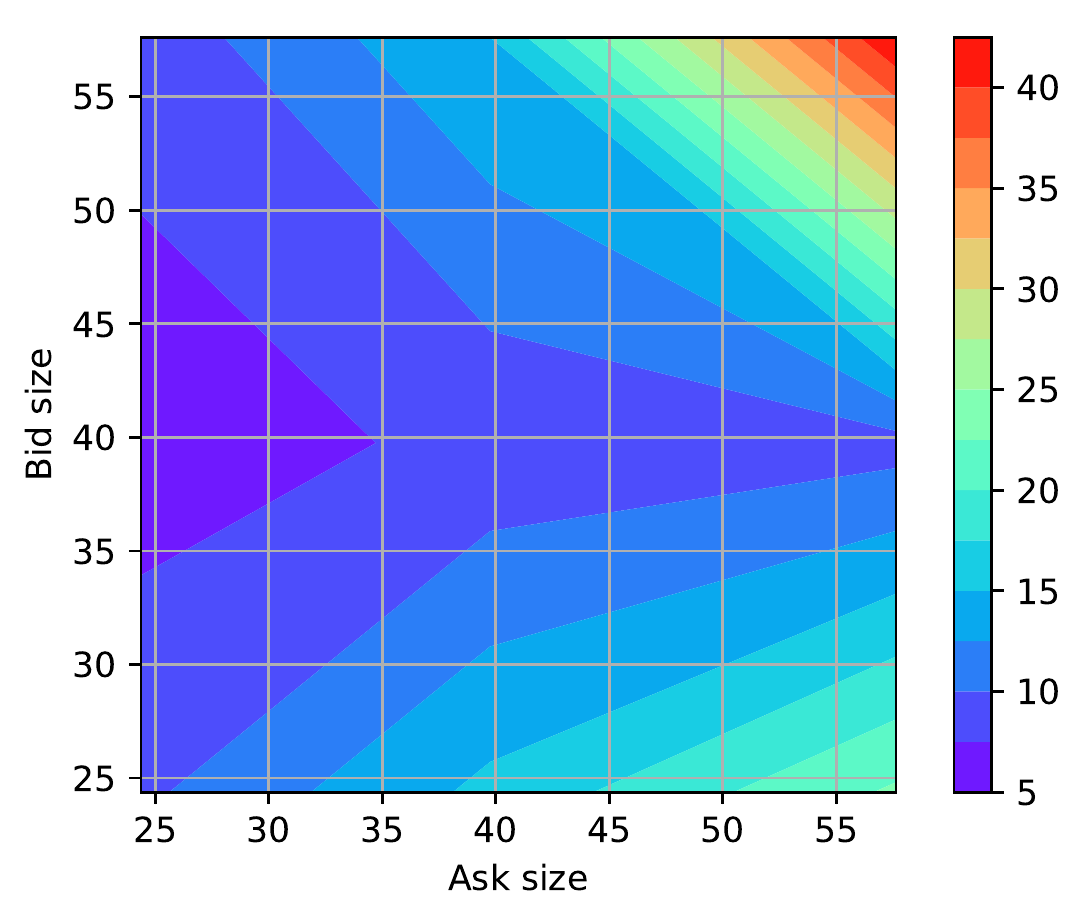}
 	\caption{(a) $Q^{New,1}$, (b) $Q^{New,2}$ and (c) $r^{+}$ for different values of  $Q^{1}$ and $Q^{2}$. $Q^{1}$ and $Q^{2}$ are divided by the average event size.}  	
  	\label{fig:QDIscInsDistrib}
\end{figure}

\paragraph{Quantities after depletion.} When one limit is depleted, we write $Q^{New,1}$ (resp. $Q^{New,2}$) for the new best bid (resp. ask). Figures \ref{fig:QDIscInsDistrib}.a, \ref{fig:QDIscInsDistrib}.b and  \ref{fig:QDIscInsDistrib}.c show respectively $Q^{New,1}$, $Q^{New,2}$ and the ratio $r^{+}(Q_1,Q_2) = \frac{Q^{New,1}}{Q^{New,2}}$ for different values of $Q^{1}$ and $Q^{2}$ before the mid price move. Since we aggregate data, the bid queue is always the depleted queue and the ask limit is the non-consumed limit. Figures \ref{fig:QDIscInsDistrib}.a and \ref{fig:QDIscInsDistrib}.b show that $Q^{New,1}$  depends mainly on $Q^{2}$  while $Q^{New,2}$ depends on both $Q^{1}$ and $Q^{2}$ . However, the interesting point is that  $r^{+}$ reach its maxima in two cases, see Figure \ref{fig:QDIscInsDistrib}.c. The first case, when the bid is low and the ask is high, can be explained by a mean reversion effect while the second one, when both queues are initially high, is due to the arrival of a large order consuming market liquidity.

\begin{figure}[!ht]
 \centering
 \footnotesize 
  \hfill Theoretical average mid price move at infinity 
  \hfill~\\
	\includegraphics[width=.4\linewidth]{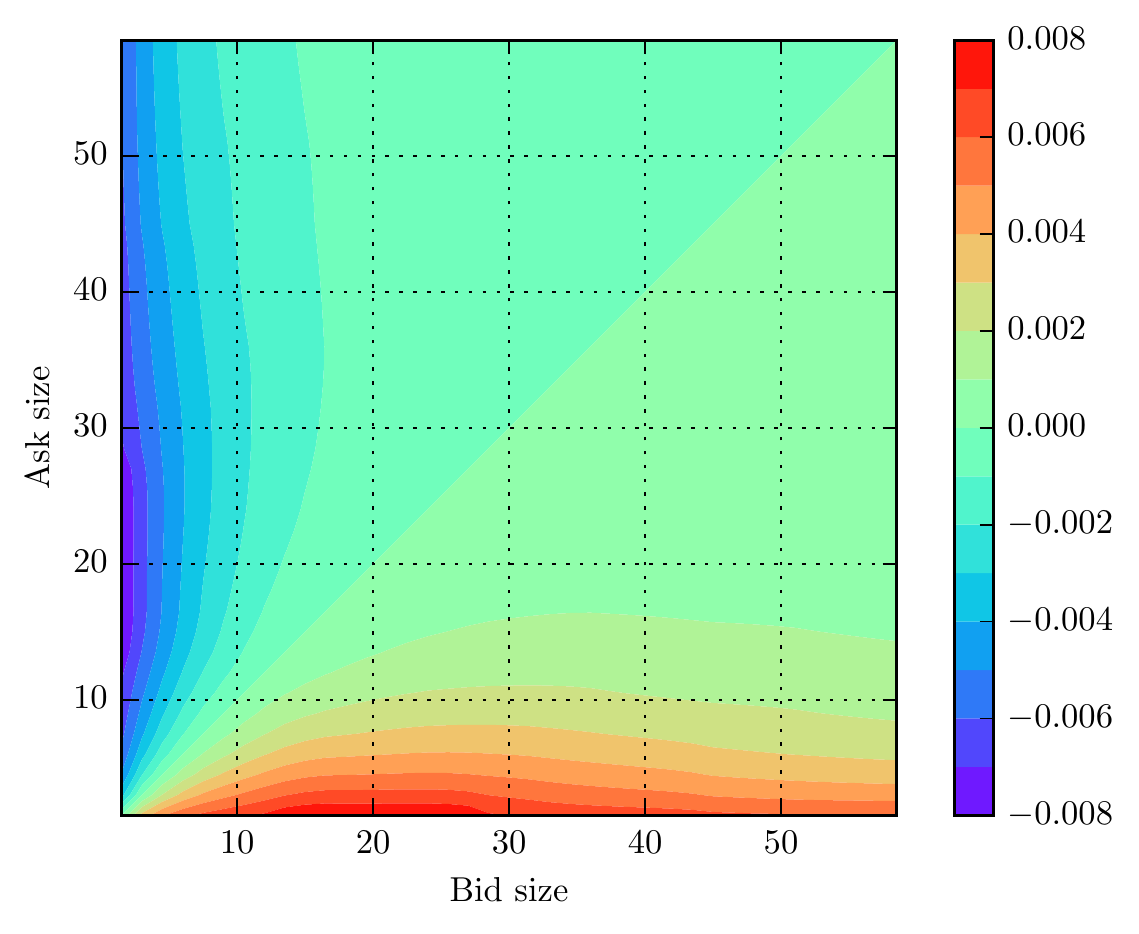}
 	\caption{ Theoretical average mid price move at infinity. $Q^{1}$ and $Q^{2}$ are divided by the average event size and the tick $\delta = 0.01 $.}  	
  	\label{fig:AveragePriceMoveValue}
\end{figure}

\begin{figure}[!ht]
	\centering
	 \hspace{0.1cm} \begin{tabular}{c}
	 (a) Empirical $Q^{2}$ distribution \\
	 after 20 events \\
	 \end{tabular}  \hfill \hspace{4cm}
	 \begin{tabular}{c}
	 (b) Theoretical $Q^{2}$ distribution \\
	 after 20 events \\
	 \end{tabular}   \hfill~\\ 
	\includegraphics[width=.4\linewidth]{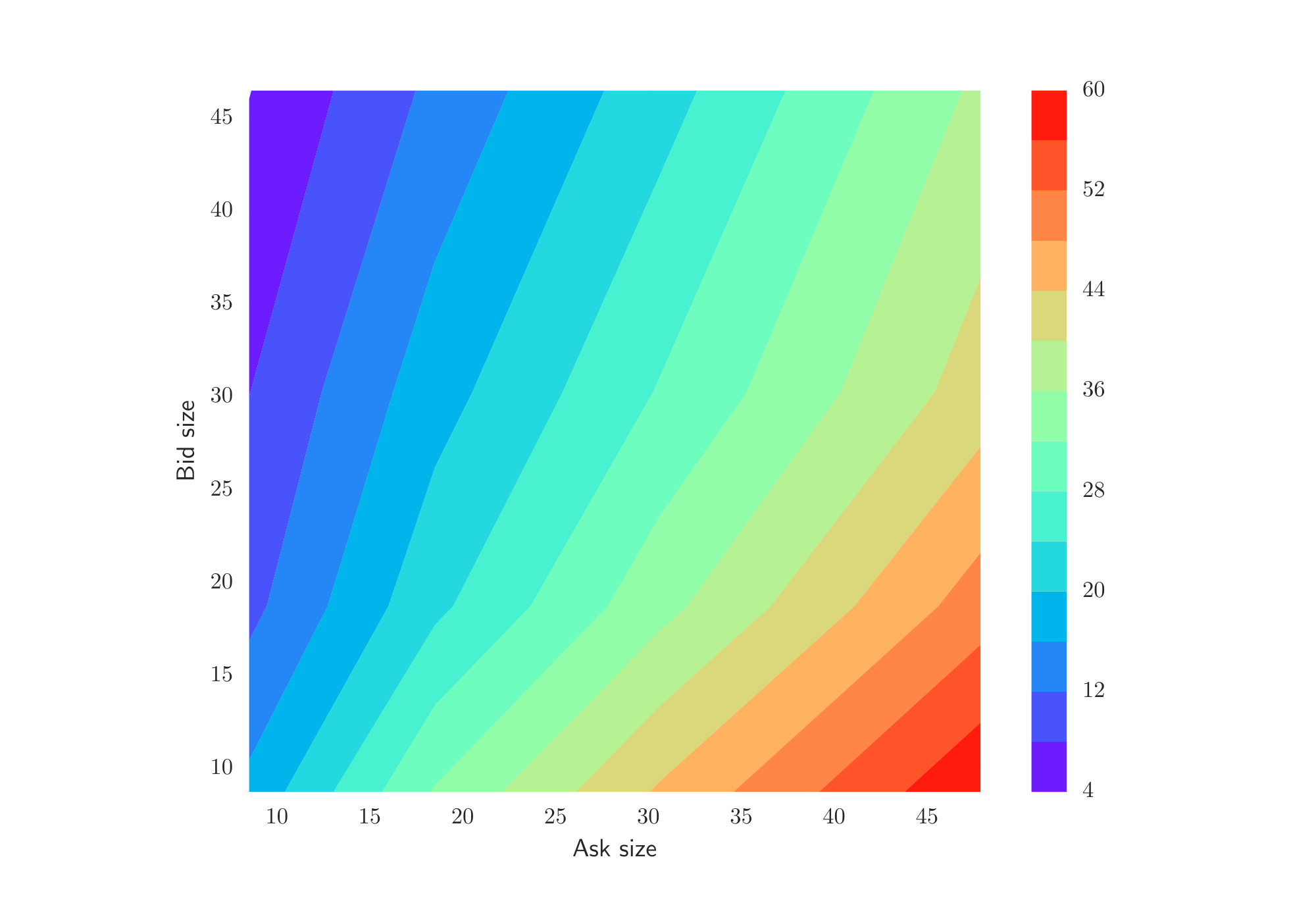}\hfill
	\includegraphics[width=.4\linewidth]{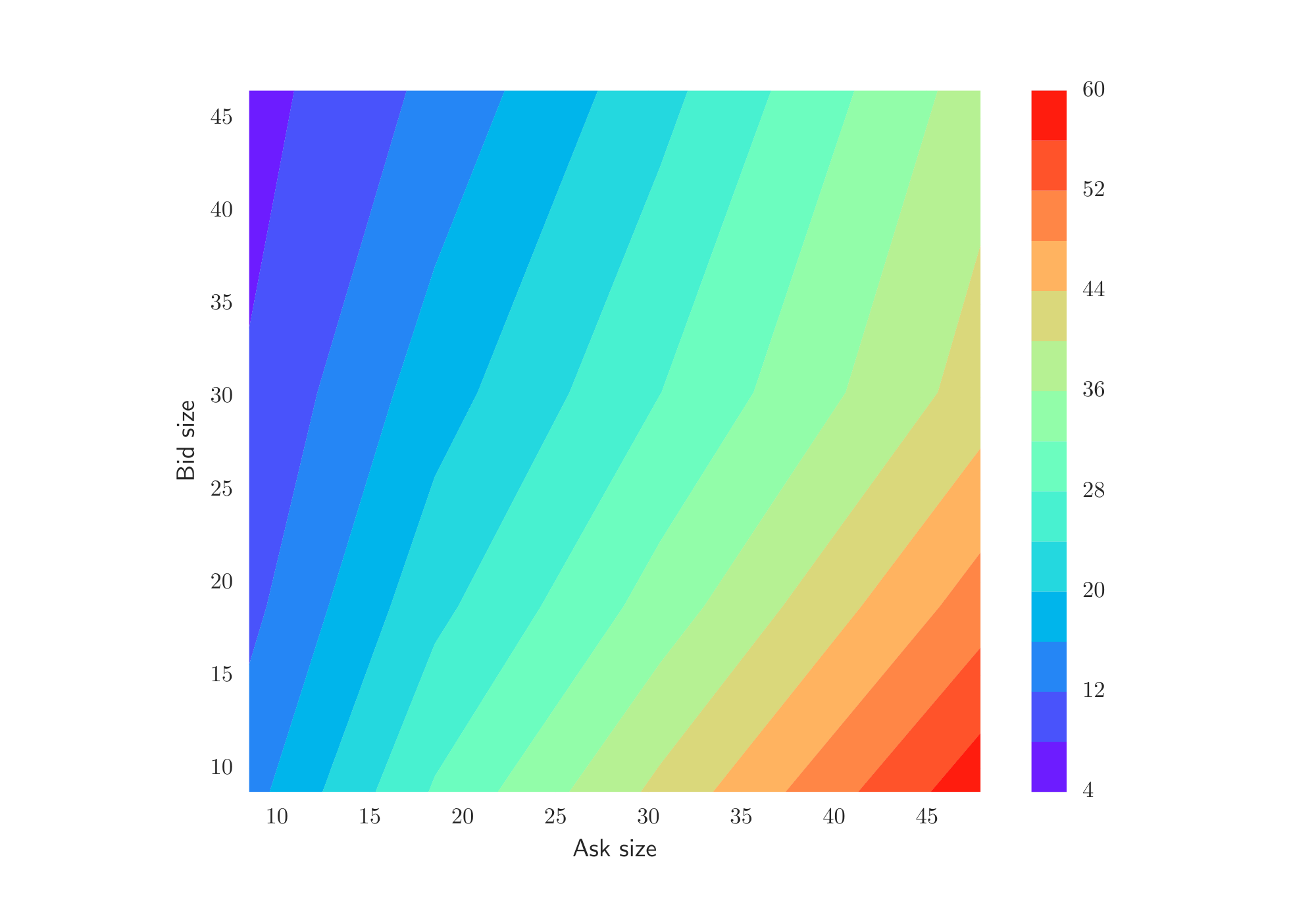}\\
	\hspace{0.cm} \begin{tabular}{c}
	 (c) Empirical $Q^{1}$ distribution \\
	 after 20 events \\
	 \end{tabular} \hspace{4cm}
	 \begin{tabular}{c}
	 (d) Theoretical $Q^{1}$ distribution \\
	 after 20 events \\
	 \end{tabular} 
	\includegraphics[width=.4\linewidth]{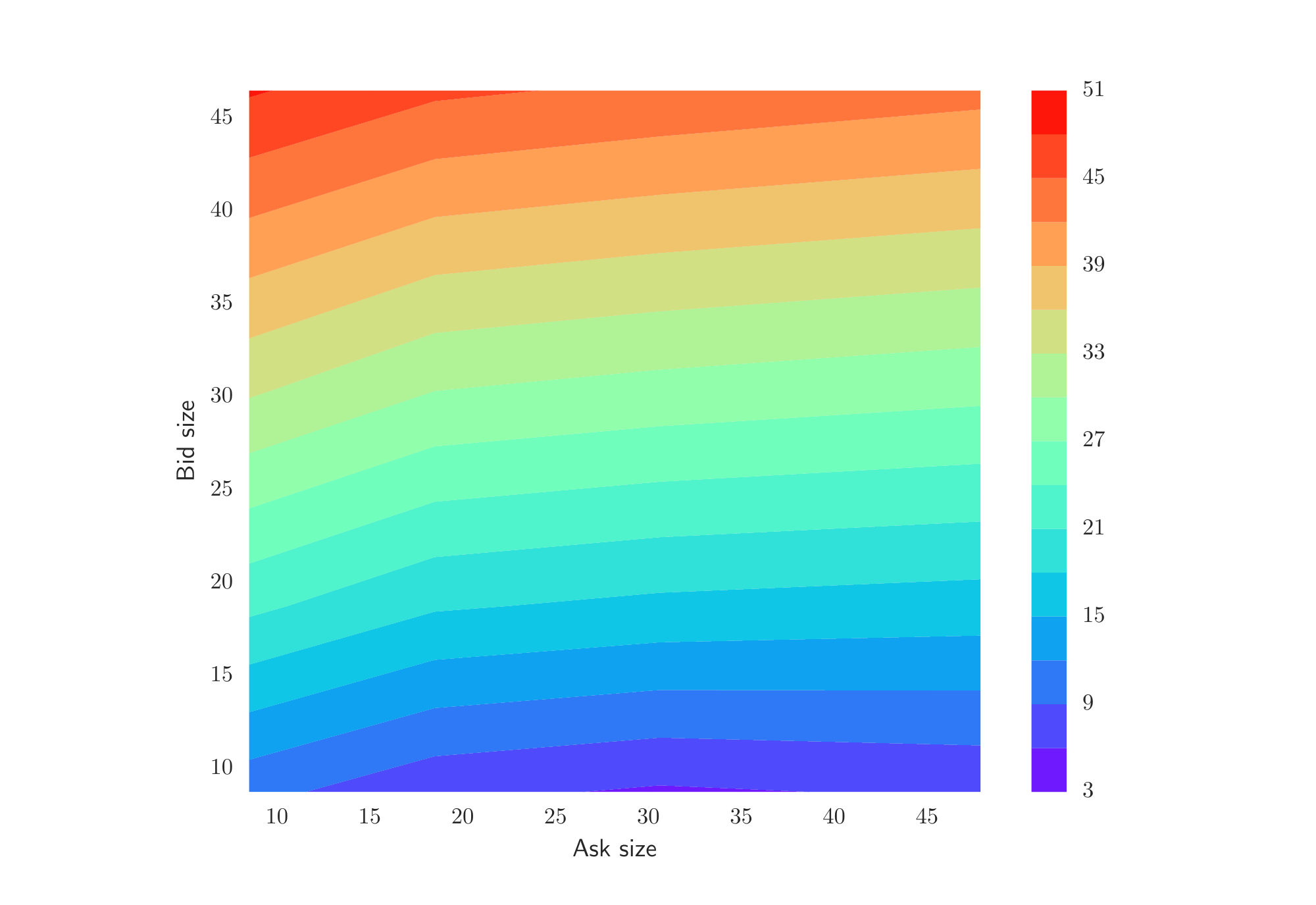}\hfill
	\includegraphics[width=.4\linewidth]{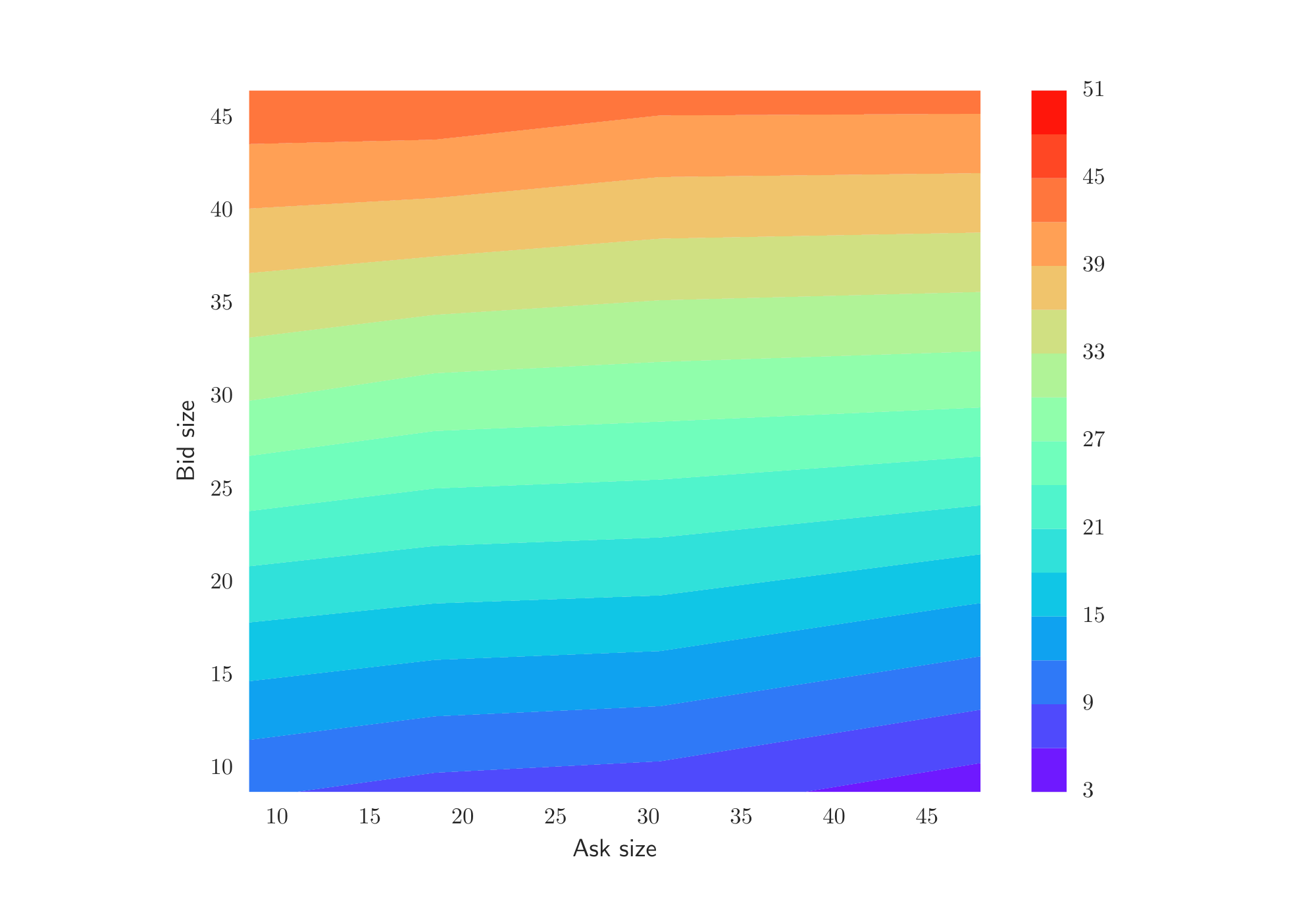}
 	\caption{(a) (resp. (c)) Empirical distribution of $Q^{2}$ (resp. $Q^{1}$) after 20 events and (b) (resp. (d)) theoretical distribution of $Q^{2}$ (resp. $Q^{1}$). $Q^{1}$ and $Q^{2}$ are divided by the average event size.}  	
  	\label{fig:EmpiricalDistrCtrlDur}
\end{figure}

\paragraph{Approximation of $\Esp_{U_0}[\Delta P_{\infty}]$.} Figure \ref{fig:AveragePriceMoveValue} shows $\Esp_{U_0}[\Delta P_{\infty}]$ defined in Section \ref{subsec:CompAverPMove} and computed using Proposition \ref{Theo:AvgPriceMoveComput}, for different values of the initial state $U_0 = (Q^{1},Q^{2},P)$. Figure \ref{fig:AveragePriceMoveValue} shows the predictive power of the imbalance : when the imbalance is positive the price increases on average and conversely. We also note that the bid-ask symmetry relation is respected. 
\paragraph{Model approximation at short-time horizon.}  Figures \ref{fig:EmpiricalDistrCtrlDur}.a, \ref{fig:EmpiricalDistrCtrlDur}.b, \ref{fig:EmpiricalDistrCtrlDur}.c and \ref{fig:EmpiricalDistrCtrlDur}.d  show respectively the empirical and theoretical distributions of $Q^{1}$ and $Q^{2}$ after 20 events. We choose 20 events since it is coherent with the duration of our control. The estimation of the theoretical distribution is based on a Monte-Carlo simulation of the order book. We can see that both distributions are close and consequently that our model is consistent with the empirical order book dynamic at least during the control duration. The model is also consistent with empirical data on long term horizon, see \cite{citeulike:12810809}.
\label{sec:ModelParamEstim}
\clearpage

\section{Ergodicity of the process $(Q_t)$}
\label{sec:ErgValFct}
\subsection{Outline of the proof}
Let $Z_t$ be a Markov process defined on the probability space $(\Omega, \mathcal{F},\mathcal{F}_t,\mathbb{P})$ and valued in $(W,\mathcal{W})$ and $P_t(x,A)$ the probability transition of $Z_t$.
\begin{defi}[Ergodicity] The process $Z_t$ is ergodic if there exists an invariant probability measure $\pi$ which satisfies
$$
\underset{t \rightarrow \infty}{\lim} ||P_t(x,.) - \pi(.) ||_{TV} = 0,
$$
where $||\mu - \mu'||_{TV} = \underset{A \in \mathcal{F}}{\sup}|\mu(A) - \mu'(A) | $.
\end{defi}
Let $\mathcal{Q}$ be the infinitesimal generator of $Q_t$. To prove that $Q_t$ is ergodic, we design a Lyapunov function $ V: \mathbb{R}_+^2 \rightarrow (0,\infty)$, on which the negative drift condition is satisfied for some $c>0$ and $d>0$:
\begin{equation*}
\mathcal{Q}V(q) \leq -cV(q)+d.
\end{equation*}
Then, using Theorem 6.1 in \cite{meyn1993stability}, the Markov process $Q_t$ is non-explosive and V-uniformly ergodic. Furthermore, by Theorem 4.2 in \cite{meyn1993stability}, it is Harris positive recurrent. 
\subsection{Proof}
The constants $\delta$, $z_0$, $z_1$, $z_2$, $C_{bound}$, $H$, $C^{Disc}$ and $L$ are defined in Assumptions \ref{Assump:Negativeindividualdrift}, \ref{Assump:Boundontheincomingflow}, \ref{Assump:RegenerationBound} and \ref{Assump:JumpsBound}. Let $0<\epsilon<\delta$, $1 < z <\min(z_0,z_1,z_2) $, and $q = (q^1,q^2) \in (\mathbb{N}^*)^2$. Using Assumptions \ref{Assump:Negativeindividualdrift} and \ref{Assump:JumpsBound}, there exists $k_0$ such that
\begin{equation*}
\left\{
\begin{array}{ll}
\displaystyle \sum_{k >k_0} \lambda^{i,\pm}(q,k)z^k < \frac{\epsilon}{2} , & \forall i \in  \left\{1, 2\right\} \\
\displaystyle \sum_{n \leq k_0}(z_0^n -1)(\lambda^{i,+}(q,n) - \lambda^{i,-}(q,n)\frac{1}{z^n_0}) \leq -\frac{\delta}{2}, & \forall q^{i} \geq C_{bound}, \forall i \in \{1,2\}.
\end{array}
\right.
\end{equation*}
Let $C'_{bound} = \max(C_{bound}, C^{Disc})$ and $Q^{Disc,i,q} = (U^{Disc,i,q}_1,U^{Disc,i,q}_2) $ with $U^{Disc,i,q}$ a random variable with law $d^i_q$ for any $i \in \{1,2\}$. We define
$$
V(q) = \sum_{i=1}^2 z^{|q^i-C'_{bound}|_+}.
$$
To simplify notations we do not write the dependence of $\lambda^{1,\pm}(k)$ and $\lambda^{2,\pm}(k)$ on $q$. 
\begin{align*}
\mathcal{Q} V(q) & = \sum_{q'\ne q} \mathcal{Q}_{q,q'}\left[ V(q') - V(q) \right] \\
				 & = \sum_{i=1}^{2}\sum_{1\leq k \leq k_0} \overbrace{\left[ \lambda^{i,+}(k) (z^{|q^i+k  - C_{bound}'|_{+}} - z^{|q^i  - C_{bound}'|_{+}}) + \lambda^{i,-}(k) (z^{|q^i-k  - C_{bound}'|_{+}} - z^{|q^i  - C_{bound}'|_{+}}) \right]}^{(1) \text{ small jumps}} \\
				 & + \sum_{i=1}^2\sum_{k > k_0} \overbrace{\left[ \lambda^{i,+}(k) (z^{|q^i+k  - C_{bound}'|_{+}} - z^{|q^i  - C_{bound}'|_{+}})\right]}^{(2) \text{ large limit orders insertion}} \\
				 & + \sum_{i=1}^2\sum_{k_0 < k < u'_i } \overbrace{\left[ \lambda^{i,-}(k) (z^{|q^i-k  - C_{bound}'|_{+}} - z^{|q^i - C_{bound}'|_{+}})\right]}^{(3) \text{ large limit orders consumption}} + \sum_{i=1}^2\sum_{k \geq u'_i} \overbrace{\left[ \lambda^{i,-}(k) (\Esp_{q}[V(Q^{Disc,i,q})] - V(q))\right]}^{(4) \text{ order book regeneration}}.\\
\end{align*}
When $u'_i \leq k_0$ with $i \in \{1,2\}$, we only need to add a constant to the expression above using Assumptions \ref{Assump:Boundontheincomingflow} and \ref{Assump:JumpsBound}. We remark that Term $(3) \leq 0 $ and $(4) \leq \lambda^{i,-}(k)L$. Furthermore for Term $(1)$ we have
\begin{align*}
(1) & = \lambda^{i,+}(k)(z^{|q^i+k-C'_{bound}|_+} - z^{|q^i-C'_{bound}|_+}) + \lambda^{i,-}(k) (z^{|q^i-k  - C_{bound}^{'}|_{+}} - z^{|q^i  - C_{bound}^{'}|_{+}})\\
	& = \lambda^{i,+}(k)\mathbf{1}_{C'_{bound}+k\geq q^i}(z^{|q^i+k-C'_{bound}|_+} - z^{|q^i-C'_{bound}|_+}) + \lambda^{i,+}(k)\mathbf{1}_{C'_{bound}+k< q^i}z^{|q^i-C'_{bound}|_+} (z^k-1) \\
	& + \underbrace{\lambda^{i,-}(k)\mathbf{1}_{C'_{bound}+k\geq q^i}(z^{|q^i-k  - C_{bound}^{'}|_{+}}- z^{|q^i-C'_{bound}|_+})}_{\leq0} + \lambda^{i,-}(k)\mathbf{1}_{C'_{bound}+k< q^i}z^{|q^i-C'_{bound}|_+} (\frac{1}{z^k}-1) \\
	& \leq \overbrace{\lambda^{i,+}(k)\mathbf{1}_{C'_{bound}+k\geq q^i}(z^{2k}-1)}^{(1.1)} + \overbrace{(z^{k}-1)z^{|q^i-C'_{bound}|_+}\mathbf{1}_{C'_{bound}+k< q^i}(\lambda^{i,+}(k)-\lambda^{i,-}(k)\frac{1}{z^k})}^{(1.2)}. \\
\end{align*}
Using Assumption \ref{Assump:Boundontheincomingflow} for (1.1), we get $(1.1) \leq H (z^{2k_0}-1)$.
Using Assumption \ref{Assump:Boundontheincomingflow} for (1.2), we also obtain
\begin{align*}
(1.2) & \leq (z^{k}-1)z^{|q^i-C'_{bound}|_+}(\lambda^{i,+}(k)-\lambda^{i,-}(k)\frac{1}{z^k}) - (z^{k}-1)z^{|q^i-C'_{bound}|_+}\mathbf{1}_{C'_{bound}+k\geq q^i}(\lambda^{i,+}(k)-\lambda^{i,-}(k)\frac{1}{z^k}) \\
&  \leq (z^{k}-1)z^{|q^i-C'_{bound}|_+}(\lambda^{i,+}(k)-\lambda^{i,-}(k)\frac{1}{z^k}) + (z^{k_0}-1)z^{k_0}(\lambda^{i,+}(k)+\lambda^{i,-}(k)).
\end{align*}
We write $M(k_0)$ for $(z^{k_0}-1)z^{k_0}$. 
Finally, by combining the above inequalities we have
\begin{align*}
\mathcal{Q}V(u') & \leq \big(\sum_{i=1}^2 \sum_{1\leq k \leq k_0}(z^{k}-1)z^{|q^i-C'_{bound}|_+}(\lambda^{i,+}(k)-\lambda^{i,-}(k)\frac{1}{z^k})+M(k_0)(\lambda^{i,+}(k)+\lambda^{i,-}(k))\big) \\
				& + \sum_{i\leq2;k>k_0}\lambda^{i,+}(k)(z^k-1)z^{|q^i-C'_{bound}|_+} + \sum_{i\leq 2;k >u'_i} \lambda^{i,-}(k) L  \\
				& \leq \sum_{i=1}^2(z^{|q^i-C'_{bound}|_+})\big(\sum_{1\leq k \leq k_0}(z^{k}-1)(\lambda^{i,+}(k)-\lambda^{i,-}(k)\frac{1}{z^k})\big) + 2HM(k_0)\\
				& \quad +\frac{\epsilon}{2} \sum_{i=1}^2(z^{|q^i-C'_{bound}|_+}) + 2HL \\
				& = \sum_{i=1}^2(z^{|q^i-C'_{bound}|_+})\big(\overbrace{\sum_{1\leq k \leq k_0}(z^{k}-1)(\lambda^{i,+}(k)-\lambda^{i,-}(k)\frac{1}{z^k})}^{\text{Assumption \ref{Assump:Negativeindividualdrift}}} + \frac{\epsilon}{2} \big)+ d	\\
				& \leq -\frac{1}{2}\delta^{(3)}V(q)+d,
\end{align*}
with $\delta^{(3)} = (\delta-\epsilon)$  and $ d = 2HM(k_0) +2HL$. Since $\epsilon <\delta$, we have $\delta^{(3)}>0 $. This completes the proof.\\
\clearpage

\section{The infinitesimal generator of the process $U^{\mu}$}
\label{sec:InftGen} 
To fully describe the dynamic of $U^{\mu}_t$, we need to define the infinitesimal generator $\mathcal{Q}^{\mu}$ of the process $U^{\mu}_t$ whithout our intervention. We write $\tilde{d}^{i}_{u} = d^{i}_{u} \iota^i_{u}$ for the joint regeneration density of the order book and the agent placement. Indeed after a regeneration, the agent's order is placed in a position $q^{bef}_0$ with a probability $\iota^i_{u}(q^{bef}_0)$. 
\paragraph{$\mathcal{Q}^{\mu}$ description:} For $\tilde{q} = (q^{bef},q^a,q^{aft},q^2,i) \in \mathbb{N}^5$, $p \in \mathbb{R}$, $\tilde{z} = (p,p^{exec}) \in \mathbb{R}^2$, $n \in \mathbb{N} $, $q^1= q^{bef}+q^a+q^{aft}$, $u=(q^1,q^2,p) \in (\mathbb{N}^{*})^2$, $\tilde{q}'=(q'^{bef},q'^a,q'^{aft},q'^2,i')\in \mathbb{N}^5$, $q'^1= q'^{bef}+q'^a+q'^{aft}$, $u'=(q'^1,q'^2,p') \in (\mathbb{N}^{*})^2$, $\tilde{z}' = (p',p'^{exec})\in \mathbb{R}^2$ and  $\lambda^{1,-}(u,k) =\big(\lambda^{1,-}_c(u,k)\mathbf{1}_{q^a = 0} +\lambda^{1,-}_m(u,k)\big)$, we have:
\begin{itemize}
\item When a limit order of size $n$ is inserted at the best bid:
\begin{equation*}
\mathcal{Q}^{\mu}_{(\tilde{q},\tilde{z});(\tilde{q}',\tilde{z}')} = \lambda^{1,+}(u,n) + \sum_{i=1}^{2}\sum_{k\geq q^i} \lambda^{i,-}(u,k)\tilde{d}^{i}_{(\tilde{q},\tilde{z})}(\tilde{q}',\tilde{z}'),
\end{equation*}
where $\tilde{q}' = (q^{bef},q^a,q^{aft}+n,q^2,i)$ and $\tilde{z}'=\tilde{z}$.
\item When a limit order of size $n$ is inserted at the best ask: 
\begin{equation*}
\mathcal{Q}^{\mu}_{(\tilde{q},\tilde{z});(\tilde{q}',\tilde{z}')} = \lambda^{2,+}(u,n) + \sum_{i=1}^{2}\sum_{k\geq q^i} \lambda^{i,-}(u,k)\tilde{d}^{i}_{(\tilde{q},\tilde{z})}(\tilde{q}',\tilde{z}'),
\end{equation*}
where $\tilde{q}' = (q^{bef},q^a,q^{aft},q^2+n,i)$ and $\tilde{z}'=\tilde{z}$.
\item When a cancellation order of size $n$ is removed from the best bid
\begin{enumerate}
\item When $n > q^{bef}+q^{aft}$:
\begin{equation*}
\mathcal{Q}^{\mu}_{(\tilde{q},\tilde{z});(\tilde{q}',\tilde{z}')} = \lambda^{1,-}_c(u,n)\mathbf{1}_{q^a \ne 0} +\sum_{i=1}^{2} \sum_{k\geq q^i} \lambda^{i,-}(u,k)\tilde{d}^{i}_{(\tilde{q},\tilde{z})}(\tilde{q}',\tilde{z}'),
\end{equation*}
where $\tilde{z}'$ and $\tilde{q}'$ satisfy 
\begin{equation*}
\left\{
\begin{array}{lcl}
p'^{exec} & = & p^{exec}\\
q'^{bef} & = & q'^{bef}_0 \mathbf{1}_{q^a = 0}\\
q'^{a}& = &  q^a\\
i'    & = &  i\\
q'^{aft} & = & (q'^1-q'^{bef})\mathbf{1}_{q^a = 0}, \\
\end{array}
\right.
\end{equation*} 
and $p'$, $q'^1$, $q'^{bef}_0$ and $q'^2$ are fixed by the regeneration law.
\item When $n < q^{bef} +q^{aft}$:
\begin{equation*}
\mathcal{Q}^{\mu}_{(\tilde{q},\tilde{z});(\tilde{q}',\tilde{z}')} = \lambda^{1,-}_c(u,n)+ \sum_{i=1}^{2} \sum_{k\geq q^i} \lambda^{i,-}(u,k)\tilde{d}^{i}_{(\tilde{q},\tilde{z})}(\tilde{q}',\tilde{z}'),
\end{equation*}
where $\tilde{z}'=\tilde{z}$ and $\tilde{q}'$ satisfies
\begin{equation*}
\left\{
\begin{array}{lcl}
q'^{bef} & = & q^{bef} \mathbf{1}_{n \leq q^{aft}} + (q^{bef}+q^{aft}-n)\mathbf{1}_{n > q^{aft}}\\
q'^{a}& = & q^a\\
i'    & = & i\\
q'^{aft} & = & (q^{aft}-n) \mathbf{1}_{n \leq q^{aft}} \\
q'^2 & = & q^2.\\
\end{array}
\right.
\end{equation*} 
\end{enumerate} 
\item When a market order of size $n$ is sent to the best bid
\begin{enumerate}
\item When $n> q^{bef}+q^a+q^{aft}$:
\begin{equation*}
\mathcal{Q}^{\mu}_{(\tilde{q},\tilde{z});(\tilde{q}',\tilde{z}')} = \sum_{i=1}^{2} \sum_{k\geq q^i} \lambda^{i,-}(u,k)\tilde{d}^{i}_{(\tilde{q},\tilde{z})}(\tilde{q}',\tilde{z}'),
\end{equation*}
where $\tilde{z}'$ and $\tilde{q}'$ satisfy 
\begin{equation*}
\left\{
\begin{array}{lcl}
p'^{exec} & = & p^{exec} + q^a (p - \frac{\psi}{2})\\
q'^{bef} & = & q'^{bef}_0\\
q'^{a}& = & 0\\
i'    & = & i \mathbf{1}_{q^a = 0}\\
q'^{aft} & = & 0 ,\\
\end{array}
\right.
\end{equation*} 
where $p'$, $q'^1$, $q'^{bef}_0$ and $q'^2$ are fixed by the regeneration law.
\item Otherwise, we have: 
\begin{equation*}
\mathcal{Q}^{\mu}_{(\tilde{q},\tilde{z});(\tilde{q}',\tilde{z}')} = \lambda^{1,-}_m(u,n)+\sum_{i=1}^{2} \sum_{k\geq q^i} \lambda^{i,-}(u,k)\tilde{d}^{i}_{(\tilde{q},\tilde{z})}(\tilde{q}',\tilde{z}'),
\end{equation*}
where $\tilde{z}'$ and $\tilde{q}'$ satisfy 
\begin{equation*}
\left\{
\begin{array}{lcl}
p' & = & p \\
p'^{exec} & = & p^{exec} + \min(n-q^{bef},q^a) (p - \frac{\psi}{2})\mathbf{1}_{n > q^{bef}}\\
q'^{bef} & = & (q^{bef}-n) \mathbf{1}_{n \leq q^{bef}} \\
q'^{a}& = & q^a-\min(n-q^{bef},q^a)\mathbf{1}_{n > q^{bef}}\\
i'    & = & i\mathbf{1}_{q^a = 0} + q'^{a}\mathbf{1}_{q^a > 0}\\
q'^{aft} & = & q^{aft} \mathbf{1}_{n < q^{bef}} + (q^{aft}+q^{bef}+q^a-n) \mathbf{1}_{n \geq q^{bef}+q^a}\\
q'^2 & = & q^2.\\
\end{array}
\right.
\end{equation*} 
\end{enumerate}
\item When there is a liquidity consumption event at the best ask
\begin{enumerate}
\item When $n> q^{2}$:
\begin{equation*}
\mathcal{Q}^{\mu}_{(\tilde{q},\tilde{z});(\tilde{q}',\tilde{z}')} = \sum_{i=1}^{2} \sum_{k\geq q^i} \lambda^{i,-}(u,k)\tilde{d}^{i}_{(\tilde{q},\tilde{z})}(\tilde{q}',\tilde{z}'),
\end{equation*}
where $\tilde{z}'$ and $\tilde{q}'$ satisfy 
\begin{equation*}
\left\{
\begin{array}{lcl}
p'^{exec} & = & p^{exec}\\
q'^{bef} & = & q'^{bef}_0\\
q'^{a}& = & q^a\\
i'    & = & i\\
q'^{aft} & = & 0, \\
\end{array}
\right.
\end{equation*} 
where $p'$, $q'^1$ $q'^{bef}_0$ and  $q'^2$ are fixed by the regeneration law.
\item Otherwise, we have: 
\begin{equation*}
\mathcal{Q}^{\mu}_{(\tilde{q},\tilde{z});(\tilde{q}',\tilde{z}')} = \lambda^{2,-}(u,n)+\sum_{i=1}^{2} \sum_{k\geq q^i} \lambda^{i,-}(u,k)\tilde{d}^{i}_{(\tilde{q},\tilde{z})}(\tilde{q}',\tilde{z}'),
\end{equation*}
where $\tilde{z}' = \tilde{z}$ and $\tilde{q}' = (q^{bef},q^a,q^{aft},q^2-n,i)$.
\end{enumerate}
\end{itemize}

\section{Computation of $\Delta P^{\mu}_{\infty}$}
\label{sec:AveragePriceMoveAfterExec}
\subsection{Proof of Proposition \ref{Theo:AvgPriceMoveComput}}
\manuallabel{subsec:CompPinft1}{.1}
Under Assumptions \ref{Assump:InsertionBound} and \ref{Assump:RegenerationLimit}, the number of reachable states $N$ is finite and we can write 
\begin{align*}
\Esp_{U_i}[\Delta P^{'}_{\infty}]& = \Esp_{U_i}[\Delta P^{'}_{\infty}\mathbf{1}_{t_2\leq t_1}] + \Esp_{U_i}[\Delta P^{'}_{\infty}\mathbf{1}_{ t_1 < t_2}] \\
								& = \Esp_{U_i}[\Esp[\Delta P^{'}_{\infty}/\mathcal{F}_{t_2}]\mathbf{1}_{t_2\leq t_1}] + \Esp_{U_i}[\Esp[\Delta P^{'}_{\infty}/\mathcal{F}_{t_1}]\mathbf{1}_{ t_1 < t_2}]\\
								& = \sum_{i'} q^{+}_ {ii'} \big(\alpha_{i'}^{+} + \sum_{k=1}^{N} d^{2}_{i',k}\Esp_{U_k}[\Delta P^{'}_{\infty}]\big) +  \sum_{i'} q^{-}_ {ii'} \big(\alpha_i^{-}+ \sum_{k=1}^{N} d^{1}_{i',k}\Esp_{U_k}[\Delta P^{'}_{\infty}]\big)  \\
 & = \sum_{i'} q^{+}_ {ii'}\alpha_{i'}^{+}+q^{-}_ {ii'} \alpha_{i'}^{-} + \sum_{k=1}^{N}\sum_{i'}(q^{+}_ {ii'}d^{2}_{i',k}+q^{-}_ {ii'}d^{1}_{i',k})\Esp_{U_k}[\Delta P^{'}_{\infty}] = q_ {i} + \sum_{k=1}^{N} p_{i,k} \Esp_{U_k}[\Delta P^{'}_{\infty}]. \numberthis \label{Eq:AveragePriceMoveFormula} 
\end{align*}
Additionally, using the symmetry relation, we have $\Esp_{U_i} \left[\Delta P^{'}_{\infty}\right] = - \Esp_{U^{sym}_i} \left[ \Delta P^{'}_{\infty}\right]$. We write $\mathcal{D}$ for the set $\mathcal{D} = \{(q^1,q^2,p) ; \; p>0 \} \cup \{(q^1,q^2,p) ; \; p=0 \, , \, q^1 \geq q^2\}$. Consequently, Equation \eqref{Eq:AveragePriceMoveFormula} reads  
\begin{align*}
\Esp_{U_{i}}[\Delta P^{'}_{\infty}](1- (p_{i,i} - p_{i,i^{sym}})) & =  q_ {i} + \sum_{k \in \mathcal{D}} (p_{i,k} - p_{i,k^{sym}})\Esp_{U_i}[\Delta P^{'}_{\infty}].
\end{align*}
Given that $0 \leq p_{i,i}<1$ (the price moves with a non-zero probability when one limit is totally consumed), we have $(1- (p_{i,i} - p_{i,i^{sym}}))>0$. This proves the result of Proposition \ref{Theo:AvgPriceMoveComput}. 
\subsection{Proof of Lemma \ref{lem:AveragePriceMoveAfterExec}}
\manuallabel{subsec:CompPinft2}{.2}
For simplification, we fix the added/cancelled quantity $q=1$. To take into account non-unitary jumps, we can simply fill the zero values of the matrix $\tilde{Q}^{*}$ with the right probabilities, see Equation (\ref{eq:MatrixA}). \vspace{2mm}\\
To compute the matrix $R$, we first fix the price $P=0$ since there is no price move before the total depletion of a limit and model the order book state by $u = (q^1,q^2) $ with $q^1$ (resp. $q^2$) the best bid (resp. ask) quantity. Then, we introduce the absorbing states $U_{0,q'}$ (resp. $U_{q',0}$) with $q'\geq 1$ associated to the cases $u'=(0,q')$(resp. $u'=(q',0)$) where $Q^{1}$ (resp. $Q^{2}$) is consumed before $Q^{2}$ (resp. $Q^{1}$). We want to compute the probabilities to visit $U_{0,q'}$ and $U_{q',0}$ with $q'\geq 1$ starting from $U'_i$. To do this, we consider the infinitesimal generator $Q^{*}$ of the Markov process $(Q^{1},Q^{2})$ (the price $P=0$ is fixed) 
$$ 
Q^{*} = \left[
\begin{array}{c@{}c}
 0_{2Q^{max}} & 0  \\
 \left[
\begin{array}{ll}
Q^{1,-} & Q^{1,+} \\
\end{array}
\right] & \tilde{Q}^* \\
\end{array}
\right],
$$
where $0_{2Q^{max}}$ is the zero square matrix of size $2Q^{max}$, $Q^{1,-}$ encodes transitions to the absorbing states $U_{0,q'}$ and $Q^{1,+}$ encodes transitions to the absorbing states $U_{q',0}$ with $1 \leq q' \leq Q^{max}$, and $\tilde{Q}^*$ is similar to the infinitesimal generator of the  process $U_t$ without regeneration. The matrix $\tilde{Q}^*$ has the following form:
\begin{equation}
\tilde{Q}^* = 
\left[ 
\begin{array}{lllll}
\tilde{Q}^{*,(1)}_1 & \tilde{Q}^{*,(1)}_0 & 0 & 0 & \hdots\\
\tilde{Q}^{*,(2)}_2 & \tilde{Q}^{*,(2)}_1 & \tilde{Q}^{*,(2)}_0 & 0 & \hdots\\
\vdots & \vdots & \vdots & \vdots & \vdots\\
\hdots & 0 & 0 & \tilde{Q}^{*,(Q^{max})}_2 & \tilde{Q}^{*,(Q^{max})}_1\\
\end{array}
\right],
\label{eq:MatrixA}
\end{equation}
where $\tilde{Q}^{*,(l)}_0$ encodes transitions from level $Q^{1} = l$ to level $Q^{1} = l+1$, matrix $\tilde{Q}^{*,(l)}_2$ encodes transition from level $Q^{1}=l$ to $Q^{1} = l-1$ and matrix $\tilde{Q}^{*,(l)}_1$ encodes transitions within level $Q^{1}=l$. $Q^{max}$ is the maximum quantity available on each limit. Within each sub-matrix $\tilde{Q}^{*,(l)}_{i}$ with $i \in \{ 0,1,2\}$, $Q^{1}$ is equal to $l$ and $Q^{2}$ vary from 1 to $Q^{max}$. The sub-matrices $\tilde{Q}^{*,(l)}_i$, for $i = 0,1$, can be written
\begin{equation*}
\begin{array}{l}
\tilde{Q}^{*,(l)}_0 =  \left(
\begin{array}{lll}
\lambda^{1,+}(l,1)  &        & \\
                       & \ddots & \\ 
                       & 		& \lambda^{1,+}(l,Q^{max})
\end{array} \right) \; \text{ and } \;
\tilde{Q}^{*,(l)}_2 =  \left(
\begin{array}{lll}
\lambda^{1,-}(l,1)  &        & \\
                       & \ddots & \\ 
                       & 		& \lambda^{1,-}(l,Q^{max})
\end{array} \right) . \\
\end{array}
\end{equation*}
Let $\lambda^{*}(l,l') = \sum_{i=1}^2 \lambda^{i,+}(l,l')+\lambda^{i,-}(l,l')$ for every $ l,l' \in \{1,\cdots,Q^{max}\}$. For $l \leq Q^{max}$, we have 
$$
\tilde{Q}^{*,(l)}_1 = \left( 
\begin{array}{lllll}
-\lambda^{*}(l,1) & \lambda^{2,+}(l,1) & 0 & 0 & \hdots \\
\lambda^{2,-}(l,2) & -\lambda^{*}(l,2) & \lambda^{2,+}(l,2) & 0 & \hdots \\
\vdots &\vdots &\vdots &\vdots &\vdots \\
\hdots & 0 & 0 & \lambda^{2,-}(l,Q^{max}) & -\lambda^{*}(l,Q^{max}) \\
\end{array}
\right).
$$
Finally, we define the matrix $Q^{1,-}$ such that $Q^{1,-}_{ii} = \lambda^{1,-}(1,i)$ for $1\leq i \leq Q^{max}$ and 0 otherwise, and the matrix $Q^{1,-}$ such that $Q^{1,+}_{iQ^{max}+1,i+1} = \lambda^{2,-}(i,1)$ for $0 \leq i \leq Q^{max}-1$ and 0 otherwise. Using Theorem 3.3.1 in \cite{citeulike:1400630}, for every absorbing state $U_{i'}$ we have 
\begin{equation*}
\left\{
\begin{array}{l}
q_{U_{0,q},U_{0,q}}^{-} = 1, \quad q_{U_{0,q},U_{i'}}^{+} = 0,  \quad q_{U_{q,0},U_{i'}}^{-} = 0, \quad q_{U_{q,0},U_{q,0}}^{+} = 1, \qquad \text{ with } q\in \{1,Q^{max} \}, \\
\sum_{j} Q^{*}_{i,j}q_{j,i'}^{\pm} = 0 \qquad \forall i \in [2Q^{max}+1,(Q^{max})^2+2Q^{max}].\\
\end{array}
\right.
\end{equation*}
In the above equations, we use a slight abuse of notation and do not differentiate the state $U_{i'}$ and the index $i'$. This reads
\begin{equation}
\tilde{Q}^{*} \tilde{R} = - z^1 \quad \text{ and } \quad R = M \tilde{R},
\label{Eq:HitTime}
\end{equation}
where $\tilde{R}_{ii'} = q_{ii'}$ for every absorbing state $U_{i'}$, $z^1 =[Q^{1,-},Q^{1,+}] $, $R =[R^-,R^+]$ is the matrix such that $R^-_{ii'} = q_{ii'}^{-}$ and $R^+_{ii'} = q_{ii'}^{+}$, and $M = [M^1,M^2]$ is the matrix such that $M^1_{i,i'} = d^{1}_{i',i}$ and $M^2_{i,i'} = d^{2}_{i',i}$. The quantities $d^{1}_{i',i}$ and $d^{2}_{i',i}$ are defined in Section \ref{subsec:CompAverPMove}. The solution of this equation is unique since $\tilde{Q}^{*}$ is invertible. When queues are independent $\tilde{Q}^{*}$ is diagonalisable, see next sub-section. In the simple case of constant intensities, $\tilde{Q}^{*}$ diagonalisation is explicitely computable. 
\subsection{Diagonalisation of $\tilde{Q}^{*}$}
\manuallabel{subsec:CompPinft3}{.3}
\subsubsection{Symmetrization of $\tilde{Q}^{*}$ under the assumption of independent queues}
\manuallabel{subsec:CompPinft31}{.1}
The idea is to find a matrix $P$ such that $P^{-1}\tilde{Q}^{*}P$ is symmetric with $P = LH$. First, we consider the bloc-diagonal matrix $H = diag \{H_1,H_2,\dots H_{Q^{max}}\}$ where every $H_i$ is a square matrix of size $Q^{max}$ such that
\begin{equation*}
\left\{
\begin{array}{l}
H_1 = I,\\
H_{i+1} = H_{i}\sqrt{\tilde{Q}^{*,(i)}_2 \left(\tilde{Q}^{*,(i-1)}_0 \right)^{-1}},\qquad \forall i \geq 1.\\
\end{array}
\right.
\end{equation*}
Here $\sqrt{.}$ refers to the square root of a matrix. The existence of such a matrix in this case is trivial since $\tilde{Q}^{*,(i)}_2$ and $\tilde{Q}^{*,(i-1)}_0$ are diagonal with strictly positive coefficients. \\
Next, we consider the bloc-diagonal matrix $L = diag \{L_1,L_1,\dots L_1\}$ where $L_1$ is a diagonal matrix with diagonal coefficients $L_1(1,1)=1$ and $L_1(i+1,i+1) = L_1(i,i)\sqrt{\frac{\tilde{Q}^{*,(0)}_1(i+1,i)}{\tilde{Q}^{*,(0)}_1(i,i+1)}}$ for all $i\geq 1$. Given that queues are independent we have $\tilde{Q}^{*,(0)}_1 = \tilde{Q}^{*,(0)}_i$ for all $i\geq 1$. Finally, we note that $P^{-1}\tilde{Q}^{*}P$, with $P =LH$, is symmetric.
\subsubsection{Diagonalisation of the symmetric matrix $P^{-1}\tilde{Q}^{*}P$: constant coefficients}
In the simple case of constant coefficients, the matrix $P$ defined in Appendix \ref{sec:AveragePriceMoveAfterExec}\ref{subsec:CompPinft3}\ref{subsec:CompPinft31} satisfies 
$$
P^{-1}\tilde{Q}^{*}P = 
\left[
\begin{array}{lllll}
 A(a,b) & V & 0 & 0 & 0 \\
 V & A(a,b) & V & 0 & 0 \\
 \vdots & \vdots & \vdots & \vdots\\
 0 & 0 & 0 & V & A(a,b)\\
\end{array}
\right] \quad \text{and} \quad 
A(a,b) = 
\left(
\begin{array}{lllll}
a & b & 0 & 0 & \hdots\\
b & a & b & 0 & \hdots\\
\vdots & \vdots & \vdots & \vdots & \vdots\\
\hdots & 0 & 0 & b & a \\
\end{array}
\right),
$$
where $V = \beta I $ with $\beta >0$ and $a$ and $b$ are some fixed constants. In such framework, the eigenvalues of $P^{-1}\tilde{Q}^{*}P$ are
\begin{equation*}
\lambda^{k,j}_{a,b,\beta} = a + 2b \cos(\frac{k\pi}{n+1}) + 2 \beta \cos(\frac{j \pi}{n+1}), \qquad \forall 1\leq k,j\leq n,
\end{equation*} 
and the associated eigenspace is generated by the eigenvector $X^{k,j} = \left(v^{j}_1X^{k}, v^{j}_2 X^{k} \cdots, v^{j}_{Q^{max}} X^{k}\right)$, where $v^{j}_{.}$ satisfies
\begin{equation*}
v^j_r = \sin(r\frac{j \pi}{n+1}), \qquad \forall 1 \leq r, j\leq n,
\label{Eq:EigenVectTridiag3}
\end{equation*}
and $X^{k}$ is a vector such that
\begin{equation*} 
X^k_l = \sin(l\frac{k \pi}{n+1}), \qquad \forall 1 \leq k, l\leq n.
\label{Eq:EigenVectTridiag1sdc}
\end{equation*}

\section{Generalities about the state process $U^{\mu}_t$ and the value function}
\label{sec:RegValFctProofExecTimeVSInitState}
In this section we allow the state process $U^{\mu}_t$ to start from an initial state valued in $ \mathcal{U} = \{u \in \mathbb{R}_{+}^5\times \mathbb{R}^2; \, \sum_{i=1}^3 u_i > 0 \; , \; u_4 >0 \}$. For simplification, we also assume that jumps are of size $1$. By replacing Assumptions \ref{Assump:RegenerationSmoothness} and  \ref{Assump:ExitDynamic} below with Assumption \ref{Assump:Boundontheincomingflow}, results of this section remain valid in the case where the state process takes values in $\mathbb{N}^5\times \mathbb{R}^2$ however the values of the constants are modified. 
\subsection{Regularity of the regenerative process $U^{\mu}_t$}
The regularity of our regenerative process is not trivial. In fact,  if we consider two processes $U^{\mu}$ and $U'^{\mu}$ satisfying the same order book dynamic (see Section \ref{subsec:ForModel}) but starting from two different initial points $u_0$ and $u'_0$, as long as there is no regeneration, for every order flow trajectory, the error $||U^{\mu}-U'^{\mu}||$ is equal to the initial error $||u_0-u'_0||$. However, when one of the two processes is regenerated before the other, the regenerated one starts a new cycle from a random position and the error $||U^{\mu}-U'^{\mu}||$ is no longer bounded by $||u_0-u'_0||$. Hence, the irregularity mainly comes from the regeneration. In our case, since the regeneration law depends on the killing state, it may introduce strong irregularities. Consequently, we need an assumption to ensure that regeneration distributions are similar when exit points are close enough. We give here a result on the regularity of the state process under two general assumptions.
\begin{Assumption}[Regeneration smoothness] There exist four positive constants $K$, $q_0$, \newline $q_1 \leq 1 \vee q_0 $ and $\beta$ such that for every $u = (q^{bef},q^{a},q^{aft},q^2,i,p,p^{exec})$ and \newline$u'=(q'^{bef},q'^{a},q'^{aft},q'^2,i',p',p'^{exec}) \in \mathcal{U}$, 
\begin{equation*}
\left\{ 
\begin{array}{ll}
\displaystyle||\tilde{d}^i_{u} - \tilde{d}^i_{u'}||_{TV} \leq  K (||u-u'||_p)^p \\
\displaystyle\tilde{d}^i_{u}(u') = 0, \text{ \normalfont when } q'^{1}\leq q_0 \text{ \normalfont or } q'^2 \leq q_0\\
\displaystyle\sum_i \big(\lambda^{i,+}(u) + \lambda^{i,-}(u)\mathbf{1}_{q^{i} > q_{1}}\big) \leq \beta ,\\
\end{array}
\right.
\end{equation*}
where $q'^{1} = q'^{bef}+q'^{a}+q'^{aft}$, $||p - p'||_{TV} = \underset{\mathcal{A}\in \mathcal{F}}{\sup} |p(\mathcal{A}) - p'(\mathcal{A})|$  is the total variation norm, $||.||_p$ is the $L_p$ norm with $p\geq1$ and $\tilde{d}^i_u$ the regeneration distribution of the process $U^{\mu}_t$.
\label{Assump:RegenerationSmoothness}
\end{Assumption}
Assumption \ref{Assump:RegenerationSmoothness} is a Lipschitz type inequality to ensure that regeneration distributions are almost similar when exit states are close enough. Furthermore, we consider a boundedness assumption and a support constraint to guarantee that regenerated limits have a size higher than a fixed minimum quantity $q_0$. We also add the following assumption.  
\begin{Assumption}[Exit dynamic] $\forall u=(q^1,q^2,p)\in \mathbb{R}_+^2 \times \mathbb{R}$ and $i \in \{1,2\}$, there exists a positive constants $\beta^{-}$ such that $\forall \epsilon >0$, $\exists q_{\epsilon} > 0$,
\begin{equation*}
 \lambda^{i,-}(u)\mathbf{1}_{q^{i} \leq q_{\epsilon}}  \geq \cfrac{\beta^{-}}{\epsilon}.
\end{equation*}
\label{Assump:ExitDynamic}
\end{Assumption}
For small size queues (i.e $q^{i} \leq q_{\epsilon}$), Assumption \ref{Assump:ExitDynamic} ensures that intensities of depletion are high (i.e $\lambda^{i,-} \geq \frac{1}{\epsilon}$) while other intensities are bounded. Such assumption avoids critical situations where the order book goes far away from an exit state after being too close to it. It is also consistent with empirical evidences since a limit disappears almost instantaneously when it becomes lower than a given bound $q_\epsilon$. We have the following result proved in Appendix \ref{sec:RegValFctProofExecTimeVSInitState}\ref{subsec:Regularitystateprocess}.
\begin{theo}[Regularity of the state process] Under Assumptions \ref{Assump:RegenerationSmoothness} and \ref{Assump:ExitDynamic},
 the process $U^{\mu}_t$ satisfies
\begin{equation}
\Esp \big[ (||U^{\mu}_t - U'^{\mu}_t ||_p)^p \big]  \leq  K_0 e^{C_0T}(||U_0-U_0^{'}||_p)^p, \quad \forall U_0,U_0^{'}\in \mathcal{U}, \forall t \in [0, T],
\label{Eq:LipschitzStateProcess}
\end{equation}
where $= ||.||_p$ is the $L_p$ norm for $p \geq 1$, $U^{\mu}_t$ (resp. $U'^{\mu}_t$) is the Markov process starting from the initial state $U_0$ (resp. $U^{'}_0$), $T$ is the final time, $K_0$ and $C_0$ are constants defined in Appendix \ref{sec:RegValFctProofExecTimeVSInitState}\ref{subsec:Regularitystateprocess}. 
\label{Th:Regstateprocess}
\end{theo}
\subsection{Regularity of the value function} 
In this section, we fix $p=1$ and write $||.||_p=||.||$. Let us assume that the function $g(u) = f(\Esp_u \left[\Delta P^{\mu}_{\infty}\right])$ is Lipschitz. When the state process is valued in $\mathbb{N}^5\times \mathbb{R}^2$ we only need $g$ to be bounded which is always satisfied Assumptions \ref{Assump:InsertionBound} and \ref{Assump:RegenerationLimit}. Then, we have the following regularity properties proved in Appendix \ref{sec:RegValFctProofExecTimeVSInitState}\ref{subsec:Regularityvaluefunction}.
\begin{prop}
The value function $V$ is 
\begin{itemize}
\item Lipschitz in space:
\begin{equation}
|V_T (t,U_1) - V_T (t,U_2) | \leq A e^{C_0(T-t)}||U_1 - U_2||, \quad \forall U_1,U_2\in \mathcal{U}, \forall t \in [0, T],
\label{Eq:LipschitzValFct}
\end{equation}
with $T$ the final time, $A$ a constant defined in Appendix \ref{sec:RegValFctProofExecTimeVSInitState}\ref{subsec:Regularityvaluefunction} and $C_0$ defined in Theorem \ref{Th:Regstateprocess}.
\item Lipschitz in time:
\begin{equation}
|V_T(t,U_1) - V_T(t^{'},U_1)| \leq  L_0|t^{'} -t|, \quad \forall U_1\in \mathcal{U}, \forall t,t' \in [0, T],
\label{Eq:HolderRegValFct}
\end{equation} 
with $L_0 = cq^a + A e^{C_0(T-t)}C$ and $C$ a constant defined in Appendix \ref{sec:RegValFctProofExecTimeVSInitState}\ref{subsec:Regularitystateprocess}. 
\end{itemize}
\end{prop}
\subsection{Execution time inequalities} 
Here again we also fix $p=1$ and write $||.||_p=||.||$. We recall that $T^{t,\mu}_{Exec}$ is defined in Section \ref{subsec:ExisUniqReg} for any control $\mu$. We provide here two execution time inequalities. First, when agent's decisions are taken at a fixed frequency $\Delta^{-1}$, we have the following inequality.
\begin{prop}
Let $U_1$, $U_2$ be two initial states and $\mu^{Opti}_1$ (resp. $\mu^{Opti}_2$) the optimal strategy for the process starting from $U_1$ (resp. $U_2$). Then, we have
\begin{align*}
\Esp \left[|T^{t,\mu^{Opti}_2}_{Exec} - T^{t,\mu^{Opti}_1}_{Exec} |\right] \leq  \Delta e^{C_1(T-t)}||U_1 - U_2|| + \Delta K_1 (T-t),\quad \forall U_1,U_2\in \mathcal{U}, \forall t \in [0, T],
\end{align*}
with $ C_1 = \frac{\log(K_0)}{\Delta} + C_0$ and $K_1 = 2 Q^{max}\beta$.
\label{Lem:ExecTimeVSInitState}
\end{prop}
Proposition \ref{Lem:ExecTimeVSInitState} shows that both initial states and agent's latency $\Delta$ affect the optimal execution time. We have a second inequality when decisions are taken at any time. 
\begin{prop}
Let $U_1$, $U_2$ be two initial states and $\mu^{Opti}_1$ (resp. $\mu^{Opti}_2$) the optimal strategy for the process starting from $U_1$ (resp. $U_2$). Then, we have
\begin{align}
\Esp \left[|T^{t,\mu^{Opti}_2}_{Exec} - T^{t,\mu^{Opti}_1}_{Exec}| \right] & \leq K'_0||U_1 - U_2||,\quad \forall U_1,U_2\in \mathcal{U}, \forall t \in [0, T],
\label{Eq:OptiTimeIneq2}
\end{align}
with $K'_0 = \log(K_0) $. The constant $K_0$ is given in Theorem \ref{Th:Regstateprocess}.
\label{Lem:ExecTimeVSInitState2}
\end{prop}
Proofs of Propositions \ref{Lem:ExecTimeVSInitState} and \ref{Lem:ExecTimeVSInitState2} are given in Appendix \ref{sec:ExecTimeVSInitStateProof}.
\subsection{Proof of Theorem \ref{Th:Regstateprocess}}
\manuallabel{subsec:Regularitystateprocess}{.4}
\paragraph{Notations:} Let $U_0,\,U_0^{'}\in \mathcal{U}=\{u \in \mathbb{R}_{+}^5\times \mathbb{R}^2; \, \sum_{i=1}^3 u_i > 0 \; , \; u_4 >0 \}$, $U^{\mu}_t$ (resp. $U'^{\mu}_t$) the process starting from $U_0$ (resp. $U'_0$), $T$ the final time horizon and $\hat{Q}^{\max} = \max(Q^{\max},\tilde{Q}^{\max})$. Let $\tau_{1,-}$ (resp. $\tau'_{1,-}$) be the first time where the best bid (resp. ask) is totally consumed, $\tau_{1,+}$ (resp. $\tau'_{1,+}$) the first time where the best ask (resp. bid) is depleted and $\tau_{1} = \tau_{1,-} \wedge \tau_{1,+}$ (resp. $\tau'_{1} = \tau'_{1,-} \wedge \tau'_{1,+}$) the first regeneration time of the process $U^{\mu}_t$ (resp. $U'^{\mu}_t$). Finally, we write $\tau_2$ for $\tau_2 = \tau_1 \vee \tau'_1 $. Let $\epsilon = (||U_0-U_0^{'}||_p)^{p}>0$, there exists $0<q_\epsilon \leq q_0$ satisfying conditions of Assumption \ref{Assump:ExitDynamic}. We fix $p\geq 1$ and write $||.||_p = ||.||$.
\paragraph{Step 1:} Assume $||U_{0}-U_{0}'||> q_{\epsilon}$. Let us show that
\begin{equation}
\Esp \big[ \underset{0 \leq t \leq T}{\sup} ||U^{\mu}_t - U'^{\mu}_t||^p \big] \leq 3||U_0-U_0^{'}||^p e^{\tilde{C}T},
\label{Eq:SubRegStateProc0}
\end{equation}
with $\tilde{C} = \frac{\hat{C} \beta}{q_{\epsilon}^p}$, $ \hat{C} = \big[5(\hat{Q}^{\max})^p+ (2\tilde{P}^{max})^p + (2\tilde{P}^{max}\hat{Q}^{\max})^p\big]$ and $\beta$ is defined in Assumption \ref{Assump:ExitDynamic}.
\paragraph{Proof of Equation (\ref{Eq:SubRegStateProc0}):} We write $\delta U^{\mu}_t = U^{\mu}_t -U_0$ and $\delta U'^{\mu}_t = U'^{\mu}_t -U'_0$.  Then, we have
\begin{equation*}
||U^{\mu}_t - U'^{\mu}_t||^p  = 3^{p-1} \left( ||\delta U^{\mu}_t||^p + ||\delta U'^{\mu}_t||^p + ||U_0-U_0'||^p\right).
\end{equation*}
Since $\Esp \big[ \underset{0 \leq t \leq T}{\sup} ||\delta U^{\mu}_t||^p \big] \leq \big[5(\hat{Q}^{\max})^p+ (2\tilde{P}^{max})^p + (2\tilde{P}^{max}\hat{Q}^{\max})^p\big]\beta T $, we have
$$
\Esp \big[ \underset{0 \leq t \leq T}{\sup} ||U^{\mu}_t - U'^{\mu}_t||^p \big] \leq C T + B,
$$
with $C = 3^{p-1}\hat{C} \beta$ and $B =3^{p-1}||U_0-U_0^{'}||^p $.  In addition $||U_{0}-U_{0}'||> q_{\epsilon}$ and so $B > 3^{p-1}q_{\epsilon}^p  $ and we have the following inequality: 
\begin{equation*}
\Esp \big[ \underset{0 \leq t \leq T}{\sup} ||U^{\mu}_t - U'^{\mu}_t||^p \big] \leq C T + B \leq B e^{\tilde{C}T},
\end{equation*} 
with $\tilde{C} = \frac{\hat{C} \beta}{q_{\epsilon}^p}$. The last inequality comes from $a x + b \leq b e^{a/b_0 x} $ when $b > b_0 $.
\paragraph{Step 2:} Assume $||U_{0}-U_{0}'||\leq  q_{\epsilon}$. First, let us show that
\begin{equation}
\Esp ||U^{\mu}_{\tau_2} - U'^{\mu}_{\tau_2} ||^p \leq \kappa ||U_0-U'_0||^{p},
\label{Eq:SubRegStateProc}
\end{equation}
where $\kappa=\kappa^1\hat{C} \beta + 2 \sqrt{2} \kappa^2 RK $, $\kappa^1 =  \frac{(1+3T\beta)}{\beta^-} $,  $\kappa^2 =  \kappa^1 \hat{C}\beta+1 $, $K$ is defined in Assumption \ref{Assump:RegenerationSmoothness} and $R$ is a constant.
\paragraph{Proof of Inequality (\ref{Eq:SubRegStateProc}):} We assume that $\tau_1 > \tau'_1 $ $a.s$. The general case uses the same lines of argument. We have 
\begin{equation*} 
||U^{\mu}_{\tau_2} - U'^{\mu}_{\tau_2} ||^p = ||U^{\mu}_{\tau_1} - U'^{\mu}_{\tau_1} ||^p \leq \overbrace{||U'^{\mu}_{\tau_1} - U'^{\mu}_{\tau'_1} ||^p}^{(1)} + \overbrace{||U^{\mu}_{\tau_1} - U'^{\mu}_{\tau'_1}||^p}^{(2)}.
\end{equation*}
\begin{itemize}
\item Part $(1)$ satisfies
\begin{equation}
\Esp [ ||U'^{\mu}_{\tau_1} - U'^{\mu}_{\tau'_1}||^p] \leq \hat{C} \beta\Esp [ |\tau_1-\tau'_1|] \leq \hat{C} \beta \kappa^1 ||U_0-U'_0 ||^{p}.
\label{Eq:SubCase1InequalityRegen}
\end{equation}
In the second inequality, we use $\Esp [ |\tau_1-\tau'_1|]\leq \kappa^1 ||U_0-U'_0 ||^p $, see Lemma \ref{Lem:IneqExecTime} below.
\item For Part (2), let $X = [0,Q^{max}]^5 \times [-\tilde{P}^{max},\tilde{P}^{max}]\times \big(Q^{max}[-\tilde{P}^{max},\tilde{P}^{max}]\big)$. There exist a Borel function $s$ valued in $[-1,1]$, see Appendix \ref{subsec:ExisS} at the end of the proof, and a positive contant $R$ such that
\begin{equation*}
||x-y||^p \leq R|s(x) -s(y)|, \qquad \forall(x,y) \in X^2.
\end{equation*}
Let us denote by $d_{x}$ the regeneration distribution of $U^{\mu}_t$ (i.e $d_{x} = d^{1}_{x}$ (resp. $d_{x} = d^{2}_{x}$) when the best bid (resp. ask) is totally depleted). Finally, we denote by $\mathcal{M}$ the set of Borel functions on $X$ that take values in $[-1,1]$. In such case, we have
\begin{align*}
\Esp \big[||U^{\mu}_{\tau_1} -U'^{\mu}_{\tau'_1} ||^p \big] & \leq  R\, \Esp \big[|s(U^{\mu}_{\tau_1}) - s(U'^{\mu}_{\tau'_1})|\big] = R \;\Esp \big[ \Esp[\int [|s(x) - s(y)|] d_{U^{\mu}_{\tau_1^-}}(x) d_{U'^{\mu}_{{\tau'}_1^-}}(y)\;dxdy/\mathcal{F}_{\tau_1}]\big]  \\
			& \leq  \sqrt{2}R\; \Esp \big[\Esp[\big|\int |s(x)| d_{U^{\mu}_{\tau_1^-}}(dx) - \int |s(y)| d_{U'^{\mu}_{{\tau'}_1^-}}(dy)\big|/\mathcal{F}_{\tau_1}]\big] \\
			& \leq \sqrt{2}R\; \Esp \big[ \Esp[\big|\underset{g \in \mathcal{M}}{\sup} \left\{ \int g d_{U^{\mu}_{\tau_1^-}}(dx) - \int g d_{U'^{\mu}_{{\tau'}_1^-}}(dx)\right\}\big|/\mathcal{F}_{\tau_1}]\big] \\
			& = 2 \sqrt{2} R\; \Esp \big[\overbrace{||d_{U^{\mu}_{{\tau}_1^-}}-d_{U'^{\mu}_{{\tau'}_1^-}}||_{TV}}^{(*)} \big] \leq 2\sqrt{2}RK\; \Esp \big[||U^{\mu}_{{\tau}_1^-}-U'^{\mu}_{{\tau'}_1^-} ||^p \big].
\end{align*}
In (*) we used the total variation norm property $2||\mu-\nu||_{TV} = \underset{g \in \mathcal{M}}{\sup}\int g \mathrm{d}\mu - \int g \mathrm{d}\nu $. Given that $\Esp [ ||U^{\mu}_{{\tau}_1^-}-U'^{\mu}_{{\tau'}_1^-}||^p] \leq \kappa^2 ||U_0-U'_0 ||^p$, see Lemma  \ref{Lem:IneqExecTime2} below, we have 
\begin{equation}
\Esp [||U^{\mu}_{\tau_1} - U'^{\mu}_{\tau'_1}||^p] \leq 2\sqrt{2}RK\kappa^2 ||U_0-U'_0 ||^{p}.
\label{Eq:SubCase2InequalityRegen}
\end{equation}
\end{itemize}
By combining Inequalities (\ref{Eq:SubCase1InequalityRegen}) and (\ref{Eq:SubCase2InequalityRegen}), we get the result.
\begin{lem} We have
$$
\Esp [ |\tau_1-\tau'_1|]\leq \kappa^1 ||U_0-U'_0 ||^{p},
$$
with $\kappa^1 = \frac{(1+3T\beta)}{\beta^-} $. 
\label{Lem:IneqExecTime}
\end{lem}
\paragraph{Proof of Lemma \ref{Lem:IneqExecTime}:}First, we assume that $\tau_1 > \tau'_1 $ $a.s$ and consider the following notation. For every state $u$, we write $u^{2,+}$ (resp. $u^{2,-}$) for the new state of the order book when a quantity $q=1$ is added to (resp. cancelled from) $Q^{2}$. The same reasoning holds for $Q^{1}$. We have 
\begin{align*}
\Esp [ |\tau_1-\tau'_1|] & = \Esp \big[\Esp[ |\tau_1-\tau'_1|/ \mathcal{F}_{{\tau'}^+_1}]\big] = \Esp \big[\Esp[ \hat{\tau}_1/ \mathcal{F}_{{\tau'}^+_1}]\big]= \Esp \big[\Esp[ \hat{\tau}_1/U^{\mu}_{{\tau'}_1}]\big] = \Esp \big[h(U^{\mu}_{{\tau'}_1})\big],
\end{align*}
where $\hat{\tau}_1 = \tau_1-\tau'_1$ is also the first regeneration time of $U^{\mu}_t$ but starting from the initial point $U^{\mu}_{{\tau'}_1}$ and $h(x) = \Esp[ \tau /U^{\mu}_0 = x] $ where $\tau$ is the first regeneration time of $U^{\mu}_t$ when $U^{\mu}_0 = x$. Since $U'^{\mu}_t$ is regenerated at time ${\tau'}_1$, then $Q'^{1,\mu}_{{\tau'}_1^-} \leq q$ or $Q'^{2,\mu}_{{\tau'}_1^-} \leq q$. Let us consider the case where $Q'^{1,\mu}_{{\tau'}_1^-} \leq q$. The case $Q'^{2,\mu}_{{\tau'}_1^-} \leq q$ is solved using the same arguments. Since  $||U^{\mu}_{{\tau'}_1^-}-U'^{\mu}_{{\tau'}_1^-} || = ||U^{\mu}_{0}-U'^{\mu}_{0}||$ (i.e the error is unchanged before the first regeneration), we have $Q^{1,\mu}_{{\tau'}_1}\leq ||U^{\mu}_{0}-U'^{\mu}_{0}||\leq q_\epsilon$.\\ 
We note $u_1=U^{\mu}_{{\tau'}_1}$. By considering the possible transitions of the process $U^{\mu}$, we have 
\begin{equation*}
1+\lambda^{1,+}(u_1)h(u_1^{1,+})+ \lambda^{2,-}(u_1)h(u_1^{2,-})+ \lambda^{2,+}(u_1)h(u_1^{2,+})-\lambda^{*}(u_1)h(u_1)=0,
\end{equation*}
with $\lambda^* = \sum_i \lambda^{i,+}+\lambda^{i,-}$. Using Assumption \ref{Assump:ExitDynamic} and $h(u)\leq T$, for every initial state $u$, we have 
\begin{align*}
h(u_1) \leq \frac{1+\lambda^{1,+}(u_1)h(u_1^{1,+})+ \lambda^{2,-}(u_1)h(u_1^{2,-})+ \lambda^{2,+}(u_1)h(u_1^{2,+})}{\lambda^{*}(u_1)} & \leq  \frac{(1+3T\beta)\epsilon}{\beta^-} \\
					& =  \kappa^1 ||U_0-U'_0 ||^{p},
\end{align*} 
with $\kappa^1 = \frac{(1+3T\beta)}{\beta^-} $. This proves the result.
\begin{lem} We have
$$
\Esp [ ||U^{\mu}_{{\tau}_1^-}-U'^{\mu}_{{\tau'}_1^-}||^p]\leq \kappa^2 ||U_0-U'_0 ||^{p},
$$
\label{Lem:IneqExecTime2}
with $\kappa^2 = \kappa^1 \hat{C}\beta+1$.
\end{lem}
\paragraph{Proof of Lemma \ref{Lem:IneqExecTime2}:}  By following the same methodology as for Lemma \ref{Lem:IneqExecTime}, we first note that 
\begin{align*}
\Esp [ ||U^{\mu}_{{\tau}_1^-}-U'^{\mu}_{{\tau'}_1^-}||^p] & \leq  \Esp \big[\Esp[ ||U^{\mu}_{{\tau}_1^-}-U^{\mu}_{{\tau'}_1^-}||^p/ \mathcal{F}_{{\tau'}_1}]\big] + \Esp \big[ ||U'^{\mu}_{{\tau'}_1^-}-U^{\mu}_{{\tau'}_1^-}||^p\big] \\
			& = \Esp \big[\Esp[ ||U^{\mu}_{\tau_1^-}- U^{\mu}_{{\tau'}_1^-}||^p/ U^{\mu}_{{\tau'}_1}]\big] + ||U_0-U'_0 ||^p \\ 
			& \leq \hat{C} \beta \Esp \big[\Esp[ |\hat{\tau}_1|/U^{\mu}_{{\tau'}_1}]\big] + ||U_0-U'_0 ||^p \leq \hat{C} \beta \Esp \big[ h(U^{\mu}_{{\tau'}_1})\big] + ||U_0-U'_0 ||^p,
\end{align*}
where $\hat{\tau}_1 = \tau_1-\tau'_1$ is the first regeneration time of $U^{\mu}_t$ but starting from the initial point $U^{\mu}_{{\tau'}_1}$ and $h(x) = \Esp[ \tau /U^{\mu}_0 = x] $ where $\tau$ is the first regeneration time of $U^{\mu}_t$ when $U^{\mu}_0 = x$. In the proof of Lemma \ref{Lem:IneqExecTime}, we show that $\Esp \big[\Esp[ h(U^{\mu}_{{\tau'}_1})]\big] \leq \kappa^1 ||U_0-U'_0 ||^{p}$. Hence, we have $\Esp [ ||U^{\mu}_{{\tau}_1^-}-U'^{\mu}_{{\tau'}_1^-}||^p]\leq \kappa^2 ||U_0-U'_0 ||^{p}$, with $\kappa^2 = \kappa^1 \hat{C}\beta+1$. This proves the result.
\paragraph{Step 3:} Assume $||U_{0}-U_{0}'||\leq  q_{\epsilon}$.  Let us show that
\begin{equation}
\Esp ||U^{\mu}_{t} - U^{',\mu}_{t} ||^p \leq \tilde{\kappa} e^{C'T} ||U_0-U'_0 ||^{p},
\label{Eq:Sub2RegStateProc}
\end{equation}
for any $t<\infty$, where $C' = (\beta(\kappa-1))_{+}$ and $\tilde{\kappa} = 1 \vee \kappa $.
\paragraph{Proof of Inequality (\ref{Eq:Sub2RegStateProc}):} Let $t<\infty$, we denote by $N^{Regen}_t$  (resp. $N'^{Regen}_t$) the random variable representing the number of regenerations of $U^{\mu}_t$ (resp. $U'^{\mu}_t$) before $t$. Let $t_1,\cdots,t_{N^{Regen}}$ be the regeneration times of $U^{\mu}_t$ and $t'_1,\cdots,t'_{N'^{Regen}}$ the regeneration times of $U'^{\mu}_t$. We build the sequence $\tilde{t}_n$ such that $\tilde{t}_1 = t_1 \vee t'_1$ and 
\begin{equation*}
\left\{
\begin{array}{lcl}
t^1_{n}  & = & \inf\{t_i > \tilde{t}_{n}; i \in \{1,\cdots,N^{Regen}_t\}\} \\
t^2_{n} & = & \inf\{t'_i > \tilde{t}_{n}; i \in \{1,\cdots,N'^{Regen}_t\}\} \\
\tilde{t}_{n+1} & = & t^1_{n} \vee t^2_{n}.
\end{array}
\right.
\end{equation*}
Let $\tilde{N}^{Regen}_t$ be the first index $n$ when one of the sets $\Omega_1^n = \{t_i > \tilde{t}_{n}; i \in \{1,\cdots,N^{Regen}_t\}\}$ or $\Omega_2^n = \{t'_i > \tilde{t}_{n}; i \in \{1,\cdots,N'^{Regen}_t\}\}$ is empty. We adopt the convention $\tilde{t}_{n} =t$, $\forall n>\tilde{N}^{Regen}_t$. Thus, we have
\begin{align*}
\Esp ||U^{\mu}_t - U^{',\mu}_t||^p & = \Esp\big[||U^{\mu}_{\tilde{t}_{\tilde{N}^{Regen}_t}} - U^{',\mu}_{\tilde{t}_{\tilde{N}^{Regen}_t}}||^p\big] \leq  \overbrace{\Esp \big[\kappa^{\tilde{N}^{Regen}_t}||U^{\mu}_{0} - U^{',\mu}_{0}||^p\big]}^{(**)} \leq ||U^{\mu}_{0} - U^{',\mu}_{0}||^p\Esp \big[\kappa^{\tilde{N}^{Regen}_t}\big],
\end{align*}
where (**) is obtained by using Inequality (\ref{Eq:SubRegStateProc}) and the conditional expectation argument. When $\kappa >1$, we denote by $N^{*}$ the Poisson process with constant intensity $\beta$. Thus, we have 
\begin{equation*}
\Esp \big[\kappa^{\tilde{N}^{Regen}_t}\big] \leq \Esp \big[e^{(N^{*}_t+1)log(\kappa)}\big] = \kappa e^{-\beta t(1-\kappa)} < \infty.
\end{equation*}
When $\kappa \leq 1$, we have $\kappa^{\tilde{N}^{Regen}_t} \leq 1 $ and consequently, $\Esp \big[\kappa^{\tilde{N}^{Regen}_t}\big] \leq 1$. This proves Inequality (\ref{Eq:Sub2RegStateProc}). By combining Inequalities (\ref{Eq:SubRegStateProc0}) and (\ref{Eq:Sub2RegStateProc}), we complete the proof of Theorem \ref{Th:Regstateprocess} with $K_0 = \tilde{\kappa}\vee 3$ and $C_0 = \tilde{C}\vee C' $.
\subsubsection{Existence of $s$} \manuallabel{subsec:ExisS}{E.4.1}Let us first consider the function $s^{-1} = -a\log(a)$ defined in $[0,1]$.  The function $s^{-1}$ is continuous in $[0,1]$ and bijection in the subinterval $[0,e^{-1}]$. Moreover, it is a H\"older function satisfying $||s^{-1}(a) -   s^{-1}(b)|| \leq R^1 ||a -   b||^{\frac{1}{p}}$, $\forall (a,b) \in [0,1]^2$. We denote by $s^1$ the inverse of $s^{-1}$ defined in $[0,e^{-1}]$. The function $s^1$ is continous and valued in $[0,1]$. Additionally, $s^1$ satisfies $||a-b||^p \leq R^1|s^1(a) -s^1(b)|, \forall(a,b) \in [0,e^{-1}]^2$.\\
Then, we define the normalization function $
s^2 : \begin{array}{rcl}
X  &  \rightarrow  &  [0,e^{-1}]^7 \\
(x^1,\cdots,x^7) & \rightarrow  & (s^2_1(x^1),\cdots,s^2_1(x^5),s^2_2(x^6),s^2_3(x^7))
\end{array}
$ where $s^2_1$, $s^2_2$ and $s^2_3$ are three auxiliary functions for normalization defined such as \newline$s^2_1(x) = \cfrac{x}{e^{-1}\hat{Q}^{\max}} $, $s^2_2(x) = \cfrac{x+\tilde{P}^{\max}}{e^{-1}\tilde{P}^{\max}}$ and $s^2_3(x) = \cfrac{x+\hat{Q}^{\max}\tilde{P}^{\max}}{e^{-1}\hat{Q}^{\max}\tilde{P}^{\max}}$. It is simple to see that $s^2$ satisfies $||x -y||^p \leq R^2 ||s^2(x) - s^2(x)||^p$, $\forall(x,y)\in X^2, $
with $R^2 = \min(\hat{Q}^{\max}, \tilde{P}^{\max},\hat{Q}^{\max}\tilde{P}^{\max})$. Finally, we define $s(x) = \sum_{i=1}^7 |s^1(s^2(x^i))|$. Using the inequality \newline$
\sum_{i=1}^N |x^i - y^i| \leq \sqrt{2}| \sum_{i=1}^N |x^i| - \sum_{i=1}^N|y^i||
$, we have $||x-y||^p \leq R^2R^1\sqrt{2}|s(x) -s(y)|, \forall(x,y) \in X^2$.
\subsection{Regularity of the value function}
\manuallabel{subsec:Regularityvaluefunction}{.5}
\paragraph{Proof of Inequality (\ref{Eq:LipschitzValFct}):} We write $U^{1,\mu}_t$ (resp. $U^{2,\mu}_t$) for the process such that $U^{1,\mu}_t =  U_1$ (resp. $U^{2,\mu}_t = U_2$). Using the fact that $g$ is Lipschitz and Inequality (\ref{Eq:LipschitzStateProcess}), we have
\begin{align*}
|V_T(t,U_1) - V_T(t,U_2)| & \leq \underset{\mu}{\sup}\, \Esp \left[|g(U^{1,\mu}_{T^{t,\mu}_{Exec_1}}) - g(U^{2,\mu}_{T^{t,\mu}_{Exec_2}})| + cq^a |T^{t,\mu}_{Exec_1} - T^{t,\mu}_{Exec_2}|\right]\\
                          & \leq \underset{\mu}{\sup}\, \Esp \bigg[|g(U^{1,\mu}_{T^{t,\mu}_{Exec_1}}) - g(U^{2,\mu}_{T^{t,\mu}_{Exec_1}})| + |g(U^2_{T^{t,\mu}_{Exec_1}}) - g(U^2_{T^{t,\mu}_{Exec_2}})| + cq^a |T^{t,\mu}_{Exec_1} - T^{t,\mu}_{Exec_2}|\bigg] \\
                          & \leq \underset{\mu}{\sup}\, \Esp \left[g_{[Lip]}K_0e^{C_0(T-t)}||U_1 - U_2|| + g_{[Lip]}C|T^{t,\mu}_{Exec_2} - T^{t,\mu}_{Exec_1}| + cq^a |T^{t,\mu}_{Exec_2} - T^{t,\mu}_{Exec_1}|  \right] \\
                          & \leq ||U_1 - U_2||\big(g_{[Lip]}K_0e^{C_0(T-t)}+ K'_0\underbar{C}\big) \leq Ae^{C_0(T-t)}||U_1 - U_2||,
\end{align*}
where $\underbar{C} = g_{[Lip]}C+cq^a$, $C$ is a constant, $K'_0 = \log(K_0) $, $\mu^{Opti}_1$ (resp. $\mu^{Opti}_2$) is the optimal control when $U_1$ (resp. $U_2$) is the starting point and $A = g_{[Lip]}K_0 + K'_0\underbar{C} $. In the penultimate inequality, we use Inequality (\ref{Eq:OptiTimeIneq2}) to complete the proof.
\paragraph{Proof of Inequality (\ref{Eq:HolderRegValFct}):} Inequality (\ref{Eq:HolderRegValFct}) is proved using the dynamic programming principle and Inequality (\ref{Eq:LipschitzValFct}). 
\subsection{Proofs of Propositions \ref{Lem:ExecTimeVSInitState} and \ref{Lem:ExecTimeVSInitState2}}
\manuallabel{sec:ExecTimeVSInitStateProof}{E.6}
\paragraph{Proof of Proposition \ref{Lem:ExecTimeVSInitState}:} We fix $\Delta >0$ and prove the result by recurrence on $n\geq 0$ for every $T \in [0,n\Delta]$.
\begin{itemize}
\item \emph{Initialisation :} Case n = 0, in this case $T^{t,\mu^{Opti}_2}_{Exec} = T^{t,\mu^{Opti}_1}_{Exec} = 0$.
\item \emph{Iteration : } Let us assume the result true for $T \in [0,n\Delta]$. Let $ T \in [0,(n+1)\Delta]$. When $ T \in [0,n\Delta]$, the result is true using the recurrence assumption. When $ T \in (n\Delta,(n+1)\Delta]$, we can write 
\begin{equation*}
T^{t,\mu^{Opti}_2}_{Exec}= T^{t,\mu^{Opti}_2}_{Exec} \mathbf{1}_{T^{t,\mu^{Opti}_2}_{Exec} \leq \Delta}  +  (T^{t,\mu^{Opti}_2}_{Exec} - \Delta )\mathbf{1}_{T^{t,\mu^{Opti}_2}_{Exec} \geq \Delta} + \Delta \mathbf{1}_{T^{t,\mu^{Opti}_2}_{Exec} \geq \Delta} .
\end{equation*}
Let $\tilde{U}^{\tilde{\mu}}$ be the process following the same dynamic as $U^{\mu}$ but with initial value $U^{\mu}_{\Delta}$ and ending at $T-\Delta$ with a control $\tilde{\mu}_t = \mu_{t+\Delta}$. Then, we have classically $T^{\tilde{\mu}}_{Exec}  =  T^{\mu,\Delta}_{Exec} - \Delta$ and $V(\Delta,u) = V^{T-\Delta}(0,u)$.
Thus, we can write  
\begin{align*}
\Esp \big[|T^{t,\mu^{Opti}_2}_{Exec} - T^{t,\mu^{Opti}_1}_{Exec} |\big]  & = \overbrace{\Esp \big[|T^{t,\mu^{Opti}_2}_{Exec} \mathbf{1}_{T^{t,\mu^{Opti}_2}_{Exec} \leq \Delta} -T^{t,\mu^{Opti}_1}_{Exec} \mathbf{1}_{T^{t,\mu^{Opti}_1}_{Exec} \leq \Delta} |\big]}^{(1)} +
\overbrace{\Esp \big[ |T^{t,\tilde{\mu}^{Opti}_2}_{Exec} - T^{t,\tilde{\mu}^{Opti}_1}_{Exec}|\big]}^{(2)} \\
                  & \qquad  + \Delta \overbrace{\Esp \big[|\mathbf{1}_{T^{t,\mu^{Opti}_2}_{Exec} \geq \Delta} - \mathbf{1}_{T^{t,\mu^{Opti}_1}_{Exec} \geq \Delta }|\big]}^{(3)}.
\end{align*}  
\begin{itemize}[leftmargin=*]
\item For Part (1), we have
\begin{align*}
\Esp \big[|T^{t,\mu^{Opti}_2}_{Exec} \mathbf{1}_{T^{t,\mu^{Opti}_2}_{Exec} \leq \Delta} -T^{t,\mu^{Opti}_1}_{Exec} \mathbf{1}_{T^{t,\mu^{Opti}_1,t}_{Exec} \leq \Delta} |\big] & \leq \Esp \big[|T^{t,\mu^{Opti}_2}_{Exec} \mathbf{1}_{T^{t,\mu^{Opti}_2}_{Exec} \leq \Delta}| \big] + \Esp \big[|T^{t,\mu^{Opti}_1}_{Exec} \mathbf{1}_{T^{t,\mu^{Opti}_1}_{Exec} \leq \Delta}| \big] \\
    & \leq \Delta \big(\Esp \big[\mathbf{1}_{T^{\mu^{Opti}_2}_{Exec} \leq \Delta} \big] +\Esp \big[\mathbf{1}_{T^{\mu^{Opti}_2}_{Exec} \leq \Delta} \big]\big) \leq \Delta^2 2 \beta. 
\end{align*}
\item For Part (2), using the recurrence assumption and Inequality (\ref{Eq:LipschitzStateProcess}), we have
\begin{align*}
\Esp \big[ (T^{\tilde{\mu}^{Opti}_2}_{Exec} - T^{\tilde{\mu}^{Opti}_1}_{Exec})\big] & \leq \Delta e^{C_1(T-(t+\Delta))} \Esp \left[ ||U^1_\Delta - U^2_\Delta|| \right] + K_1 \Delta (T-t-\Delta)\\ 							  & \leq  \Delta K_0 e^{C_1(T-(t+\Delta))+C_0\Delta}||U_1-U_2||+K_1 \Delta (T-t-\Delta) . 
\end{align*}
\item For Part (3), using the same arguments of Part (2), we have
\begin{align*}
\Delta \Esp \big[|\mathbf{1}_{T^{t,\mu^{Opti}_2}_{Exec} \geq \Delta} - \mathbf{1}_{T^{t,\mu^{Opti}_1}_{Exec} \geq \Delta }|\big] \leq \Delta^2 2\beta .
\end{align*}
\end{itemize}
Finally, since $C_1 = \frac{\log(K_0)}{\Delta}+C_0 $, we have
\begin{align*}
\Esp \big[T^{t,\mu^{Opti}_2}_{Exec} - T^{t,\mu^{Opti}_1}_{Exec} \big] & \leq 2\Delta^2 Q^{max}\beta  +  K_0 \Delta e^{C_1(T-(t+\Delta))+C_0\Delta}||U_1-U_2|| + K_1 \Delta (T-t-\Delta) \\
		& \leq   K_1 \Delta (T-t) +  K_0 \Delta e^{C_1(T-(t+\Delta))+C_0\Delta}||U_1-U_2|| \\
        & = \Delta \big( K_1  (T-t) +  e^{C_1(T-t)}||U_1-U_2||\big).
\end{align*}
\end{itemize}
\paragraph{Proof of Proposition \ref{Lem:ExecTimeVSInitState2}:} Using Proposition \ref{Lem:ExecTimeVSInitState}, we have  
\begin{align*}
\Esp \big[T^{\mu^{Opti}_2,t}_{Exec} - T^{\mu^{Opti}_1,t}_{Exec} \big] & \leq \Delta (||U_1 - U_2||e^{C_1(T-t)} + K_1  (T-t)) \\
        &  \underset{0}{\sim} C_0\Delta ||U_1 - U_2|| + \frac{\log(K_0)}{\Delta}\Delta ||U_1 - U_2|| + K_1 \Delta (T-t) \\
        & \underset{0}{\sim} K'_0||U_1 - U_2||.
\end{align*}
\section{Resolution of the optimal control problem}
\label{sec:OptiProDynEq}
\subsection{Proof of Theorem \ref{Eq:EqDynProgPrin} }
First, let us assume that the time derivative $\partial_t V$ is continuous is each sub-interval $(k\Delta, (k+1)\Delta)$. Then, we can show classically that $V$ satisfies the equations of Theorem \ref{Eq:EqDynProgPrin} by applying It$\bar{\text{o}}$'s formula. Thus, it suffices to exhibit a solution and use a verification argument to conclude.
 Let us exhibit a solution $V$ by solving equations of Theorem \ref{Eq:EqDynProgPrin} step by step 
\begin{itemize}
\item \emph{Step 1 - Initialisation:} Since we know the value of $V$ at time $T$ and $V$ satisfies
\begin{equation}
0 = -cq^a\mathbf{1} + \tilde{g} + \mathcal{A}V  , \, \forall t \in (k_1\Delta,T],
\label{Eq:SolveEqFinalTime}
\end{equation}
where $k_1 = \lfloor \frac{T}{\Delta} \rfloor $, $k_1\Delta$ the first decision time and the vector $\tilde{g}$ encodes the execution time effect. Indeed, for every $\tilde{q} = (q^{bef},q^a,q^{aft},q^2, i)\in \mathbb{N}^5$, $q^1 = q^{bef}+	q^a+q^{aft}$, $u = (q^1,q^2,p)$, $z =(p,p^{exec})\in \mathbb{R}^2$, $\tilde{q}' = (q'^{bef},q'^a,q'^{aft},q'^2, i')\in \mathbb{N}^5$, $q'^1 = q'^{bef}+	q'^a+q'^{aft}$, $u' = (q'^1,q'^2,p')$ and $z' =(p',p'^{exec})\in \mathbb{R}^2$, we have $\tilde{g} = \sum_{n\geq 0} \tilde{g}_n$ where $\tilde{g}_n$ is defined such that
\begin{itemize}
\item When $i \ne 0$ and $ q^{bef}+q^a \leq n < q^1$
\begin{equation*}
\tilde{g}_n(\tilde{q},z) =  \lambda^{1,-}_m(u,n) g(\tilde{q'},z'),
\end{equation*}
with $z' = z$ and $\tilde{q}'$ such that  $q'^{bef} =  0$, $
q'^{a} = 0$, $q'^{aft} = q^{1}-n$, $q'^{2} = q^{2}$ and $i' = 0$.\\
\item When $i \ne 0$ and $ n \geq  q^1$
\begin{equation*}
\tilde{g}_n(\tilde{q},z) = \lambda^{1,-}_m(u,n) \sum d^{1,-}_{(q,p),(q',p')}g(\tilde{q'},z').
\end{equation*}
with $z'$ and $\tilde{q}'$ such that  $
q'^{bef} =q'^1$, $q'^{a} = 0$, $q'^{aft} =  0$ and $i'= 0$ and $p'^{exec} =p^{exec} + q^{a}(p- \frac{\psi}{2})$, where $q'^1$, $q'^2$ and $p'$ are fixed by the regeneration distribution.
\item In the remaining cases, we have $\tilde{g}_n(\tilde{q},z) = 0$.
\end{itemize}
We know explicitely the solution of (\ref{Eq:SolveEqFinalTime})
\begin{equation*}
V_t  = e^{(T-t)Q}g+ (T-t)\big[-cq^a\mathbf{1}+ \tilde{g}\big]  , \, \forall t \in (k_1\Delta,T],
\end{equation*}
where $g$ is the vector such that $g_i = g(U_i)$ and $k_1 = \lfloor \frac{T}{\Delta} \rfloor $. 
\item \emph{Step 2 - Iteration:} At time $k_1 \Delta$, the agent can take a decision. So he compares expressions of Equation (\ref{Eq:EqDynProgBetStep0}) and takes the maximum. When the optimal control is market the agent stops the execution otherwise he reiterates Step 1 with new initial values.
\end{itemize}
Since the exhibited solution satisfies the required regularity of $\partial_t V$, we conclude with a verification theorem as in Theorem 4.1 in \cite{oksendal2005applied}.
\subsection{Proof of Theorem \ref{lem:EqContDynProg}}
Let $G = ([0,\underline{Q}^{\max}])^5\times [-\tilde{P}^{\max},\tilde{P}^{\max}]\times [-\tilde{I}^{\max},\tilde{I}^{\max}]$, $G_0 = ([0,\underline{Q}^{\max}])^4\times \{0\} \times [-\tilde{P}^{\max},\tilde{P}^{\max}]\times [-\tilde{I}^{\max},\tilde{I}^{\max}]$, and $g$ a Lipschitz function representing the final constraint. We denote by $\underline{Q}^{\max} = \max(Q^{\max},\tilde{Q}^{max}) $ and $\tilde{I}^{\max} = \tilde{P}^{\max}\underline{Q}^{\max}$. Equations satisfied by $V$ can be formally derived by assuming that $V$ is smooth and using the dynamic programming principle
\begin{equation}
\left\{ 
\begin{array}{ll}
\max \bigg( \partial_t v + \mathcal{Q} v,\mathcal{K}^l v - v, \mathcal{K}^c v - v, g - v\bigg) = 0 & \text{ on } [0,T)\times G, \\
v = g & \text{ on }  [0,T)\times G_0,   \\
v(T,.) = g & \text{ on }  G,  \\
\end{array}
\right.
\label{Eq:ViscSolEq}
\end{equation}
with $ \mathcal{Q}f(u) = \int f(u') - f(u) \mathrm{d}Q(u';u)$ the infinitesimal generator of the state process, $\mathcal{K}^{r} f(u) = \int f(u') \mathrm{d}k^r(u';u) $ for every continuous and bounded function $f$, state $u$ and control $r \in \{l,c\}$. Since a control $r$ may lead to several states, we write $k^r(u';u)$ for the probability to reach the state $u'$ starting from $u$ after taking the decision $r$. 
\paragraph{Existence, uniqueness of the solution:} Uniqueness of the solution comes from a standard comparison principle using the same arguments as in \cite[Theorem 2.2]{biswas2010viscosity}. Existence of the solution can also be derived following \cite[Theorem 2.3]{biswas2010viscosity}.
\paragraph{Regularity of the solution:} Let us show that $\partial_t V$ is continuous except on the boundary of $\{V = g \}$. We denote by $V$ the continuous and Lipschitz viscosity solution of (\ref{Eq:ViscSolEq}).\\
Let $r$ be the control which modifies the agent's state when it exists. Let $O$ be the open set $O =\{V > \max(\mathcal{K}^r V, g)\} \cup \{(t,u); \, V>g, \, k^r(u;u) = 1\}$. On $O$, we have $\partial_t V = - \mathcal{Q} V$ in the viscosity sense. Hence, by considering a sequence of smooth functions converging uniformly towards $V$, we have $\partial_t V$ is continuous on  $O$, see \cite[Corollary 5.6]{bouchard2011weak} for a close construction.\\
Let $O_1 = \{\mathcal{K}^r V = V, \, V>g\}$ and $\overset{\circ}{O}_1$ its interior assumed non-empty, otherwise there is nothing to prove. Since $V$ is Lipschitz, $\partial_t V$ is essentially bounded.  To show that $\partial_t V$ is uniquely defined on $\overset{\circ}{O}_1$, we assume the opposite and consider a point $x_0 = (t_0,u_0)$ where $\partial_t V$ admits two possible values. We have
\begin{equation*}
V(x_0) = \mathcal{K}^r V = \int V(t_0,u') \mathrm{d}k^r(u';u_0).
\end{equation*}
There exists at least one $u'_0$ satisfying $k^r(u'_0;u_0)> 0$ and $(t_0,u'_0) \in O$. To see this, let us take \newline$u'_0 = \argmax\{V(t_0,u'),\, k^r(u';u_0) >0 \}$. Since $V(t_0,u'_0) \geq V(x_0) > g$, we have $(t_0,u'_0) \notin \{V=g\} $. If $ (t_0,u'_0) \in O$, it is exactly the needed result. If $(t_0,u'_0) \notin O$, then $V(t_0,u'_0) = \mathcal{K}^r V = \int V(t_0,u') \mathrm{d}k^r(u';u'_0) \leq  V(t_0,u'_0)$. Hence, the only possibility is $k^r(u'_0;u'_0) =1$ which provides the needed contradiction.\\
Since $u'_0 \in O$, then $\partial_t V(t_0,u'_0)$ is uniquely defined and $\partial_t V$ is continuous around $(t_0,u'_0)$. Hence, the function $\partial_t \tilde{V}$, with $\tilde{V} = V - k^r(u'_0;.) V(.,u'_0) $, is not uniquely defined in $x_0$ and $\tilde{V}$ satisfies 
\begin{equation*}
\tilde{V}(x_0) = \tilde{\mathcal{K}}^r V = \sum_{u', u' \ne u'_0} V(t_0,u') k^r(u';u_0) .
\end{equation*}
Since the sum in the above equation is finite, we can apply the same arguments as before several times to find that the null function is not uniquely defined which provides the needed contradiction. Hence $\partial_t V$ is uniquely defined on $\overset{\circ}{O}_1$. Furthermore, since $\partial_t V$ is continuous on $O$, we can prove by contradiction and using the same arguments that $\partial_t V$ is continuous on $\overset{\circ}{O}_1$ and thus on $\bar{O}_1$.\vspace{3mm}\\
Let $O_2 = \bar{O} \cap \bar{O}_1$, where $\bar{O}_1$ is the closure of $O_1$ and $x$ be a point on $O_2$. Thus, $x$ is the limit point of $(x_n)_{n\geq 0}$ and $(x_n^1)_{n\geq 0}$, such that $(x_n)_{n} \in O$ and $(x_n^1)_{n}\in O_1$. Let $l$ (resp. $l_1$) be the limit value of $\underset{n \rightarrow \infty}{\lim} \partial_t V(x_n)$ (resp. $ \underset{n \rightarrow \infty}{\lim} \partial_t V(x^1_n)$). Hence, we can check
$$
l = \underset{n \rightarrow \infty}{\lim} \partial_t V(x_n) =  \underset{n \rightarrow \infty}{\lim}  QV(x_n) = QV(x) = \underset{n \rightarrow \infty}{\lim} QK^rV(x^1_n) =   K^r \underset{n \rightarrow \infty}{\lim} QV(x^1_n) = K^r \underset{n \rightarrow \infty}{\lim} \partial_t V(x^1_n) = l_1.
$$
Thus, $\partial V$ is continuous on $O_3 = \bar{O} \cup \bar{O}_1$. On the set $O_4= \{V = g \} $, $\partial_t V$ is clearly continuous since $\partial_t V = 0$.\\
Finally, we consider the set $O_5 = \partial O_4 $ and $x$ a point on $O_5$. Here again, $x$ is the limit point of $(x_n)_{n\geq 0}$ and $(x_n^1)_{n\geq 0}$, such that $(x_n)_{n} \in O_3$ and $(x_n^1)_{n}\in O_4$. Let $l$ (resp. $l_1$) be the limit value of $\underset{n \rightarrow \infty}{\lim} \partial_t V(x_n)$ (resp. $ \underset{n \rightarrow \infty}{\lim} \partial_t V(x^1_n)$). Thus, we have
$$
l = l_1  \qquad \Leftrightarrow \qquad   Qg = 0.
$$
This relation is not necessarily satisfied.
\paragraph{Conclusion:} Equation (\ref{Eq:ViscSolEq}) is satisfied almost everywhere by $V$. Since $\partial_t V$ is continuous except on the set $ O_5 = \partial \{V=g\}$, Equation (\ref{Eq:ViscSolEq}) is satisfied pointwise except on $O_5$.
\subsection{Optimal strategy}
Let $\tau^{T}_{i}:=\tau_{i}\wedge T^{\mu}_{Exec}$. Since $V$ satisfies Equation (\ref{Eq:ContEqDynProgBetStep0}), we have
\small
\begin{align*}
\Esp[g(U^{\mu}_{T^{\mu}_{Exec}})-cq^aT^{\mu}_{Exec} ] &=\Esp [V(T^{\mu}_{Exec},U^{\mu}_{T^{\mu}_{Exec}})] \\
&=V(0,U_{0})+\sum_{i \geq 0} \Esp\bigg[\int_{\tau_{i}^{T}}^{\tau_{i+1}^{T}} \left[\mathcal{A} V(s,U^{\mu}_{s}) - cq^a\right] ds  + \left[V(\tau_{i},(U^{\mu}_{\tau_{i}-})^{\beta_i}) - V(\tau_{i},U^{\mu}_{\tau_{i}-})\right]  \bigg]\\
&=V(0,U_{0}).
\end{align*}
\normalsize
Since, by construction,  $\Esp\big[V(\tau_{i},(U^{\mu}_{\tau_{i}-})^{\beta_i}) - V(\tau_{i},U^{\mu}_{\tau_{i}-})\big] =0$, and $\mathcal{A} V(.,U) - cq^a=0$, this shows that this policy satisfies $\Esp[g(U^{\mu}_{T^{\mu}_{Exec}})-cq^aT^{\mu}_{Exec} ]=V(0,U_{0})$, and is therefore  optimal, by definition of $V(0,U_{0})$. 

\section{Proof of Theorem \ref{lem:ConvergenceDiscreteFrameRes} }
\label{sec:ProofConvergenceDiscreteFrameRes}
\subsection{Proof of Inequality (\ref{Eq:ErrEstimPartCont})} Let us fix $\Delta $ and show the result by recurrence on $n$ for every $T \in [0,n\Delta]$.\\
\textbf{Initialisation:} in this case we have $V^{'} =V = g $.\\
\textbf{Iteration:} let us assume the result true for $n$. Let $T \in [0,(n+1)\Delta)$.
\begin{itemize}
\item When  $T \in [0,n\Delta]$: the result is true using the recurrence assumption.
\item When  $T \in (n\Delta,(n+1)\Delta]$: let $t \in [0,T]$. When $t \in (\Delta,T]$, the result is true by using $V(t,U) = V_{T-t}(0,U) $, $ \tilde{V}^\Delta(t,U) = \tilde{V}^\Delta_{T-t}(0,U)$ and the recurrence hypothesis. 
Let us take $t \in[0,\Delta,T)$. Using the dynamic programming principle, we have
\begin{align*}
| \tilde{V}^\Delta(t,u) - V(t,u)| & \leq  \bigg|\underset{\mu}{\sup} \, \Esp \left[ cq^a ([\tilde{T}^{t,\mu}_{Exec}-t]\mathbf{1}_{\tilde{T}^{t,\mu}_{Exec}\leq t+\Delta} - [T^{t,\mu}_{Exec}-t]\mathbf{1}_{T^{t,\mu}_{Exec}\leq t+\Delta}) \right. \\
						 & \left. \qquad + \, cq^a \Delta (\mathbf{1}_{\tilde{T}^{t,\mu}_{Exec}>t+ \Delta} - \mathbf{1}_{T^{t,\mu}_{Exec}> t+\Delta}) + \big(\tilde{V}^\Delta_{T-t}(\Delta,\tilde{U}^{\mu,\Delta}_{\Delta}) - V_{T-t}(\Delta,U^{\mu}_{\Delta})\big) \right]\bigg|	. 
\end{align*}
\begin{itemize}[leftmargin=*]
\item First we have
\begin{equation*}
\Esp \big[ |(\tilde{T}^{t,\mu}_{Exec}-t)\mathbf{1}_{\tilde{T}^{t,\mu}_{Exec}\leq t+\Delta} - (T^{t,\mu}_{Exec}-t)\mathbf{1}_{T^{t,\mu}_{Exec}\leq t+\Delta} |\big] \leq \Delta \Esp \big[\mathbf{1}_{\tilde{T}^{t,\mu}_{Exec}\leq t+\Delta} +  \mathbf{1}_{T^{t,\mu}_{Exec}\leq t+\Delta}\big] \leq \Delta^2 2H .
\end{equation*}
\item Second, using (\ref{Eq:DiscProbTransRel}), we have
\begin{align*}
\big|\Esp \big[\big(\tilde{V}^\Delta_{T-t}(\Delta,\tilde{U}^{\mu,\Delta}_{\Delta}) - V_{T-t}(\Delta,U^{\mu}_{\Delta})\big) \big]\big|  &  = \big|\sum_{u'} \big[P_{u,u'} \tilde{V}^\Delta_{T-t}(\Delta,u') - \mathbb{P}\big[ U_{\Delta} = u'| U_{0}=u\big]V_{T-t}(\Delta,u')\big]\big| \\
 										& \leq  \sum_{u'} \big[P_{u,u'} \big|(\tilde{V}^\Delta_{T-t}(\Delta,u') - V_{T-t}(\Delta,u')\big| \big] \\
 										& \leq R(T-t-\Delta)\Delta. 
\end{align*}
\normalsize
\item Finally, we have 
\begin{align*}
cq^a \Delta \Esp \big[ |\mathbf{1}_{\tilde{T}^{t,\mu}_{Exec}>t+ \Delta} - \mathbf{1}_{T^{t,\mu}_{Exec}> t+\Delta} |\big] & \leq cq^a \Delta \Esp \big[|\mathbf{1}_{T^{t,\mu}_{Exec}\leq t+\Delta} +  \mathbf{1}_{\tilde{T}^{t,\mu}_{Exec}\leq t+\Delta}|\big] \leq cq^a \Delta^2 2 H.
\end{align*}
\end{itemize}
By combining above inequalities, we conclude 
\begin{align*}
| \tilde{V}^\Delta_{T-t}(t,u) - V_{T-t}(t,u) | & \leq R (T-t-\Delta)  \Delta + R\Delta^2 \leq R (T-t) \Delta.
\end{align*}
\end{itemize}
\begin{rem}
We can prove Inequality (\ref{Eq:ErrEstimPartCont}) for the finite difference scheme (i.e $P = I + \Delta Q$) by adding an error term $C \Delta^2 $ since $e^{\Delta Q} - (I + \Delta Q) = \cfrac{\Delta^{2}}{2}Q^2 + o(\Delta^{2})$.
\label{Rem:FDiff}
\end{rem}
\subsection{Proof of Equation (\ref{Eq:ConvergenceControlDiscrete})}
Let $\mu^{Opti,\Delta}$ be the piecewise constant optimal control associated to the process $\tilde{U}^{\mu,\Delta}_t$. We say that a sequence of functions $f^n$ converges to $f$ in a stationary way when $\exists n_0 $ such that $\forall n\geq n_0$, $f^n = f $.
\paragraph{Outline of the proof:} First, we prove the existence of a subsequence $(\phi_n)_{n\geq 0}$ such that \newline$\mu^{Opti,\Delta_{\phi(n)}}(\omega) \underset{n \rightarrow \infty}{\rightarrow} \bar{\mu}(\omega)$ in a stationary way, where $\bar{\mu}(\omega)$ is a piecewise constant function.\\
Then, using $\tilde{V}^{\Delta}(t,U) \underset{\Delta \rightarrow 0}{\rightarrow} V(t,U)$ and the stationary convergence $\mu^{Opti,\Delta_{\phi(n)}}$, there exists $n_0$ such that $\forall n \geq n_0$, $\Esp[g(\tilde{U}^{\Delta_{\phi(n)},\bar{\mu}}_{\tilde{T}^{\bar{\mu}}_{Exec}})-cq^a \tilde{T}^{\bar{\mu}}_{Exec}] = \Esp[g(\tilde{U}^{\Delta_{\phi(n)},\mu^{Opti,\Delta_{\phi(n)}}}_{ \tilde{T}^{\mu^{Opti,\Delta_{\phi(n)}}}_{Exec}})-cq^a  \tilde{T}^{\mu^{Opti,\Delta_{\phi(n)}}}_{Exec}]  \underset{\Delta \rightarrow 0}{\rightarrow} V(t,U)$. Since $U^{\mu}_t$ is right continuous, $\bar{\mu}$ is optimal.

\paragraph{Proof of the stationary convergence :} First, let us prove that there exists $\epsilon > 0$  such that for every $a \in [0,T]$, we can find a subsequence $\mu^{Opti,\Delta_{\phi^a(n)}}(\omega)$ which is stationary in $[a,a+\epsilon)$. Let $a \in [0,T]$, since the space $\mathcal{C} = \{l,c,m\}$ is compact, we can extract a subsequence $\phi^{a}(n)$ such that $\mu^{Opti,\Delta_{\phi^a(n)}}(\omega)(a)$ converges towards a given limit  $\mu(\omega)(a)$. Since $\mathcal{C}$ is finite the sequence $\mu^{Opti,\Delta_{\phi^a(n)}}(\omega)(a)$ is stationary. Let $\epsilon(\omega)>0 $ be the minimum time between two consecutive jumps in $[0,T]$. Hence, $\mu^{Opti,\Delta_{\phi^a(n)}}(\omega)(a)$ is constant in $[a,a+\epsilon)$. Consequently, $\mu^{Opti,\Delta_{\phi^a(n)}}(\omega)(x) \underset{n \rightarrow \infty}{\rightarrow} \mu(\omega)(a), \; \forall x \in 	[a,a+\epsilon)$ in a stationary way. \\
Let $m_\epsilon = \lfloor \cfrac{T}{\epsilon} \rfloor $. For every $i \in \{0,\cdots,m_\epsilon \}$, there exists $\phi^{i\epsilon}$ such that \newline$\mu^{Opti,\Delta_{\phi^{i\epsilon}(n)}}(\omega)(x) \underset{n \rightarrow \infty}{\rightarrow} \mu(\omega)(i\epsilon),$ $ \forall x \in [i\epsilon,(i+1)\epsilon)$.\\
We define the piecewise constant limit function $\bar{\mu}(\omega)$ such that
$$
\bar{\mu}(\omega)(x) = \mu(\omega)(i\epsilon), \quad  \forall x \in [i\epsilon,(i+1)\epsilon)
$$
By construction, there exists $\phi(n)$ (constructed by a finite number of $\phi^{i\epsilon}(n)$ compositions), the sequence $\mu^{Opti,\Delta_{\phi(n)}}(\omega) \underset{n \rightarrow \infty}{\rightarrow}\bar{\mu}(\omega)$ in a stationary way.
\clearpage

\end{document}